\documentclass[fleqn]{2023SCGE}
\usepackage{graphicx}% Include figure files
\usepackage{dcolumn}% Align table columns on decimal point
\usepackage{bm}% bold math
\usepackage[mathlines]{lineno}% Enable numbering of text and display math
\usepackage{multirow}
\usepackage{comment}
\usepackage{enumerate}
\usepackage{color}
\usepackage{amsmath}

\allowdisplaybreaks

\usepackage[dvipsnames]{xcolor} 
\usepackage{orcidlink} % for orcid
\usepackage{ulem}
\newcommand{\Aemulus}{\texttt{Aemulus}-$\nu$ }
\newcommand{\BACCO}{\texttt{BACCO} }

\newcommand{\Kun}{\textsc{Kun} }
\newcommand{\Jiutian}{\texttt{JIUTIAN} }
\newcommand{\CLASS}{\texttt{CLASS} }
\newcommand{\N}{$N$}
\newcommand{\velocileptors}{\texttt{velocileptors} }
\newcommand{\ZeNBu}{\texttt{ZeNBu} }
\newcommand{\scikitlearn}{\texttt{scikit-learn} }

\newcommand{\rmd}{{\rm d}}
\newcommand{\hatc}{\hat{c}}

\newcommand{\bfDelta}{{\bf\Delta}}

\newcommand{\bfk}{{\bf k}}
\newcommand{\hatk}{\hat{k}}
\newcommand{\mM}{\mathcal{M}}
\newcommand{\mK}{\mathcal{K}}
\newcommand{\hatn}{\hat{n}}
\newcommand{\bfq}{{\bf q}}
\newcommand{\hatq}{\hat{q}}

\newcommand{\mO}{\mathcal{O}}

\newcommand{\bfx}{{\bf x}}
\newcommand{\hatx}{\hat{x}}

\newcommand{\haty}{\hat{y}}

\newcommand{\mPsi}{{\mathbf{\Psi}}}

\newcommand{\la}{{\langle}}
\newcommand{\ra}{{\rangle}}
\newcommand{\rmloop}{{\rm loop}}

\begin{document}

\ensubject{subject}

%%%%%%%%%%%%%%%%%%%%%%%%%%%%%%%%%%%%%%%%%%%%%%%%%%%%%%%
%%% Authors do not modify the information below
%%%
%Letter to the Editor
\ArticleType{Article}
\SpecialTopic{SPECIAL TOPIC: }%???????
\Year{2025}
\Month{January}
\Vol{66}
\No{1}
\DOI{??}
\ArtNo{000000}
\ReceiveDate{February 16, 2025}
\AcceptDate{*****}
%\OnlineDate{January 1, 2016}
%%%%%%%%%%%%%%%%%%%%%%%%%%%%%%%%%%%%%%%%%%%%%%%%%%%%%%%
%Solve the problem of large white space when inserting big figures.
\renewcommand\floatpagefraction{.9}
\renewcommand\topfraction{.9}
\renewcommand\bottomfraction{.9}
\renewcommand\textfraction{.1}
\setcounter{totalnumber}{50}
\setcounter{topnumber}{50}
\setcounter{bottomnumber}{50}

\title{CSST Cosmological Emulator III: Hybrid Lagrangian Bias Expansion Emulation of Galaxy Clustering}
%of Biased Tracer in Real Space

\author[1,2,3,4]{Shuren Zhou\orcidlink{0000-0002-7060-8236}}{}
% \email{zhoushuren@sjtu.edu.cn}
\author[1,2,3,4]{Zhao Chen\orcidlink{0000-0002-2183-9863}}{}
% \email{chyiru@sjtu.edu.cn}
\author[2,3,4]{Yu Yu\orcidlink{0000-0002-9359-7170}}{}
% \email{yuyu22@sjtu.edu.cn}

\thanks{Corresponding author (email:
~\href{zhoushuren@sjtu.edu.cn}{zhoushuren@sjtu.edu.cn};
~\href{chyiru@sjtu.edu.cn}{chyiru@sjtu.edu.cn};
~\href{yuyu22@sjtu.edu.cn}{yuyu22@sjtu.edu.cn}
)}

%%% Author information for page head.
\AuthorMark{Shuren Zhou}

%%% Authors for citation.
\AuthorCitation{Shuren Z, Zhao C, Yu Yu, et al.}

\address[1]{Tsung-Dao Lee Institute $\&$ School of Physics and Astronomy, Shanghai Jiao Tong University, Shanghai 200240, China}
\address[2]{Department of Astronomy, School of Physics and Astronomy, Shanghai Jiao Tong University, Shanghai 200240, China}
\address[3]{State Key Laboratory of Dark Matter Physics, School of Physics and Astronomy, Shanghai Jiao Tong University, Shanghai 200240, China}
\address[4]{Key Laboratory for Particle Astrophysics and Cosmology (MOE)/Shanghai Key Laboratory for Particle Physics and Cosmology, Shanghai 200240, China}

\date{\today}

% ---------------------------------------------------------------------------------------
% ---------------------------------------------------------------------------------------

\abstract{
%In modern galaxy survey, 
%\orange{YY: CSST Emu I paper had a similar title in the orignal version but Pengjie suggested us to highlight it by adding ``to k=10''. We can think about the similar improvement here.}
Galaxy clustering is an important probe in the upcoming China Space Station Telescope (CSST) survey to understand the structure growth and reveal the nature of the dark sector. However, it is a long-term challenge to model this biased tracer and connect the observable to the underlying physics. In this work, we present a hybrid Lagrangian bias expansion emulator, combining the Lagrangian bias expansion and the accurate dynamical evolution from $N$-body simulation, to predict the power spectrum of the biased tracer in real space. We employ the \texttt{Kun} simulation suite to construct the emulator, emulating across the space of 8 cosmological parameters including dynamic dark energy $w_0$, $w_a$, and total neutrino mass $\sum m_{\nu}$. The sample variance due to the finite simulation box is further reduced using the Zel'dovich variance control, and it enables the precise measurement of the Lagrangian basis spectra up to the quadratic order. The emulation of basis spectra realizes 1\% level accuracy, covering wavelength $ k \leq 1 \,{\rm Mpc}^{-1}h$ and redshift $0\leq z\leq 3$ up to the quadratic order field. To validate the emulator, we perform a joint fit to the halo auto power spectrum and the halo-matter cross power spectrum measured from 46 independent simulations. Depending on the choice of counterterm, the joint fit is unbiased up to $k_{\rm max}\simeq 0.7\,{\rm Mpc}^{-1}h$ within $1\sim 2$ percent accuracy, for all the redshift and halo mass samples. As part of the CSST cosmological emulator series, this emulator is expected to provide accurate theoretical predictions for the galaxy power spectrum in upcoming CSST survey.
}



%\keywords{}

\maketitle

%\tableofcontents
%\newpage

% ---------------------------------------------------------------------------------------
% ---------------------------------------------------------------------------------------
% ---------------------------------------------------------------------------------------
% ---------------------------------------------------------------------------------------
\begin{multicols}{2}

\section{Introduction}

Galaxy clustering is an important cosmological probe in ongoing and upcoming galaxy surveys, such as the Dark Energy Spectroscopic Instruction (DESI) survey \cite{adame2024desiV, adame2024desiVII, karim2025desi}, Roman mission \cite{akeson2019wide}, Euclid mission \cite{aussel2025euclid} and the China Space Station Telescope (CSST) survey \cite{gong2019cosmology}. It contains valuable information to understand the structure formation across the cosmic time, reveal the nature of dark matter and dark energy, and answer various fundamental questions \cite{adame2024desiV, adame2024desiVII, karim2025desi, lodha2025extended, elbers2025constraints, chaussidon2024constraining}. Combined with weak gravitational lensing from cosmic shear or cosmic microwave background (CMB) distortion, we can further break the degeneracy between the matter fluctuation amplitude and the biased nature of galaxy distribution, accessing the tight cosmological constraint \cite{chen2024analysis, kim2024atacama, sailer2024cosmological, sailer2025evolutionstructuregrowthdark, xiong2024cosmological, gong2019cosmology, luo2025photometric}. 

However, the theoretical prediction of galaxy power spectrum and its cross-correlation with other probes is always a challenging task, due to the highly non-linear matter clustering, galaxy bias, and the stochasticity resulting from the non-local astrophysical processes \cite{seljak2004large, seljak2009suppress, baldauf2013halo, desjacques2018large, zhou2023principal, zhou2024parametrization}. Halo occupation distribution (HOD) is one of the empirical frameworks used to model the complex galaxy distribution. 
%\orange{YY: to me this assumption is irrelevant to HOD; ZC: this seems to be the assumption of halo model?YY: yes, halo model is not a necessary condition of HOD. of course halo model + HOD can predicton galaxy clustering, but HOD can be directly applied on simulated halos without halo model.}
It assumes the relationship between galaxy and dark matter halo, where the galaxy is populated within the halo statistically determined by a few phenomenological parameters, particularly the halo mass. 
Another well-motivated framework, free from the assumption of galaxy-halo connection, is the bias expansion, also known as the effective field theory of large-scale structure (EFTofLSS). 
%The theoretical origin starts with the equivalence principle, implying that the only local observable in the non-relativistic limit is the second order derivative of gravitational potential, so we can 
It expands the functional dependence of the biased tracer distribution as a series of physical quantities constructed by the gravitational potential derivatives, where the coefficient is the galaxy bias, and all the non-perturbative impacts arising from small-scale processes are absorbed by a few counterterms \cite{desjacques2018large, perko2016biased, senatore2015bias, porto2014lagrangian}. 
It is shown to be a powerful framework for modeling the non-linear galaxy clustering in both Eulerian perturbation theory (EPT) and Lagrangian perturbation theory (LPT) prescription \cite{maus2024analysis, lai2024comparison, maus2025comparison, noriega2024comparing, ramirez2024full, masot2025full}, and successfully applied to the analysis of survey data for interpreting the cosmological information \cite{ivanov2020cosmological, colas2020efficient, d_Amico_2020, adame2024desiV, adame2024desiVII, pellejero2024cosmological}. 
Apart from the standard cosmology analysis, the EFTofLSS presents an advantageous approach to capture the physics feature in the galaxy power spectrum, such as primordial oscillations \cite{chen2020modeling, calderon2025primordialfeatureslighteffective}.
And it also provides a new perspective to investigate the connection between halo clustering and galaxy formation \cite{kokron2022priors, ivanov2024full,ivanov2025simulation, maus2024analysis, liu2025fastbaryonicfieldpainting}. 

An essential step in the EFTofLSS is the resummation of long-wavelength modes, especially the baryon acoustic oscillations (BAO). Compared to the traditional EPT prescription, the LPT takes advantage of natural long-wavelength perturbation resummation \cite{carlson2013convolution, wang2014analytic}, but the validity is still limited to a mild non-linear region due to the perturbation expansion of the displacement. To accurately describe the dynamical evolution in the bias expansion, Ref.~\cite{modi2020simulations} proposes to utilize the displacement obtained from cosmological $N$-body simulations to replace the perturbation displacement, and find that it significantly improves model fitting to smaller scales compared to 1-loop theory.
% with the quadratic order Lagrangian bias expansion. 
This hybrid Lagrangian bias expansion method, known as hybrid effective field theory (HEFT), is adopted to various kinds of emulator construction subsequently, and also applied to the analysis of cosmological survey data widely \cite{kokron2022priors, zennaro2022priors, zennaro2023bacco, pellejero2023bacco, kokron2021cosmology, derose2023aemulus, pellejero2024hybrid, hadzhiyska2021hefty,sailer2024cosmological,chen2024analysis, shiferaw2024uncertainties, kim2024atacama, sailer2025evolutionstructuregrowthdark}. 
The \BACCO simulation suite \cite{zennaro2022priors, zennaro2023bacco, pellejero2023bacco} employs the cosmology-rescaling algorithm to generate various cosmology realizations in the presence of massive neutrinos and $w_0w_a$ dark energy, based on which they emulate the biased tracer spectra in both real and redshift space, respectively. 
\Aemulus simulation suite \cite{kokron2021cosmology, derose2023aemulus} adopt the Zel'dovich variance control technique \cite{kokron2022accurate, derose2023precision, hadzhiyska2023mitigating} to further reduce the sample variance. They achieve precise measurement of Lagrangian basis spectra in the presence of massive neutrinos and dynamical $w$ dark energy, emulating the basis spectra at the one percent level of accuracy. 
Both emulators are validated consistently, and confirmed to fit the power spectrum and inferring the unbiased cosmological parameters up to the maximum scale $k=0.4\,{\rm Mpc}^{-1}h$ \cite{nicola2024galaxy}. 
Besides, in current surveys, HEFT becomes an appealing choice for cosmological interpretation because of its excellent performance. It is adopted to model the non-linear clustering of DESI luminous red galaxies, combined with the cross-correlation with the cosmic shear from DES \cite{chen2024analysis}, CMB lensing from ACT \& Planck PR4 \cite{sailer2024cosmological,kim2024atacama}, and cosmic magnification from DESI imaging galaxy \cite{sailer2025evolutionstructuregrowthdark}. The HEFT approach enables the joint analysis of the tomographic measurements of matter fluctuation amplitude reaching low redshift $z\sim 0.1$, and allows the utilization of non-linear spectrum up to scale $k= 0.6\,{\rm Mpc}^{-1}h$. 
Therefore, HEFT is expected to be a promising biased tracer model in the upcoming surveys.

The forthcoming CSST is designed to operate surveys featuring multi-wavelength imaging and slitless spectroscopic observation simultaneously over a ten-year timescale \cite{zhan2011consideration, zhan2021wide}. It covers approximately $17500\,\deg^2$ sky coverage, with seven imaging bands and three slitless spectroscopic bands, mainly targeted to probe the weak gravitational lensing and galaxy clustering \cite{gong2019cosmology}. It will further advance the understanding of the nature of the dark sector, large-scale tests of gravity, and other investigations on the important cosmological questions. 
To meet the scientific goals, a substantial number of cosmological simulations are required to support various kinds of purposes, including emulator training and cosmological inference \cite{han2025jiutian}. The \Kun simulation suite, consisting of 129 high-resolution simulations, is designed to predict the precise matter clustering with one percent level accuracy down to extremely non-linear scale \cite{chen2025csst}. It covers a broad eight-dimensional cosmological parameter space, including massive neutrinos and dynamical dark energy $w_0w_a$, and prepares for the construction of various accurate emulation tools for cosmological analysis. While for the galaxy clustering emulation, we need an additional model to describe how an observed galaxy sample traces the underlying dark matter. 
In this work, we utilize the HEFT to model the biased tracers in real space, and construct the power spectrum emulator based on the \Kun suite simulation. This is part of the galaxy clustering extension of CSST cosmological emulators \cite{han2025jiutian, chen2025csst}, and also the first attempt to emulate the biased tracer spectra based on \Kun simulation suite.

In the following sections, we detail the construction of the emulator and validate its accuracy and flexibility. 
%The emulation accuracy of the basis spectra is presented in Fig.~\ref{fig:LOO}. 
Further details on the perturbation theory description are provided in \ref{appendix:CLEFT-intro} and corresponding references. 
%{Throughout this paper, we refer the terminology \textit{matter} to the sum of cold dark matter and baryon, since the simulations do not actually treat neutrino as extra particle species.}
Throughout this paper, we denote the integration in real space as $\int_\bfq \equiv \int {d^3\bfq}$, and the integration in Fourier space as $\int_\bfk \equiv \int {d^3\bfk\over (2\pi)^3}$. 
%\orange{YY: ``And'' after period dot is a bad habit in oral language}And 
The $\delta^D$ is the Dirac-$\delta$ function.

% ---------------------------------------------------------------------------------------
% ---------------------------------------------------------------------------------------
% ---------------------------------------------------------------------------------------
% ---------------------------------------------------------------------------------------

\section{Bias Expansion in Lagrangian Prescription}

In this section, we introduce the fundamental principle of the hybrid Lagrangian biased expansion, and outline the theoretical spectrum required for the emulator construction.

%%%%%%%%%%%%%%%%%%%%%%%%%%%%%%%%%%%%%%%%%%%%%%%%%%%%%%%%%
%%%%%%%%%%%%%%%%%%%%%%%%%%%%%%%%%%%%%%%%%%%%%%%%%%%%%%%%%
\begin{figure*}[htb!]
\centering
\includegraphics[width=0.85\textwidth]{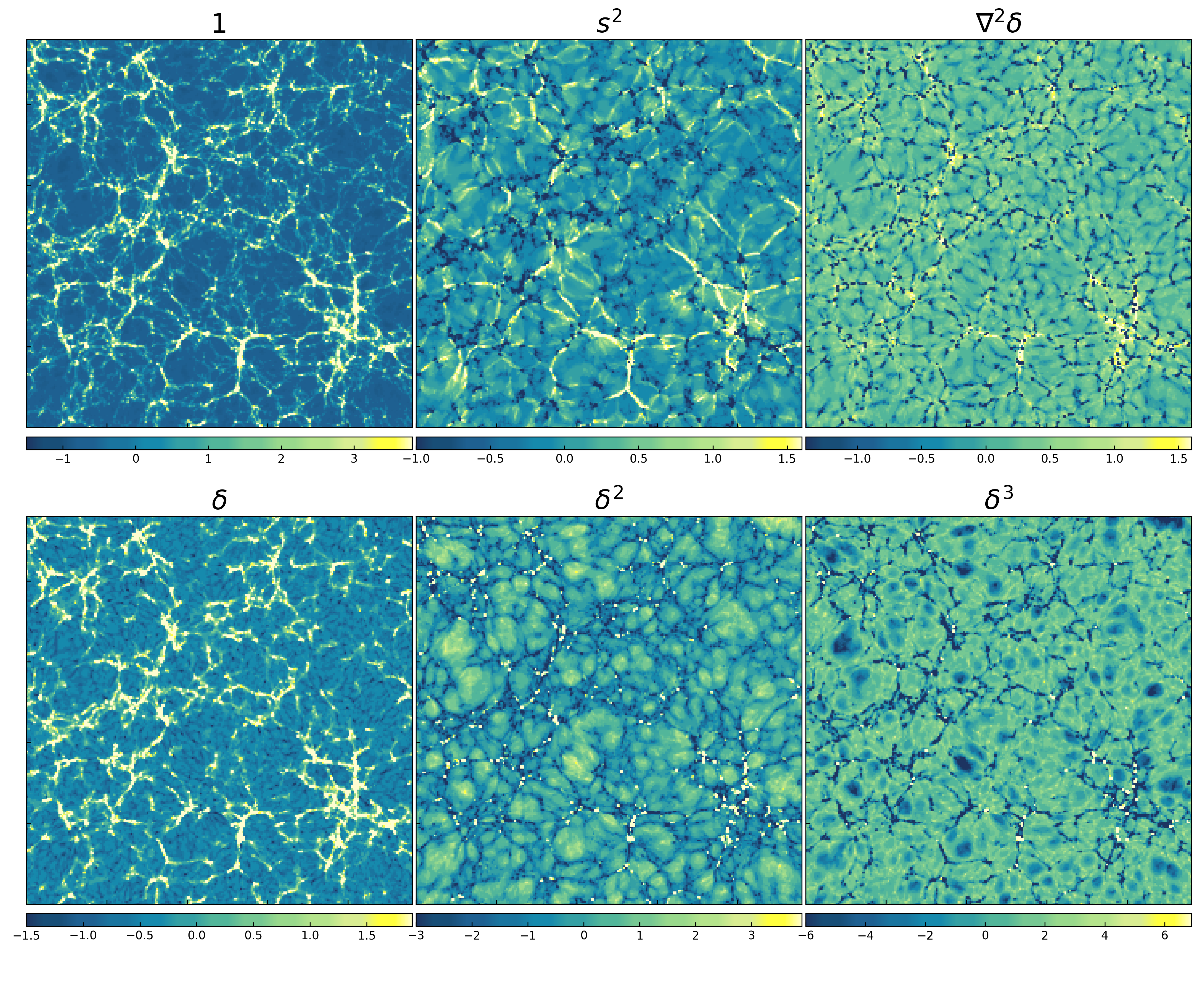}
\caption{ \label{fig:field_maps}
Lagrangian basis fields $\mO_i(\bfx, z)$ in Eulerian space at redshift $z=1$, with physical size of the region $200\times 200\,h^{-2}{\rm Mpc}^2$ and projected depth $5\,h^{-1}{\rm Mpc}$. They are advected from the initial redshift $z_{\rm ini}=127$. The constant Lagrangian field $\mO_i(\bfq)=1$ is the matter overdensity field $\mO_i(\bfx,z)=\delta_m(\bfx,z)$ in Eulerian space. Notice that all the Lagrangian basis fields are subtracted the zero-lag before assigned to the targeted redshift, e.g., $\delta^2 \rightarrow \delta^2-\sigma^2_L$. 
}
\end{figure*}
%%%%%%%%%%%%%%%%%%%%%%%%%%%%%%%%%%%%%%%%%%%%%%%%%%%%%%%%%
%%%%%%%%%%%%%%%%%%%%%%%%%%%%%%%%%%%%%%%%%%%%%%%%%%%%%%%%%

\subsection{Lagrangian Bias Expansion}

On large scales, the equivalence principle implies that the only measurable quantity in the non-relativistic limit is the second-order derivative of the gravitational potential $\partial_i\partial_j\Phi$, therefore the galaxy distribution is a general function $F[\partial_i\partial_j\Phi](\bfq)$, where we adopt the Lagrangian prescription and coordinate $\bfq$ labels the fluid element \cite{desjacques2018large}. We can expand the dependence of the galaxy overdensity in Lagrangian space as  
\begin{equation}
\begin{aligned}
&\; 1 + \delta_g(\bfq)      \\
=&\; 1 +b_1\delta(\bfq) + {1\over2}b_2\left(\delta^2(\bfq)- \la\delta^2\ra\right) + b_s \left(s^2(\bfq)- \la s^2\ra\right)      \\
&+ b_\nabla \nabla^2 \delta(\bfq)  + {b_3\over 6} \left( \delta^3(\bfq) -3\sigma^2_L\delta(\bfq) \right) + \varepsilon(\bfq)   \ ,
\end{aligned}
\label{equ:deltag_q}
\end{equation}
where $\delta(\bfq) = \nabla^2\Phi$ is the linear {cold dark matter and baryon (\textit{cb})} overdensity, and $s^2=s_{ij}s_{ij}$ is the scalar tidal field from tidal tensor $s_{ij}=\left( \partial_i\partial_j\,\nabla^{-2} - {1\over 3}\delta^D_{ij}\right)\delta$. The counterterm $\nabla^2 \delta(\bfq)$ is the leading correction of the impact arising from the non-local process on small scales. $\varepsilon(\bfq)$ is the stochastic term to capture the high order terms and residual stochastic noise.  
Additionally, we consider one of the third-order terms $\delta^3$ to investigate the possible effect of higher order terms, where all the third-order terms are degenerated in 1-loop order and only $\delta^3$ has a constant contribution on large scales \cite{schmittfull2019modeling}. Not all cubic terms are included to avoid a vast number of nuisance parameters. 
We have subtracted the zero-lag in Eq.~\ref{equ:deltag_q}, by replacement $\delta^2\rightarrow\delta^2-\la\delta^2\ra$, $s^2\rightarrow s^2-\la s^2\ra$ and $\delta^3 \rightarrow \delta^3-3\sigma^2_L\delta$, where $\la\delta^2\ra=\sigma^2_L$, $\la s^2\ra={2\over 3}\sigma^2_L$, and it is the default in the following context. 

To match the late-time observation, we advect the basis fields from Lagrangian space to Eulerian space, 
\begin{align}
\label{eau:Lagrangian-x}
\mO_i(\bfx, z) &= \int_\bfq \mO_i(\bfq)\,\delta^D(\bfx - \bfq - \mPsi(\bfq, z) )  \ ,\\
\label{eau:Lagrangian-k}
\mO_i(\bfk, z) &= \int_\bfq \mO_i(\bfq)\,e^{ - i\bfk\cdot\left( \bfq + \mPsi(\bfq, z) \right) } \ ,
\end{align}
where $\mO_i \in\{1,\delta,\delta^2,s^2,\nabla^2\delta,\delta^3\}$. The $\mO_i(\bfx, z)$ is obtained by advecting the Lagrangian field $\mO_i(\bfq)$ to Eulerian space at redshift $z$, and $\mO_i(\bfk, z)$ is the corresponding Fourier transform. 
Here $\mPsi$ is the displacement of the {\textit{cb}} particle resting at $\bfq$ in the initial time. In the analytical basis spectrum calculation, it is given by perturbation expansion $\mPsi = \mPsi^{(1)} +\mPsi^{(2)} +\mPsi^{(3)} + \cdots$, and solved order by order. In the hybrid method, the displacement is given by the accurate particle motion under gravitational evolution in the N-body simulation, as displayed in Fig.~\ref{fig:field_maps}. 
In late-time coordinate Eq.~\ref{eau:Lagrangian-x}, the galaxy distribution $n_g(\bfx,z)=\bar{n}_g\left[1+\delta_g(\bfx,z)\right]$ at redshift $z$ is given by the Lagrangian bias expansion 
\begin{align}\label{equ:bias_expansion}
\delta_g(\bfx, z) = \sum_i b_i \mO_i(\bfx, z) + \varepsilon(\bfx,z) \ .
\end{align}
The galaxy auto power spectrum and the cross power spectrum with the underlying matter field are
\begin{align} 
\label{equ:bias_expansion_Pgg}
P_{gg}(k) &= \sum_{ij} b_ib_j\, P_{ij}(k) + P_{\varepsilon\varepsilon}(k) \ ,  \\
\label{equ:bias_expansion_Pgm}
P_{mg}(k) &= \sum_{i} b_i\, P_{mi}(k) + P_{m\varepsilon}(k) \ ,
\end{align}
where $P_{ij}$ is the cross power spectrum between fields $\mO_i(\bfx, z)$ and $\mO_j(\bfx, z)$. 
{Notice that $\mO_i(\bfq)=1$ is the \textit{cb} component overdensity field $\delta_\textit{cb}(\bfx)$ in Eulerian expression, with a fixed value of bias parameter $b=1$. 
The total matter (\textit{tot}) overdensity field $\delta_m$ is the sum of \textit{cb} and neutrino components, $\delta_m(\bfx) = (1-f_\nu)\, \delta_\textit{cb}(\bfx) + f_\nu\, \delta_\nu(\bfx)$, where $f_\nu\equiv\Omega_\nu/\Omega_m$ is neutrino density fraction. 
As \Kun suite simulates single cold species \textit{cb}, we can only measure the \textit{cb} clustering directly from simulation. The massive neutrino impact is included in the homogeneous background, but not actually the neutrino perturbation. 
However, because neutrinos do not cluster as the cold species below the free-streaming scale $k_{\rm fs}\sim H(z)/v_{\rm th}$, where $v_{\rm th}$ is the thermal velocity, it is accurate to assume $r_{\textit{cb},m} \equiv P_{\textit{cb},m} /\sqrt{P_{\textit{cb},\textit{cb}}P_{mm}} \rightarrow 1$, the massive neutrinos tracing the clustering of cold dark matter and baryon \cite{lesgourgues2012neutrino, saito2008impact, agarwal2011effect, upadhye2014large, castorina2015demnuni, heitmann2016mira, chen2022cosmological}. The deviation of the assumption is mere $r_{\textit{cb},m}\simeq 1 + \mO(f_\nu^2)$. Thus, we can convert \textit{cb} field into \textit{tot} field by a transfer function $\delta_m(\bfk) = T_{m,\textit{cb}}(k)\, \delta_\textit{cb}(\bfk)$, with
\begin{align} 
\label{equ:tot-cb-convert-1}
T_{m,\textit{cb}}(k) &\equiv  \sqrt{P_{mm}(k) / P_{\textit{cb},\textit{cb}}(k) }   \quad ,
\end{align} 
where, 
\begin{align} 
\label{equ:tot-cb-convert-2}
% P_{\textit{cb},\textit{cb}}(k) &=& P_{11}(k)  \quad,  \\
P_{mm}(k) &=  \left[ (1 -f_\nu) \sqrt{P_{\textit{cb},\textit{cb}}(k)}  + f_\nu \sqrt{P_{\nu\nu}(k)} \right]^2    \;.
\end{align} 
Here $T_{m,\textit{cb}}$ has been efficiently emulated in our previous work \cite{chen2025csst}. The cross power spectrum of \textit{tot} field is obtained from \textit{cb} field directly, 
\begin{align} 
\label{equ:tot-cb-convert-3}
P_{mi}(k) \,=\, T_{m,\textit{cb}} P_{1i}(k)   \;,\quad 
P_{mm}(k) \,=\, T_{m,\textit{cb}}^2 P_{11}(k)   \;.
\end{align} 
Therefore, in the following context, we are not necessary to distinguish between \textit{cb} and \textit{tot} fields, since they are interconvertible. 
}

In the auto power spectrum, the leading order approximation for the stochastic contribution is constant, $P_{\varepsilon\varepsilon} = \alpha_0 /\bar{n}_g$ with a free parameter $\alpha_0$ accounting for the non-Poisson effect in the galaxy clustering. While for the cross power spectrum, the stochastic term $P_{m\varepsilon}$ in the cross power spectrum is always neglected, with $P_{m\varepsilon}=0$. We explicitly write it down here to demonstrate that, except for stochasticity, $\varepsilon$ also accounts for any higher order clustering correlated with the underlying $\delta(\bfx)$ \cite{cabass2020likelihood}. In brief, in the prediction of galaxy clustering, we need the additional bias and stochastic parameter to describe the galaxy properties, $\{b_1, b_2, b_s, b_\nabla, b_3, \cdots, \alpha_0, \cdots\}$, where the higher order stochastic parameters $\alpha_2, \alpha_4, \cdots$ are allowed when considering scale-dependent stochasticity. 
Our fiducial choice is the quadratic order bias expansion with constant stochastic noise amplitude, $\{b_1, b_2, b_s, b_\nabla, \alpha_0\}$, and alternately, the cubic order $b_3$ term is included to investigate the potential impact.

% ---------------------------------------------------------------------------------------
% ---------------------------------------------------------------------------------------

%%%%%%%%%%%%%%%%%%%%%%%%%%%%%%%%%%%%%%%%%%%%%%%%%%%%%%%%%
%%%%%%%%%%%%%%%%%%%%%%%%%%%%%%%%%%%%%%%%%%%%%%%%%%%%%%%%%
\begin{figure*}[htb!]
\includegraphics[width=0.99\textwidth]{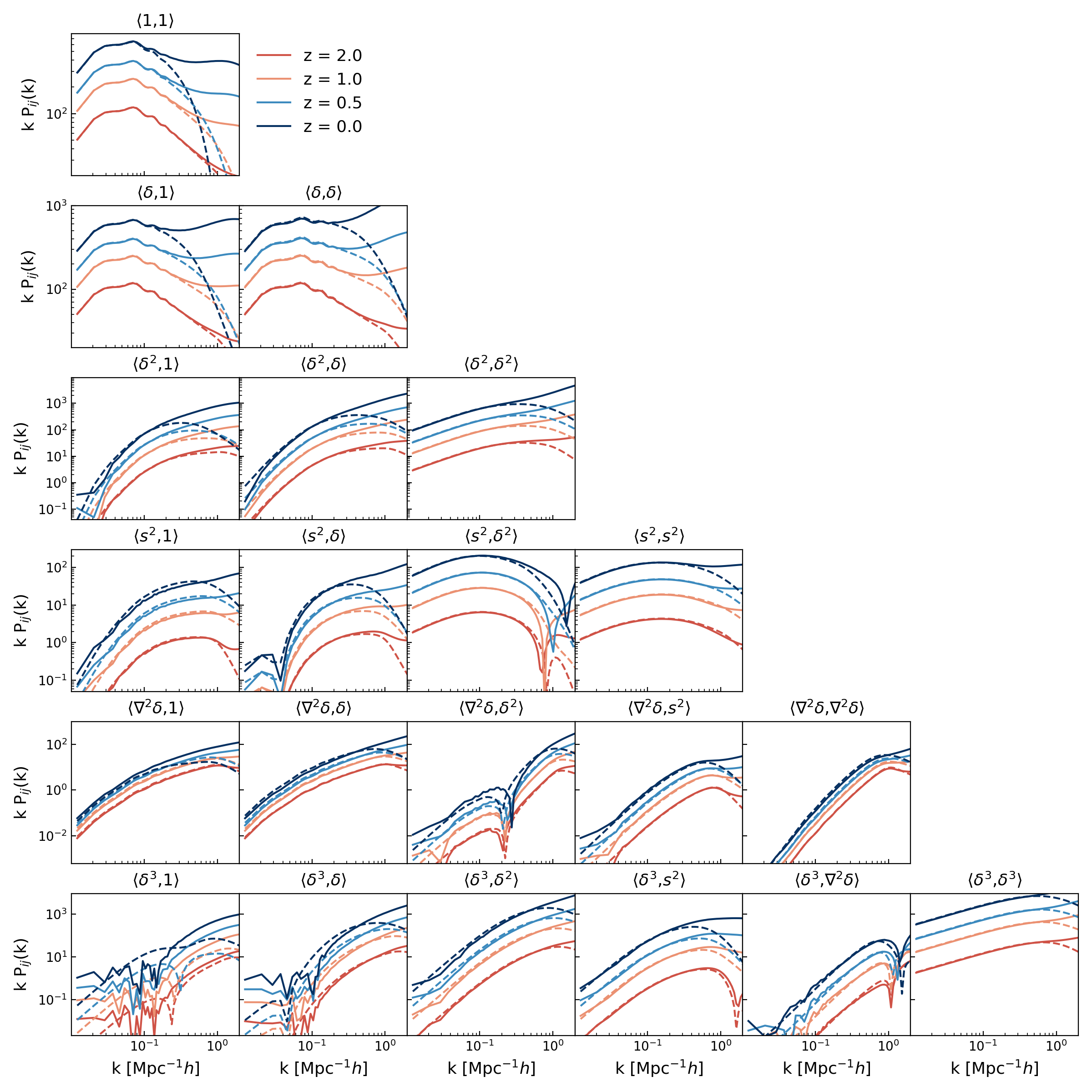}
\caption{ \label{fig:simu_1loop}
The Lagrangian basis spectra measured from simulations (solid lines) are compared to the 1-loop theoretical calculations (dashed lines), using samples randomly selected from the 83 cosmologies employed in the emulator construction. 
The four selected redshift bins, $z=2.0$, $1.5$, $1.0$, and $0.0$, are indicated by different colors. The Zel'dovich variance control has been applied to the simulation spectra, and the absolute values of all power spectra are taken for visualization purposes. 
The simulation results are consistent with the 1-loop theory on large scales, particularly for the leading  and quadratic order basis spectra. However, significant deviations appear at low redshift and small scales. The $\nabla^2\delta$ basis field is sensitive to the small-scale modes, and the $\delta^3$ field is not fully captured by the 1-loop perturbation theory. 
Specifically, the $P_{1\delta^3}(k)$ spectra measured from simulations are positive over the range $k \gtrsim 0.1\, \mathrm{Mpc}^{-1}h$ across all four redshift bins shown. In contrast, the 1-loop prediction remains negative throughout the entire $k$-range at $z=0.0$, despite exhibiting a similar trend in the figure. 
}
\end{figure*}
%%%%%%%%%%%%%%%%%%%%%%%%%%%%%%%%%%%%%%%%%%%%%%%%%%%%%%%%%
%%%%%%%%%%%%%%%%%%%%%%%%%%%%%%%%%%%%%%%%%%%%%%%%%%%%%%%%%

\subsection{Theoretical Templates of Lagrangian Basis Spectra}
\label{sec:cleft_theory}

We utilize the fully non-linear displacement from simulation to calculate the basis spectra and emulate them, but the power spectrum obtained from the analytical perturbation theory is also necessary in this work. 
The emulator construction requires the theoretical spectra for three purposes. 
(i) We need to divide the simulation spectrum by a theoretical reference to suppress the dynamic range, and then reduce the cosmological dependence of the surrogate model, therefore improving the interpolation accuracy. 
(ii) We employ the technique of Zel'dovich variance control to reduce the sample variance, which requires accurate calculation of the Zel'dovich power spectrum. 
(iii) At the scale around the simulation box size, some spectra are not handled by the simulation because of the sample variance coupling with the beyond-linear evolution, so we replace the basis spectra at $k\lesssim 0.01\, {\rm Mpc}^{-1}h$  with resummated 1-loop theory spectra. 

We employ the convolution Lagrangian effective field theory (CLEFT) to calculate the analytic power spectrum \cite{carlson2013convolution}, implemented in the efficient Python code \velocileptors \cite{chen2021redshift,chen2020consistent}. The CLEFT formulation involves the non-perturbation resummation as an exponential, which resummates the long-wavelength modes such as the BAO features. 
For the basis spectrum that cross-correlating fields $\mO_i$ and $\mO_j$ in Eq.~\ref{equ:bias_expansion_Pgg} \& \ref{equ:bias_expansion_Pgm}, we express it as Fourier integral, 
\begin{align}
& P_{ij}(k) = \int_\bfq e^{-i\bfk\cdot\bfq} \, \mM_{ij}(\bfk,\bfq) \ ,  \\
& \mM_{ij}(\bfk,\bfq) = \la \mO_i(\bfq_1) \mO_j(\bfq_2) \,e^{i\bfk\cdot\bfDelta}\ra \,|_{\bfq=\bfq_1-\bfq_2}  \ ,
\end{align}
where $\bfDelta \equiv \mPsi(\bfq_1) - \mPsi(\bfq_2)$, and the displacement $\mPsi = \mPsi^{(1)} + \mPsi^{(2)} + \cdots$ is treated by the perturbation expansion in the theoretical calculation. 
To illustrate the resummation, we express the integration kernel as $\mM_{ij} = e^{-{1\over2}\,k_ik_jA_{ij}^L} \mK_{ij}$, where $A_{ij}^L \equiv \la\Delta_i^{(1)}\Delta_j^{(1)}\ra_c$ is the two-point correlation of linear displacement.
The leading order contribution of particle displacement is resummated in the exponential, and all the non-linear corrections are expanded and captured in the redefined kernel $\mK_{ij}$. We {present the results in Fig.~\ref{fig:simu_1loop}, and} list the kernel expressions up to 1-loop order in the Appendix~(\ref{appendix:CLEFT-1loop}). 

Although the 1-loop spectrum with full resummation of linear displacement achieves higher accuracy in mild non-linear scale, it pays the price of damping the spectrum on small scales. For example, the matter power spectrum $P_{11}(k)$ damps and crosses zero value at $k\sim 0.3\, {\rm Mpc}^{-1}h$ in some cosmologies within our designed parameter space, and this scale is far from the emulation range we require. For the purpose of reducing the dynamical range of measured spectrum, we need to maintain stable behavior on small scales. Similar to the previous study \cite{chen2020consistent, derose2023aemulus}, for the spectrum diverging in high $k$, we expand the infrared resummation and keep the first few correction terms. While for those that exhibit stable behavior with infrared resummation, specifically, 
$\mM_{\delta^2\delta^2}$, $\mM_{\delta^2s^2}$, $\mM_{\delta^2\nabla^2}$, $\mM_{s^2\nabla^2}$, $\mM_{\nabla^2\nabla^2}$, $\mM_{\delta^3 s^2}$, $\mM_{\delta^3\nabla^2}$ and $\mM_{\delta^3\delta^3}$, we use the resummated spectrum as reference in surrogate model. The detailed kernel expressions are listed in Appendix~(\ref{appendix:CLEFT-kexpand}). 

The last class of the analytical basis spectrum is the Zel'dovich power spectrum. For the purpose of variance suppression (Sec.~\ref{sec:zvc}), we generate simulated realizations with linear displacement $\mPsi^{(1)}$, namely Zel'dovich approximation. The linear displacement only sources two-point correlation, and any higher-order connections vanish. So there are finite terms of correlator in the expansion expression of $\mM_{ij}$, and we can obtain the analytically exact results. The efficient computation is realized in \ZeNBu \cite{derose2023aemulus}, and we list expressions of additional $\nabla^2\delta$ $\&$ $\delta^3$ basis spectra in Appendix~(\ref{appendix:CLEFT-zel}).

In the actual calculation, we also need to account for the slight distinction between the Lagrangian density field and structure growth in the simulation, where we utilize the initial fluctuation of the simulation as the former Lagrangian density field.
%in simulations, where the latter sources the dynamical evolution in the simulation. 
In details, to incorporate the impact of radiation and neutrino perturbation in the gravity-only simulation, the \Kun suite initializes the simulation at $z_{\rm ini}=127$ with the backscaled spectrum $P_{\rm back}(k,z_{\rm ini})$, specifically the linear power spectrum at $z_{\rm low}=1$ backscaled to redshift $z_{\rm ini}$ by a scale-independent growth factor $D(z)$ (Sec.~\ref{sec:simulation}), where the linear power spectrum $P_L(k,z)$ is given by the Boltzmann solver \CLASS. Consequently, the Lagrangian density, or the initial fluctuation in simulation, differs from the true fluctuation at the initial redshift 
\footnote{To illustrate further, we take the Zel'dovich approximation as an example. We denote the Lagrangian density fluctuation as $\delta^\prime(\bfq)$, which is the initial fluctuation in the simulation backscaled from $z_{\rm low}$. But the correct latter-time matter clustering in simulation or the dynamical evolution of the particle, given by the displacement $\mPsi^{(1)}(\bfq, z)=D(z)\nabla\nabla^{-2}\delta(\bfq)$, is sourced by the true linear fluctuation $\delta(\bfq)$ at $z_{\rm ini}$ multiplied the Boltzmann solver output $D(z)$. 
%differs from the $\delta(\bfq)$ which determines the dynamical evolution.
}
, but the dynamical evolution of the late universe is maintained correctly in the simulation \cite{zennaro2017initial, adamek2023euclid, chen2025csst}. 
Though the Lagrangian fluctuation amplitude is slightly different for most of the cosmologies, we take it into account to better match the simulation measurement and theory prediction. 
Specifically, we utilize the backscaled power spectrum $P_{\rm back}(k,z)$ to seed the Lagrangian density fluctuation at redshift $z$, but utilize the \CLASS output $P_L(k, z)$ for the power spectrum in displacement integration, where $P_L(k, z)$ describes the correct matter clustering. 
%\begin{equation}\label{equ:Pk_back}
%P_{\rm back}(k,z) = {D^2(z)\over D^2(z_{\rm low})} P_L(k, z_{\rm low})
%\end{equation}
Operationally, we use the backscaled $P_{\rm back}(k,z)$ in the integration of the correlator constructed by two Lagrangian fields, e.g. $\mK_{\delta\delta}\ni\la\delta\delta\ra$. For those constructed by two displacement fields, e.g. $A_{ij} = \la\Delta_i\Delta_j\ra$, we use the direct output $P_L(k,z)$. For those constructed by both Lagrangian density and displacement field, e.g. $\mK_{1\delta}\ni\la\delta\Delta\ra$, we approximate the power spectrum as $P_{\rm x}(k,z) = \sqrt{P_{\rm back}P_L}$ in integration. 
All the modifications mentioned above are implemented in the real-space modules of the code \velocileptors .

% ---------------------------------------------------------------------------------------
% ---------------------------------------------------------------------------------------
% ---------------------------------------------------------------------------------------
% ---------------------------------------------------------------------------------------

\section{Emulator Construction}

In this section, we introduce the procedure for measuring and emulating basis spectra, including an overview of simulations (Sec.~\ref{sec:simulation}), steps of basis spectra measurement (Sec.~\ref{sec:measure_Pkij}), technique of variance suppression for measurement (Sec.~\ref{sec:zvc}) and the emulation approach (Sec.~\ref{sec:emulation}).

% ---------------------------------------------------------------------------------------
% ---------------------------------------------------------------------------------------

\subsection{\Kun simulations suite}
\label{sec:simulation}

The \Kun suite, consisting of 129 high-resolution \N-body simulations, is designed for emulator training targeting the upcoming CSST survey \cite{chen2025csst}. 
It is part of the \Jiutian simulation suite, which serves as simulation preparation for various supporting purposes, including observation analysis pipelines validation and cosmological inferences \cite{han2025jiutian}. 
In \Kun suite, each simulation has $3072^3$ particles in a $1\,h^{-3}{\rm Gpc}^3$ box volume, where the high mass resolution enables precise evaluation of cold dark matter and baryon clustering down to extremely non-linear region, predicting the power spectrum covering $0.00628 \leq k \leq 10\,{\rm Mpc}^{-1}h$ and $0\leq z \leq 3$ \cite{chen2025csst}. 
The simulation parameterizes the equation of state of the dynamical dark energy as $w(a)=w_0 + w_a(1-a)$ and incorporates massive neutrino impacts. 
It includes 128 sets of $w_0w_a$CDM$\nu$ cosmologies derived from Sobol sequence sampling, plus a fiducial Planck 2018 best-fit cosmology, with a total of 129 sets of simulations\footnote{All cosmologies can be found at \url{https://kunsimulation.readthedocs.io/en/latest/cosmologies.html}.}.
The cosmological parameters cover the ranges
\begin{align*}
\Omega_b &\in [0.04, 0.06]\ ,  \\
\Omega_{cb} &\in [0.24, 0.40]\ ,  \\
n_s &\in [0.92, 1.00]\ ,  \\
H_0 &\in [60, 80] \; {\rm km}\,s^{-1} {\rm Mpc}^{-1}\ ,  \\
A_s &\in [1.70, 2.50] \times 10^{-9}\ ,  \\
w_0 &\in [-1.30, -0.70]\ ,  \\
w_a &\in [-0.50, 0.50]\ ,  \\
\quad \sum m_\nu &\in [0.00, 0.30]  \;{\rm eV}  \ ,
\end{align*}
where we consider only a single massive neutrino species. 
{We do not treat the massive neutrinos as extra particle species, but incorporate their effects} in the simulations by the recent Newtonian gauge transformation \cite{2020JCAP...09..018P,2022JCAP...09..068H}, {where it has been shown that modifying Hubble rate and applying backscaling technique already capture the full neutrino impact on matter clustering with sufficient precision. }

We utilize the modified Gadget-4 $N$-body solver to run all the simulations. 
The initial linear power spectrum is generated by backscaling the \CLASS output from $z_{\rm low}=1$ to initial redshift $z_{\rm ini}=127$ with a scale-independent growth factor, specifically, 
\begin{align}\label{equ:simu-initial-Pk}
P_{\rm back}(k,z_{\rm ini}) = {D^2(z_{\rm ini})\over D^2(z_{\rm low})} P_L(k, z_{\rm low})\ ,
\end{align}
where the linear growth factor $D(z)$ accounts for the Newtonian evolution of cold dark matter and massive neutrino perturbative fluids, and the linear power spectrum $P_L$ is given by \CLASS solver output.
{The growth factor is scale-independent because we do not source the neutrino perturbation growth in the Poisson equation \cite{upadhye2014large, heitmann2016mira, 2020JCAP...09..018P,2022JCAP...09..068H}. In our previous paper~\cite{chen2025csst}, this simulation strategy was validated to achieve accuracy at the sub-percent level required for emulator construction.}
All particles are initialized by the second-order Lagrangian perturbation at the fixed redshift $z = 127$ with the fixed amplitude density field, and an isotropic glass-like distribution gives the pre-initial loads. 
We store the particle snapshot outputs at $12$ selected redshifts, in detail, 
$
z \;=\; \{ 3,\; 2.5,\; 2,\; 1.75,\; 1.5,\; 1.25,\; 1,\; 0.8,\; 0.5,\; 0.25,\; 0.1,\; 0 \}
$. 
The \textsc{FoF+SubFind}~\cite{2001MNRAS.328..726S} and \textsc{Rockstar}~\cite{2013ApJ...762..109B} halos and subhalos are also saved at these redshifts.
The simulation details and the convergence tests are presented in the previous paper~\cite{chen2025csst}. 

Because a large amount of computational resources is required for calculation, we only calculate the basis spectra for the first 83 sets of simulation to train the emulator, where the simulation labels range from \texttt{c0000} to \texttt{c0082}. The other 46 simulations, ranging from \texttt{c0083} to \texttt{c0128}, are treated as independent simulations to validate the model fitting performance using their {\textit{cb}} halo samples. The halo samples are given by the \textsc{FoF} halo catalogs, with a fixed linking length of $b=0.2$ across all cosmologies. We select the halo with mass ranging $11<\log(M)<14$ for the following tests, where the halo mass is defined as the total mass of particles, in the unit of $h^{-1}M_\odot$. 
{Although the halo finder is applied to the \textit{cb} particles, the halo properties such as halo bias and halo mass differ from those obtained using the full \textit{cb}+neutrino by less than $0.5\%$ for reasonable neutrino mass \cite{villaescusa2014cosmology,castorina2014cosmology}. }
%We divide the whole halo sample into five subsamples according to their mass: $[10^{11.0}, 10^{11.5}]\, ,[10^{11.5}, 10^{12.0}]\, ,[10^{12.0}, 10^{12.5}]\, ,[10^{12.5}, 10^{13.0}]\, ,[10^{13.0}, 10^{14.0}]\, ,$.
%\red{Need some comments on the mass binning? May connect to the ELG samples for the very low mass $10^{11}$?}

% ---------------------------------------------------------------------------------------
% ---------------------------------------------------------------------------------------

\subsection{Measurement of Basis Spectra}
\label{sec:measure_Pkij}

Our measurement procedures in the simulation are listed as follows. The general scheme is similar to previous works Refs.~\cite{zennaro2023bacco,kokron2021cosmology}, but the technical details are different. 
\begin{itemize}
\item[(i)] 
Initialization of the density field. 
Given the simulation initial condition, which is a random field with unit amplitude, we rescale the amplitude by the smoothed linear power spectrum $\sqrt{W(k)P_{\rm back}(k)}$ to obtain the Lagrangian density field $\delta(\bfq)$. The smoothed window is chosen to keep the shape of power spectrum above the pivotal scale $k_{\rm piv}$ unchanged with $W(k\le k_{\rm piv})=1$, while dampen the spectrum below by $W(k \geq k_{\rm piv}) = \exp\left[ -{1\over 2} R^2(k-k_{\rm piv})^2 \right]$. The smoothing parameters are chosen as $k_{\rm piv}=1\, {\rm Mpc}^{-1}h$ and $R=5\,h^{-1}{\rm Mpc}$. 
\item[(ii)] 
Construction of the Lagrangian basis fields. 
We compute the rest of Lagrangian fields $\delta^2$, $s^2$, $\nabla^2\delta$ and $\delta^3$ from smoothed density field $\delta(\bfq)$. For the particle with initial position $\bfq_P$, we associate it with the nearest grid point at $\bfq_G$, and assign the field values at grid $\bfq_G$ to the particle $\bfq_P$. 
Moreover, the field values are compensated by the corresponding derivative corrections up to quadratic order, $\mO_i(\bfq_P) = \mO_i(\bfq_G) + \partial_\bfq\mO_i|_{\bfq_G} (\bfq_P-\bfq_G) + {1\over 2}\partial_\bfq^2\mO_i|_{\bfq_G} (\bfq_P-\bfq_G)^2 $. 
\item[(iii)] 
Advection from Lagrangian space to late-time coordination. 
For each redshift snapshot, we assign the particle to the mesh by its late-time coordinate with the cloud-in-cloud method, weighted by the associated field value $\mO_i$ from its initial position, to construct the late-time basis fields. 
{The mesh size is $1200^3$, with Nyquist frequency $k_{\rm Nyq}= 3.77 \,{\rm Mpc}^{-1}h$.}
Then we rescale the values of basis fields by the scale-independent growth factor $\left({D(z) \over D(z_{\rm ini})}\right)^n$, where the power index $n$ is the order of the field.
\item[(iv)] 
Compute the cross power spectrum $P_{ij}(k)$ for all the combinations of basis fields, and compensate the window function from the previous cloud-in-cloud mass assignment. 
{To bin the power spectrum, we adopt $11$ uniformly spaced linear bins within $0.008 < k < 0.1\, {\rm Mpc}^{-1}h$, and $50$ logarithmically spaced bins within $0.1 < k < 3.1\, {\rm Mpc}^{-1}h$, totally $44$ bins for $k<1.0\, {\rm Mpc}^{-1}h$. }
\end{itemize}

We introduce the smoothing window for the initial density field. 
Because of the finite resolution of the grid, there is always a cutoff in the high-frequency modes, thus, there is a degree of freedom for the choice of modes included in the bias expansion. We maintain the shape for the modes $k < 1\, {\rm Mpc}^{-1}h$, which is expected to retain main contribution at $k < 1\, {\rm Mpc}^{-1}h$ for the late-time matter clustering. 
Meanwhile, we prefer to dampen the small-scale modes for better matching between the simulation result and the theoretical prediction on large scales. 
%for better theoretical control on the large scales, which is one of the prominent advantages of bias expansion for galaxy clustering modeling.
On the one hand, the emulator performance is insensitive to the choice of smoothing scale \cite{zennaro2022priors}. 
On the other hand, the convolution nature of fields like $s^2$ and $\delta^2$ couples the modes from different scales, therefore without the small scales cutoff, the theoretical calculation is given by the triple Fourier integration over three-dimensional box $(k_x,k_y,k_z)$, even for just simple zero-lag $\sigma_L^2=\int_{|k_i|<{\pi\over L}}\, P_L(\bfk)$. 
%we have to perform triple integration over the simulated modes within the square box in the theoretical prediction, even for just simple zero-lag $\sigma_L^2=\int_{|k_i|<{\pi\over L}}\, P_L(\bfk)$. 
Moreover, in the actual computation, we adopt the mesh size $N_{\rm ini}^3=6144^3$, and pad the initial random phase to low-$k$ cells while the high-$k$ cells are padded by zero. Our small-scale cutoff restricts the non-vanishing modes to $k\lesssim 2\, {\rm Mpc}^{-1}h$, far from the grid resolution limit, and it avoids suffering from spurious aliasing contribution in discrete grid calculation \cite{taruya2018grid}. 
In brief, we apply the smoothing window to the Lagrangian density field for robust theoretical control and numerical stability, and we emphasize that the choice of small-scale cutoff is always subject to individual preference rather than dictated by rigorous physics.

After the Lagrangian basis field construction, we have to make compensation for assigning the Lagrangian field value from the regular grid to the particle. In \Kun simulation, rather than pre-initialization in a uniform grid, the particle is initially placed in a glass-like distribution, so we can not associate each particle with a grid cell exactly. If we directly assign the value of the nearest cell to the particle, the mismatch between particle and nearest cell position effectively introduces random position shuffling, suppressing the power of cross-correlation, especially for high order basis fields. The mismatch is alleviated by adding the first few correction terms in the Taylor expansion. We implement it in Fourier space, 
\begin{align}
\mO_i(\bfq_P) &= \int_\bfk e^{i\bfk\cdot\bfq_P} \mO_i(\bfk)    \label{equ:compensate_expand}  \\
&= \int_\bfk e^{i\bfk\cdot\bfq_G} \mO_i(\bfk) 
+ {\bf\Delta}_{\rm PG}\cdot \int_\bfk e^{i\bfk\cdot\bfq_G}\, i \bfk\,\mO_i(\bfk)       \nonumber \\
&\quad - {1\over 2} {\bf\Delta_{\rm PG}\Delta_{\rm PG}} {\;:} \int_\bfk e^{i\bfk\cdot\bfq_G}\, \bfk\bfk\,\mO_i(\bfk) +\cdots\ ,    \nonumber 
\end{align}
where ${\bf\Delta}_{\rm PG}=\bfq_P-\bfq_G$. The first term is the value of the nearest cell, while the next two terms are the first and second order field moments. The compensation method up to the second order moment achieves an accurate enough correction in our validation (Appendix~\ref{append:validation_taylor_compensation}), so we truncate at the second order and discard the higher order terms. An intuitive understanding of the effectiveness of the compensation method is that, the typical minimum mode of the field spans size $\sim 3\, h^{-1}{\rm Mpc}$, much larger than $|\bf\Delta_{\rm PG}|_{\rm max}\sim$ cell size $\sim 0.16\, h^{-1}{\rm Mpc}$. Thus, the field is differentiable within the scale we consider, and the expansion Eq.~\ref{equ:compensate_expand} is a well-convergent series. 

In principle, the presence of massive neutrinos induces the scale dependence of the linear growth factor, $D(z)\rightarrow D_{\nu}(z, k)$, since neutrinos do not cluster below their free-streaming scale. However, the gravity-only simulations account for the neutrino impact in the homogeneous background, but not the gravitational interaction during advancing particles. Therefore, the linear evolution is scale-independent in the \Kun simulation suite, and the linear growth of the Lagrangian fields is also scale-independent. We simply multiply the basis fields with the ratio of scale-independent growth factors, and it is fully consistent with the procedure used to generate the simulation initial conditions (see Eq.~\ref{equ:simu-initial-Pk}). This approximation has been validated to reproduce matter clustering at low redshift accurately \cite{chen2025csst, 2020JCAP...09..018P,2022JCAP...09..068H}.

% ---------------------------------------------------------------------------------------
% ---------------------------------------------------------------------------------------

%%%%%%%%%%%%%%%%%%%%%%%%%%%%%%%%%%%%%%%%%%%%%%%%%%%%%%%%%
%%%%%%%%%%%%%%%%%%%%%%%%%%%%%%%%%%%%%%%%%%%%%%%%%%%%%%%%%
\begin{figure*}[htb!]
\includegraphics[width=0.99\textwidth]{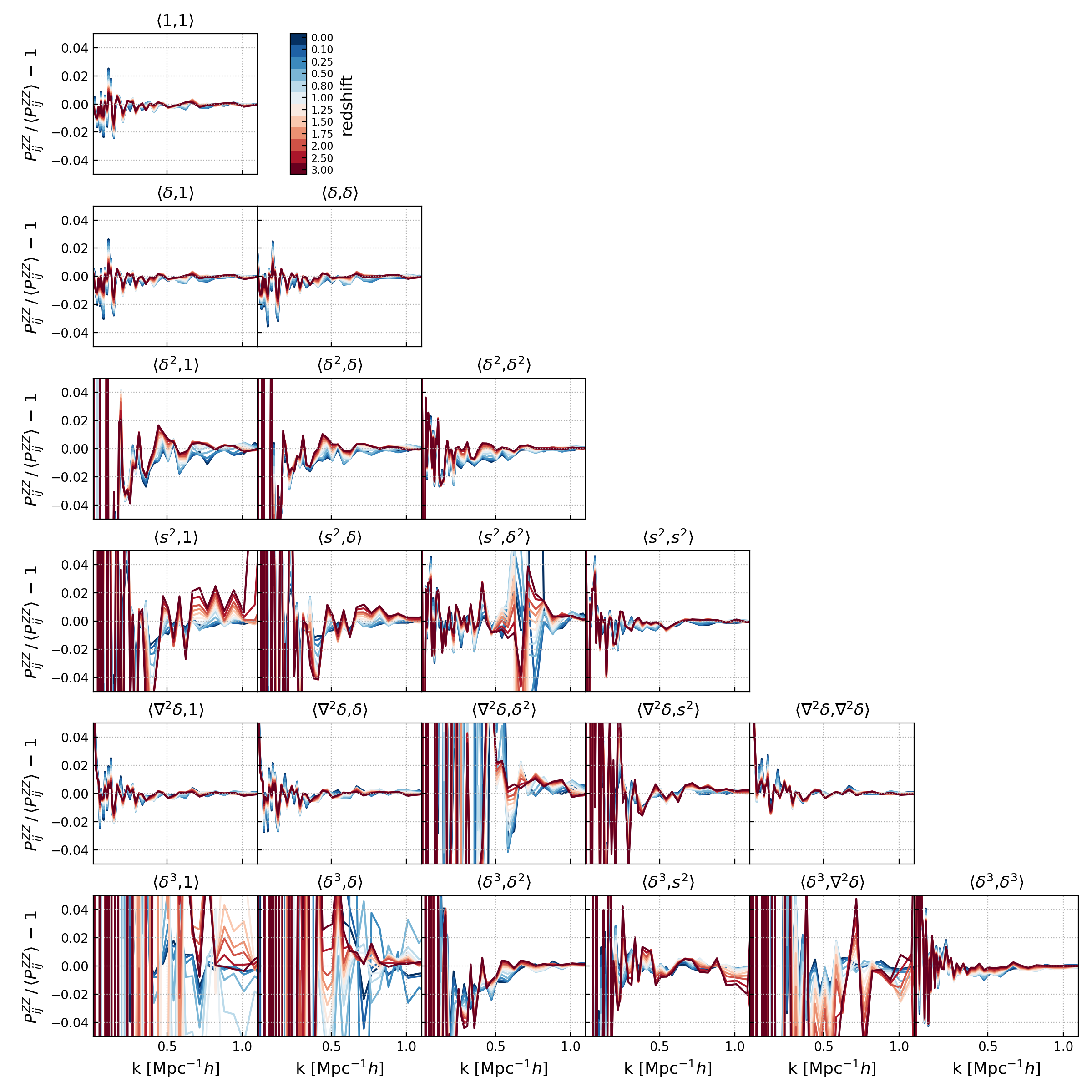}
\caption{ \label{fig:ZVC_variation}
The sample variance fluctuations derived from Zel'dovich variance control are shown for a random selection of the 83 cosmologies used in the emulator construction. 
The power spectrum $P_{ij}^{ZZ}$ is computed from the corresponding Zel'dovich simulations with $3072^3$ particles, while the ensemble average $\la P_{ij}^{ZZ}\ra$ is obtained from the analytical theory. Line colors range from red to blue, indicating redshifts from $z=3.0$ to $z=0.0$. Across all redshifts and basis spectra, the extracted linear order noise remains unbiased because of the exact agreement between the Zel'dovich realizations and the precise analytical prediction. 
}
\end{figure*}
%%%%%%%%%%%%%%%%%%%%%%%%%%%%%%%%%%%%%%%%%%%%%%%%%%%%%%%%%
%%%%%%%%%%%%%%%%%%%%%%%%%%%%%%%%%%%%%%%%%%%%%%%%%%%%%%%%%

\subsection{Zel'dovich Variance Control}
\label{sec:zvc}

The raw simulation measurements of the basis spectra are noisy because of the sample variance in a single realization. However, as the phase of the initial field is known, the fluctuation sourced by the sample variance can be effectively mitigated using the variance control technique. 
The general idea of variance control is suppressing the variance of the observable $\hatx$ by constructing an estimator $\haty$, aided by the control variate $\hatc$.
\begin{align}\label{equ:zvc}
\haty \equiv \hatx - \beta\, (\hatc - \mu_c)\ ,
\end{align}
where we denote the realization expectation $\mu_c=\la\hatc\ra$. For any weight $\beta$ value, we always have expectation $\la\haty\ra=\la\hatx\ra$, so $\haty$ is unbiased estimator of $\hatx$. 
The optimal weight with minimized $\haty$ variance is 
\begin{align}
\nonumber \beta^* = { {\rm Cov}[\hatx,\hatc] \over {\rm Var}[\hatc] }\ ,
\end{align}
and the minimized variance is 
\begin{align}
\nonumber   {\rm Var}(y) = {\rm Var}(x)  \left[ 1 - r_{xc}^2 \right]  
\;,\quad   r_{xc}  = { {\rm Cov}[\hatx,\hatc]  \over \sqrt{ {\rm Var}[\hatx]{\rm Var}[\hatc] } }\ .
\end{align}
Therefore, the higher the correlation between the observable and the control variate, the greater the suppression of sample variance. 

We utilize the basis spectra from Zel'dovich approximation as the control variate, because the matter clustering from linear displacement already presents high cross-correlation with $N$-body simulation even up to scale $k\sim 1\, {\rm Mpc}^{-1}h$ \cite{kokron2022accurate,derose2023precision,hadzhiyska2023mitigating}. Specifically, we repeat our basis spectra calculation as same as done in the N-body simulation, initializing the density field with the same initial condition. But the displacement is given by the Zel'dovich approximation, 
\begin{align}
\mPsi^{(1)}(\bfq, z) = D(z)\int_\bfk e^{i\bfk\cdot\bfq} {i\bfk\over k^2}\delta(\bfk) \ ,
\end{align}
instead of the non-linear displacement. The particles are initialized in the regular grid to avoid the aforementioned issues. In this way, we can generate an inexpensive yet highly correlated realization for a given cosmology and redshift. 
To obtain the variance suppressed basis spectra, we rephrase Eq.~\ref{equ:zvc} as
\begin{align}
\hat{P}_{ij} = P_{ij}^{NN} - \beta^*_{ij}\left( P_{ij}^{ZZ} - \la P_{ij}^{ZZ}\ra \right)\ ,
\end{align}
where the $P_{ij}^{NN}$ is the raw simulation output, and $\hat{P}_{ij}$ is the variance suppressed basis spectrum, with expectation $\la\hat{P}_{ij}\ra = \la P_{ij}^{NN}\ra$. 
For the optimizing weight $\beta^*_{ij}$, we only consider the disconnected contribution \cite{kokron2022accurate}, 
\begin{align}
\label{equ:zel_weight}
\beta^*_{ij}(k) = 
{ P^{NZ}_{ii} P^{NZ}_{jj} + P^{NZ}_{ij} P^{ZN}_{ij}  \over
 P^{ZZ}_{ii} P^{ZZ}_{jj} + \left(P^{ZZ}_{ij}\right)^2 }   \ , 
\end{align}
where the upper index $N$ denotes the field from the $N$-body realization and $Z$ denotes the field from the Zel'dovich realization. That is, $P^{ZZ}_{ij}$ is the spectrum cross-correlating Zel'dovich realizations $\mO_i^Z$ and $\mO_j^Z$.

In the actual computation, for a given cosmology and redshift, we generate the Zel'dovich realization to calculate the basis spectrum $P^{ZZ}_{ij}$, and also calculate the corresponding analytical prediction $\la P^{ZZ}_{ij}\ra$. 
While for the optimized weight Eq.~\ref{equ:zel_weight}, we fix the $\beta^*_{ij}$ value as those obtained from the first simulation. Notice that the variance suppression is always unbiased as long as the analytical prediction matches the realization averaging exactly.

% ---------------------------------------------------------------------------------------
% ---------------------------------------------------------------------------------------

\subsection{Parameter Space Interpolation}
\label{sec:emulation}

After obtaining the variance-suppressed spectra, we perform the basis spectra interpolation across the cosmological parameters and redshifts. 
We construct the surrogate model \cite{kokron2021cosmology, derose2023aemulus, zennaro2023bacco}
\begin{align}
\label{equ:surrogate}
\Gamma^{ij}(\bfk,z,{\bf\Omega}) = \log\left[ P_{ij}(\bfk, z,{\bf\Omega})\over P_{ij}^T(\bfk, z,{\bf\Omega}) \right]\ ,
\end{align}
where $P_{ij}$ is the simulation measurement after Zel'dovich variance control, and $P_{ij}^T$ the theoretical template to reduce the dynamical range of the spectra. ${\bf\Omega} = \{\Omega_b, \Omega_{cb}, n_s, H_0, A_s, w_0, w_a, \sum m_\nu\}$ is 8 cosmological parameters as prediction input. 
To further suppress the measurement noise apart from Zel'dovich variance control, we apply the Savitzky-Golay filter with an order-2 polynomial to smooth the ratio before emulation. Because of the wide smoothing window, we do not smooth $P_{11}$, $P_{1\delta}$, and $P_{\delta\delta}$ where the BAO features are significant. While for some basis spectra, there is still substantial residual noise on the large scales due to the decorrelation of Zel'dovich and simulation realization, and we have to remove these $k$ regions to prevent filtering biasing small-scale spectra. More details on smoothing and other fine-tuning settings are listed in \ref{appendix:fine_tuning_parameters}. 

For the dimensionality reduction of data, we perform the singular value decomposition (SVD), 
\begin{align}
\label{equ:PCA}
\Gamma^{ij}(\bfk,z,{\bf\Omega}) = \sum_{n}^{N_{PC}} \alpha_n(z,{\bf\Omega}) \phi_n(\bfk)\ ,
\end{align}
where $\alpha_n(z,{\bf\Omega})$ is the coefficients capturing the cosmology and redshift dependence, and $\phi_n(\bfk)$ is the basis functional. Operationally, we stack the data into a matrix with size $(N_{\rm cosmo}\times N_z, N_k)$. Within SVD, $\phi_n(\bfk)$ is the right eigenvector and $\alpha_n(z,{\bf\Omega})$ is the matrix product of left eigenvectors and singular values. We select only the first $N_{PC}=12$ principal components as the basic choice for emulation, and the impact of residual eigencomponents is significantly lower than the interpolation error. Exceptionally, due to the noisy output of $\Gamma^{1\delta^3}$ and $\Gamma^{\delta\delta^3}$ measurements, we need to include all the eigencomponents for emulating these two. 

Furthermore, some basis spectra may change the sign, and the transition points vary with redshift and cosmology, resulting in a negative value in the ratio between $P_{ij}$ and $P_{ij}^T$. Specifically, for some $k$-range in basis spectra $P_{\delta^2\nabla^2}$, $P_{s^2\delta^3}$, and $P_{\nabla^2\delta^3}$, directly taking the logarithm value of the ratio is invalid, while without logarithm rescaling, the dynamical range is too large to emulate. So we replace the spectra in surrogate model Eq.~\ref{equ:surrogate} with
\begin{align}
\label{equ:surrogate_rescaled}
P_{ij}^{(T)}  \;\rightarrow\;   P_{ij}^{(T)} + \tilde{r}_{ij}\sqrt{P_{ii}^TP_{jj}^T}\ ,
%\Gamma^{ij}(\bfk,z,{\bf\Omega}) = 
%\log\left[ 
%    P_{ij} + \tilde{r}_{ij}\sqrt{P_{ii}^TP_{jj}^T} \over 
%    P_{ij}^T + \tilde{r}_{ij}\sqrt{P_{ii}^TP_{jj}^T} \right]
\end{align}
where the auxiliary term $\tilde{r}\sqrt{P_{ii}^TP_{jj}^T}$ is introduced to ensure the positive value for $k$-range we interest in. The $\tilde{r}_{ij}$ is roughly estimated from the cross-correlation coefficient between basis fields $\mO_i$ and $\mO_j$, which is fine-tuned to avoid the auxiliary term overwhelming $P_{ij}$ or $P_{ij}^T$ terms. The auxiliary $\tilde{r}_{ij}$ is fixed for all cosmologies and redshifts. 

For the high-dimensional interpolation of coefficient $\alpha_n(z,{\bf\Omega})$, we utilize Gaussian process regression, a method widely adopted in the emulator construction \cite{chen2025csst, zhai2019aemulus, ho2022multifidelity}. It assumes the correlation between the data points following the multi-dimensional Gaussian distribution, and their covariance is given by the parameterized kernel. By maximizing the marginal likelihood of the training data, one can obtain the optimal hyperparameters of the kernel. 
Given the hyperparameters, the interpolation result at a test point is expressed as the linear combination of training data, weighted by the kernel function. 
%Given the hyperparameters and assuming the noise level of the data, we can predict the values at the test point from the conditional distribution. 
In the basis spectrum emulation, we use the commonly parameterized kernel, the product of a constant kernel and a radial basis kernel. Besides, we transform redshift $z$ to the scale factor $a=1/(1+z)$ as the time variable, then stack it with cosmological parameters as a 9-element input vector. The output data vector $\alpha_n(z,{\bf\Omega})$ is normalized to zero-mean and unit standard deviation before model training.

% ---------------------------------------------------------------------------------------
% ---------------------------------------------------------------------------------------
% ---------------------------------------------------------------------------------------
% ---------------------------------------------------------------------------------------

\section{Emulation Performance}

%%%%%%%%%%%%%%%%%%%%%%%%%%%%%%%%%%%%%%%%%%%%%%%%%%%%%%%%%
%%%%%%%%%%%%%%%%%%%%%%%%%%%%%%%%%%%%%%%%%%%%%%%%%%%%%%%%%
\begin{figure*}[htb!]
\centering
\includegraphics[width=0.99\textwidth]{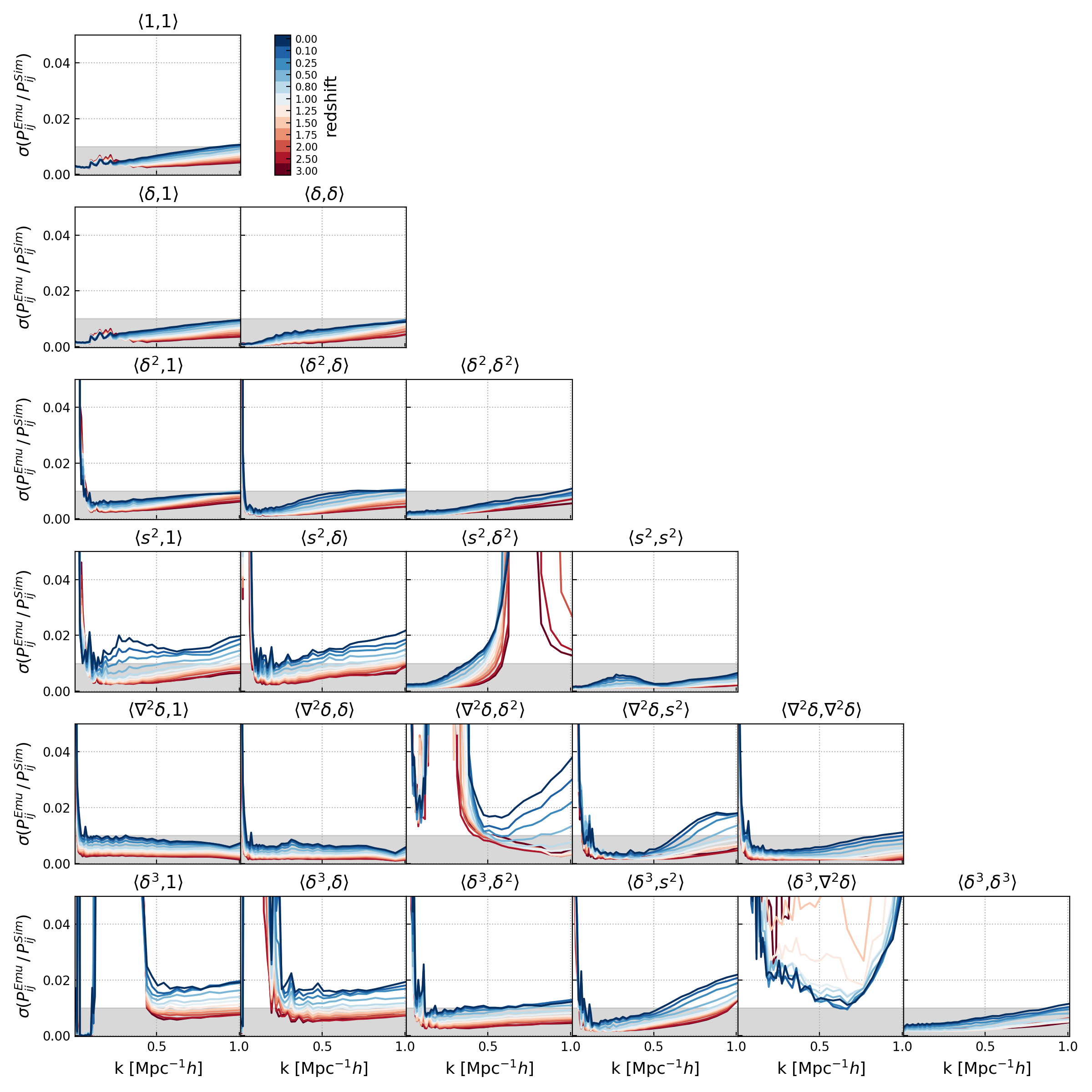}
\caption{ \label{fig:LOO}
Leave-one-out 68th percentile fractional deviation of the emulator. 
Line colors range from red to blue, corresponding to redshifts from $z=3.0$ to $z=0.0$. 
The gray shaded regions indicate the $1\%$ deviation, where most of the quadratic order basis spectra fall below this threshold. For the basis spectra $P_{\delta^2s^2}$, the apparent divergence beyond the plot range is due to the denominator approaching zero, and does not imply a large absolute fractional error. 
}
\end{figure*}
%%%%%%%%%%%%%%%%%%%%%%%%%%%%%%%%%%%%%%%%%%%%%%%%%%%%%%%%%
%%%%%%%%%%%%%%%%%%%%%%%%%%%%%%%%%%%%%%%%%%%%%%%%%%%%%%%%%

% ---------------------------------------------------------------------------------------
% ---------------------------------------------------------------------------------------

In this section, we present the result of basis spectrum emulation. In Sec.~\ref{sec:result_basis_spectra}, we examine the measurement results from the $N$-body simulation. In Sec.~\ref{sec:result_validation}, we present the emulation accuracy and the emulator performance, and validate the bias expansion modeling with the emulator based on the halo sample power spectrum.

\subsection{Measurement Results from Simulation}
\label{sec:result_basis_spectra}

In Fig.~\ref{fig:field_maps}, we present the visualization of the Lagrangian fields obtained from the simulation measurement. As expected, different Lagrangian fields emphasize different clustering regions of the large-scale structure \cite{zennaro2023bacco}. 
Compared to the matter overdensity field $\mO_i=1$, the scalar tidal field $s^2$ appears to have much thicker filaments. The Laplacian field $\nabla^2\delta$ presents weaker large-scale clustering, but exhibits small-scale fluctuation since it is dominated by non-linear scale power. 
The linear density field $\delta$ presents remarkable filament and cluster features, while the massive halo structures are much more emphasized in the $\delta^2$ field. Only extremely massive halos are highlighted in the cubic density $\delta^3$. The distinctive halo structure emphasized in the $\delta$ weighting fields, especially the high order $\delta$ fields, is sourced by the sensitivity of these fields to the proto-dense region. 
Another prominent property of $\delta^2$ and $\delta^3$ fields is the under-dense feature in the cosmic filament region, which is sourced by the collapse of the under-dense region around the primordial density peak. In detail, the high order $\delta$ field magnifies the high-density region, then the middle-density region around the dense region appears to be under-dense because of mass conservation. During the hierarchical structure formation, the dense regions collapse into massive halos, while the under-dense regions successively collapse into filaments. The observed field values coming from the under-dense Lagrangian region are further enhanced by the local number density. Therefore, we observe the significant under-density in both the filaments and the regions around the massive halo. One can understand other features intuitively with the same scenario. 

The basis spectrum measurement in the finite simulation volume suffers from the sample variance, and we alleviate it by Zel'dovich variance control. In Fig.~\ref{fig:ZVC_variation}, we present the power spectrum ratio between the Zel'dovich realization $P_{ij}^{ZZ}$ and the ensemble averaging $\la P_{ij}^{ZZ}\ra$, equivalently the fluctuation sourced by sample variance correlated with linear displacement. The fractional fluctuation is tightly confined to the region around ${ P_{ij}^{ZZ} / \la P_{ij}^{ZZ}\ra } -1 \sim 0$, especially for the leading order spectrum like $P_{11}(k)$. It confirms that our Zel'dovich simulation result and theoretical calculation exactly match, therefore the variance reduction is unbiased. Large fluctuations appear in the linear scale of some basis spectra, such as the $k < 0.1\,{\rm Mpc}^{-1}h$ of $P_{s^2\delta}(k)$, because the signal is tiny and the measurements are dominated by the sample variance. While for the region like $k\sim 0.6\, {\rm Mpc}^{-1}h$ of $P_{s^2\delta^2}(k)$, the apparently large fractional difference is caused by the crossing zero-value of the theoretical calculation in the denominator, rather than truly large noise. 

Since the Zel'dovich variance control only accounts for the correlated noise with linear displacement, any fluctuation caused by the high order mode-coupling is not reduced. For the basis spectra $P_{1\delta^2}$, $P_{1 s^2}$, $P_{\delta\delta^2}$ and $P_{\delta s^2}$, where these quadratic order spectra are generated by mode-coupling, the Zel'dovich variance control is disabled for the residual fluctuation arising from loop orders, though a large amount of noise correlated with linear displacement has been removed. 
For the spectra of third order terms, such as $P_{1\delta^3}$ and $P_{\delta\delta^3}$ of which the beyond-linear contribution dominates on large scales, the disconnected term only accounts for part of the fluctuation and the noise correlated with beyond-linear terms is still significant after variance suppression. 
It is shown in the Fig.~\ref{fig:simu_1loop}, the spectra $P_{1\delta^3}$ and $P_{\delta\delta^3}$ are still noisy at scale $k\lesssim 0.2\, {\rm Mpc}^{-1}h$ due to the decorrelation. 
%Nevertheless, the cubic order spectra are extremely sub-dominated in the scale we interest, and the large 

In Fig.~\ref{fig:simu_1loop}, we compare the basis spectra after variance suppression to the 1-loop theory calculation. Most of the simulation outputs agree with theoretical results well on the large scales, particularly for the linear and quadratic order spectra. 
If the spectrum in the perturbation calculation consists solely of disconnected terms, which are accurate components in perturbation theory compared to next-to-leading-order corrections, the 1-loop spectra consistently describe non-linear spectra well, such as $\mK_{\delta^2\delta^2}\ni \la\delta\delta\ra^2$ and $\mK_{s^2\delta^2}\ni \la\delta s_{ij}\ra\la\delta s_{ij}\ra$. 
If both the linear and 1-loop terms dominate, the theoretical spectra are accurate at high redshift while gradually worsening at low redshift. For example, we have $\mK_{1s^2} = -{1\over 2}k_ik_j \la\Delta_i s_{mn}\ra \la\Delta_js_{mn}\ra + i k_i\la\Delta_i s^2\ra$ up to 1-loop order, where the 1-loop correction $\la\Delta_i s^2\ra$ become important in low redshift and results in a mismatching of overall amplitude at $k \lesssim 0.5\,{\rm Mpc}^{-1}h$. 
We also cross-validate $P_{1s^2}$ with the grid-based 2LPT calculation, which fully resummates all the second order displacement contribution. We find that the degeneracy disappears and the 2LPT calculation matches with non-linear spectra on the large scales almost perfectly. Thus, we conclude that the slight mismatch between simulation and 1-loop theory $P_{1s^2}$ results from the missing contribution from 2LPT. 

However, we caution that further high order LPT corrections do not improve the accuracy in the highly non-linear matter clustering, since the perturbation expansion breaks down. We calculate the grid-based 1LPT, 2LPT, and 3LPT power spectra and compare them to the fully non-linear simulation results, with details listed in Appendix~\ref{appendix:nLPT}. We find that the 2LPT is able to reproduce non-linear $s^2$ spectra up to $k\sim 0.8\,{\rm Mpc}^{-1}h$ at $z\sim 1$. But for some basis spectra such as $P_{1\delta^2}$, the 2LPT and 3LPT improve the high redshift $z\sim 2$ predictions while worsening the low redshift $z\sim 0.5$ results. Interestingly, though the high order LPT breaks for predicting the spectra like $P_{\delta^2\delta}$ and $P_{\delta^2\delta^3}$, the 1LPT still characterizes the linear scale power well, because the Zel'dovich approximation captures the primary structures of clustering \cite{white2014zel, schmittfull2019modeling}.

% ---------------------------------------------------------------------------------------
% ---------------------------------------------------------------------------------------

\subsection{Emulation Validation}
\label{sec:result_validation}

In Fig.~\ref{fig:LOO}, we present the emulation accuracy quantified by the Leave-one-out (LOO) error. 
We pick one simulation from the total $N$ simulations as the validation set and utilize the rest of $N-1$ simulations as the training set to construct the emulator. The LOO error is defined as the 68th percentile error across all validation samples after repeating $N$ times LOO estimation, where $N=83$ is the total simulations adopted for emulator training. Our emulation realizes $1\%$ accuracy for the cross-correlation of Lagrangian fields $1$, $\delta$, and $\delta^2$ at all redshifts $0\leq z\leq 3$, which are the dominations in the bias expansion model. Another quadratic order $s^2$ terms also reach $1\sim 2\%$ accuracy, especially the auto-power spectrum, that well restricted to $<1\%$ error region. Our $1\%$ level emulation accuracy satisfies the requirement of current and upcoming galaxy surveys \cite{derose2023aemulus}, including the design target of the CSST emulators \cite{chen2025csst}.
Notice that the apparent large fractional error in some scales, such as $k \lesssim 0.1 \,{\rm Mpc}^{-1}h$ of $P_{\delta^2 1}$ and $k \sim 0.7 \,{\rm Mpc}^{-1}h$ of $P_{s^2\delta^2}$, is simply due to the tiny value in the denominator, where the power spectra value is small and the corresponding error in bias expansion is also negligible. 
As for the counterterm $\nabla^2\delta$ and the third-order $\delta^3$ spectra, their suboptimal emulation does not affect the galaxy clustering modeling. 
The counterterm $\nabla^2\delta$ term is introduced to absorb the non-local effects, and it is degenerated with behaviors such as the baryonic physics and finite halo size. Thus, it is expected to be insensitive to cosmological information, and one can also adopt alternative choice of replacing $P_{\nabla^2X}$ with $-k^2 P_{1X}$ without degradation. 
While the $\delta^3$ terms are shown to be small and degenerated with counterterms in the current survey analysis \cite{maus2024analysis}, we include them in our emulation, aiming at investigating possible effects from high order bias terms down to extremely non-linear scale, for theoretical interest. So, the potential emulation error does not impact the application. 
%What's more, there is slightly large prediction error of $P_{\nabla^2\delta^2}$ and $P_{\delta^3\nabla^2}$, while these apparent large fractional error do not impact on final expansion modeling. 
What's more, we fail to emulate $P_{\delta^31}$ at $k\lesssim 0.4\, {\rm Mpc}^{-1}h$ and $P_{\delta^3\delta}$ at $k\lesssim 0.2\, {\rm Mpc}^{-1}h$ because of the noisy measurement from simulation, but the cross-correlation coefficients in these regions are at the level of only a few percent, e.g. $r^2_{\delta^31}(k\lesssim 0.4\, {\rm Mpc}^{-1}h) \sim 1\%$, so we can safely mask these regions and set the value as zero. 
There is similar case for the $P_{\delta^3\nabla}$ with  $r^2_{\delta^3\nabla} \lesssim 10\%$. 
%The $\nabla^2\delta$ term is introduced to absorb the non-local effects like small scales clustering and galaxy formation, therefore, we expect these terms are effective in the galaxy physics domination scales rather than cosmological information. Besides, it is often a reasonable approximation to replace terms like $P_{\nabla^2X}$ with $k^2 P_{\delta X}$, 

\begin{figure}[H]
\centering
\includegraphics[width=0.95\columnwidth]{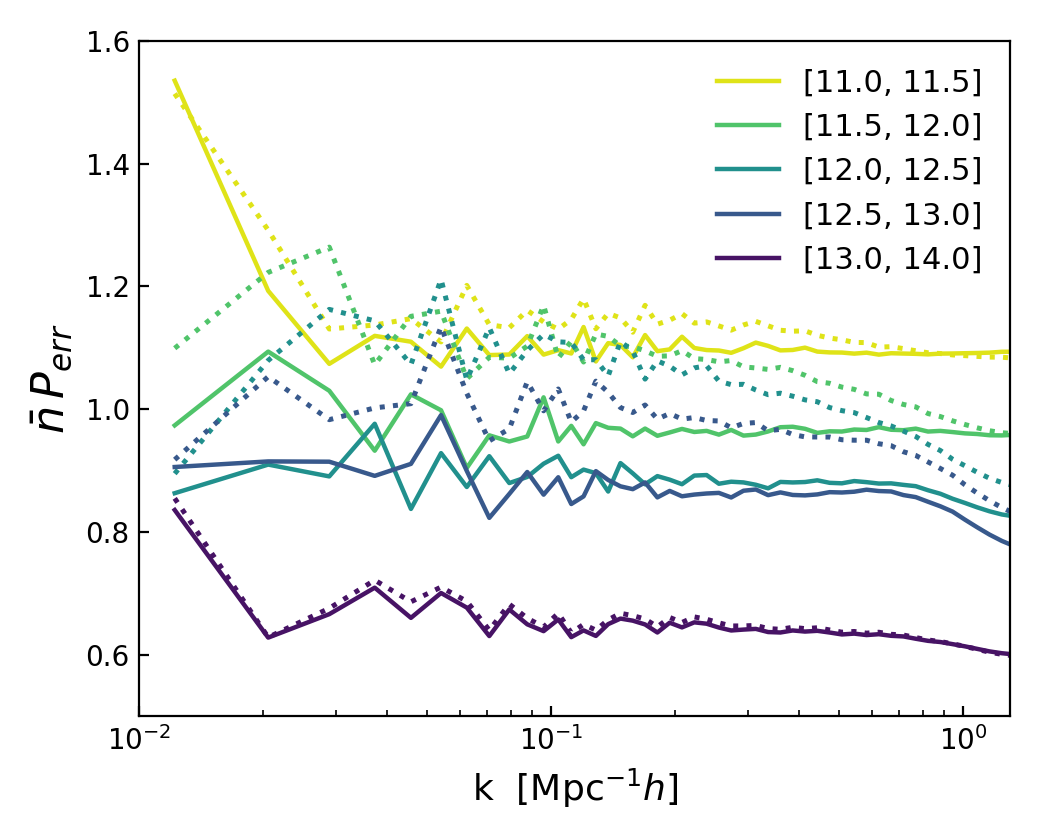}
\caption{ \label{fig:halo_fit_Perr}
In the field-level modeling of halo overdensity field, the fitting performance is quantified by the power spectrum of residual field $\hat\varepsilon(\bfk)\equiv \delta_h(\bfk) - \hat\delta^{\rm HEFT}(\bfk)$. Here, $\delta_h(\bfk)$ is the true halo overdensity measured from the simulation and $\hat\delta^{\rm HEFT}(\bfk)$ is the Lagrangian bias expansion model. 
They are normalized by the mean number density $\bar{n}$, where $\bar{n}P_{\rm err}=1$ corresponds to the Poisson expectation. 
The results are shown for the Planck cosmology simulation at $z=1$, with 5 colors indicating the 5 halo mass bins within $11<\log(M)<14$. 
The dotted lines indicate the results with 4 bias parameters $\{b_1, b_2, b_s, b_\nabla\}$, while the solid lines indicate the results with 5 bias parameters $\{b_1, b_2, b_s, b_\nabla, b_3\}$. We seek the optimal bias parameters by fitting the large-scale $k$ modes up to $|\bfk|_{\rm max}=0.2\, {\rm Mpc}^{-1}h$, and as shown, the bias expansion models are consistent with the true halo distribution at $k\lesssim 1\,{\rm Mpc}^{-1}h$ in the field level. 
Further, the inclusion of $\delta^3$ field reduces the super-Poisson stochasticity arising from the small-scale clustering, and this reduction is significant for the low mass samples. 
}
\end{figure}

To verify the bias expansion performance and validate our smoothing scheme for the Lagrangian density field, we present bias expansion fitting of Eq.~\ref{equ:bias_expansion} on the halo samples. We first perform the fitting in the field level to avoid the sample variance, and estimate the galaxy bias by minimizing the EFT likelihood \cite{schmittfull2019modeling,cabass2020likelihood,kokron2022priors,shiferaw2024uncertainties}. The optimal solution of the bias parameters $\hat{b}_i$ are 
\begin{align}\label{equ:field_level_bias}
& \hat{b}_i = A_{ij}^{-1} B_j  \ , \\
& A_{ij} = \int_{|\bfk|<k_{\rm max}} { \mO_i(\bfk)\mO_j(-\bfk) \over \sigma^2_{\varepsilon}(k)  } \ , \\
& B_{j} = \int_{|\bfk|<k_{\rm max}} { \mO_j(\bfk) \left[\delta_h-\delta_m\right](-\bfk)  \over \sigma^2_{\varepsilon}(k)  }  \ ,
\end{align}
where $\sigma^2_{\varepsilon}(k)$ is variance of stochastic noise field $\varepsilon(\bfx)$, generally depending on the scale of the residual field. Though the field level modeling is expected to work extended to extremely non-linear scale, we choose the conservative minimum fitting scale $k_{\rm max}=0.2\, {\rm Mpc}^{-1}h$ to avoid the stochastic noise varying with scale, approximating $\sigma^2_{\varepsilon}$ as constant. 
We quantify the residual noise $\hat\varepsilon \equiv \delta_h(\bfk) - \hat\delta^{\rm HEFT}(\bfk)$ between the halo overdensity and bias expansion model with power spectrum, $P_{\rm err} = \la|\hat\varepsilon|^2\ra$, and present the result at redshift $z=1$ for Planck cosmology in Fig.~\ref{fig:halo_fit_Perr}. 
It shows that the bias expansion model describes the halo overdensity in the field level across a wide range $k\lesssim 1\,{\rm Mpc}^{-1}h$, with nearly flat residual noise power spectrum. The non-Poisson behavior switches from super-Poisson to sub-Poisson as the halo mass increases, consistent with the non-linearity and exclusion effect \cite{baldauf2013halo}. 
It also shows that utilizing 4 bias parameters $\{b_1, b_2, b_s, b_\nabla\}$ already captures most of the large-scale structure, leaving nearly scale-independent noise. The inclusion of $\delta^3$ field can further improve the non-linear clustering modeling on small scales, therefore reducing the residual noise for the low mass samples. While it has a negligible effect on massive samples, since the halo exclusion dominates the residual noise, and such a discrete feature is not well captured by the long-wavelength expansion. 
What's more, differing from the transfer function approach \cite{schmittfull2019modeling, schmittfull2021modeling}, where the biased tracer fields can be uniquely separated into the combination of orthogonal basis fields and residual stochastic field, we adopt Eq.~\ref{equ:field_level_bias} to estimate the constant bias parameter to match the power spectrum level modeling. It relies on the $k_{\rm max}$ which retains the validity of constant stochasticity. The alternative choice $k_{\rm max}=1\, {\rm Mpc}^{-1}h$ degrades the fitting by worsening the large-scale fitting, since it underestimates the stochasticity in small-scale modes and over-penalizes the large scales modes. We emphasize that conservative $k_{\rm max}$ is to ensure the validity of constant noise, but not limit the bias expansion modeling range, as presented in Fig.~\ref{fig:halo_fit_Perr}, where the residual field is nearly white spectra over a wide $k$-range. 
%Besides, the 4 bias fitting is insensitive to minimum scale $k_{\rm max}$ we adopt to fit the optimal bias parameters, where the alternative choice $k_{\rm max}=1\, {\rm Mpc}^{-1}h$ change the residual noise spectra slightly. While for the 5 bias fitting, the small scales modes degrade the fitting by worsening the large scales noise. It implies although the $\delta^3$ field improves the non-linear clustering modeling and suppress the residual noise, it potentially increases the scale-dependence of stochasticity since the residual high order bias terms are modified. 
Nevertheless, it manifests that our biased tracer modeling with hybrid Lagrangian bias expansion is possible to extend to scale $\sim 1\, {\rm Mpc}^{-1}h$.

In the above analysis, we are only concerned with the auto correlation of the residual noise. While the cross correlation between $\varepsilon(\bfx)$ and underlying matter may not vanish, and potentially impact the joint modeling in the case of combining galaxy clustering with weak lensing. 
To evaluate the joint analysis of galaxy clustering and weak lensing, we perform the joint fitting of the halo auto power spectrum and the halo-matter cross power spectrum, where the halo samples are obtained from 46 cosmologies that are not adopted for emulator construction, with 12 redshift bins and 5 halo mass bins for each simulation. 
%We adopt the Gaussian covariance approximation to build the likelihood, and vary the maximum fitting $k$ edge to seek the minimum scale at which the bias expansion can reach unbiased fitting. 
{We vary the maximum fitting $k$ edge to seek the minimum scale at which the bias expansion can reach unbiased fitting. The covariance used in the likelihood construction includes the emulator uncertainty and data covariance (see details in Appendix~\ref{append:gaussian_sample_variance}), where the data covariance consists of the diagonal components with Gaussian covariance approximation and the off-diagonal components estimated from 25 pairs of FastPM simulations.}
Across all the cosmology and halo mass samples, the bias model with 4 bias parameters plus a shot noise amplitude $\{b_1, b_2, b_s, b_\nabla, \alpha_0\}$ can reach scale $k_{\rm max} = 0.57\, {\rm Mpc}^{-1}h$ within $1\sim 2$ percent accuracy at $z\gtrsim 0.1$, except for few extremely massive halo bins at $z\simeq 0$ with slightly lower $k_{\rm max}$.
Notice that $1\%$ accuracy is already the emulator limit, and the realized fitting $k_{\rm max}$ is expected by previous works Refs.~\cite{hadzhiyska2021hefty, zennaro2023bacco, derose2023aemulus, chen2024analysis, kim2024atacama, sailer2024cosmological, nicola2024galaxy}.

%%%%%%%%%%%%%%%%%%%%%%%%%%%%%%%%%%%%%%%%%%%%%%%%%%%%%%%%%
%%%%%%%%%%%%%%%%%%%%%%%%%%%%%%%%%%%%%%%%%%%%%%%%%%%%%%%%%
%%%%%%%%%%%%%%%%%%%%%%%%%%%%%%%%%%%%%%%%%%%%%%%%%%%%%%%%%
%%%%%%%%%%%%%%%%%%%%%%%%%%%%%%%%%%%%%%%%%%%%%%%%%%%%%%%%%

%%%%%%%%%%%%%%%%%%%%%%%%%%%%%%%%%%%%%%%%%%%%%%%%%%%%%%%%%
%%%%%%%%%%%%%%%%%%%%%%%%%%%%%%%%%%%%%%%%%%%%%%%%%%%%%%%%%
\begin{figure*}[phtb!]
\centering
\includegraphics[width=0.99\textwidth]{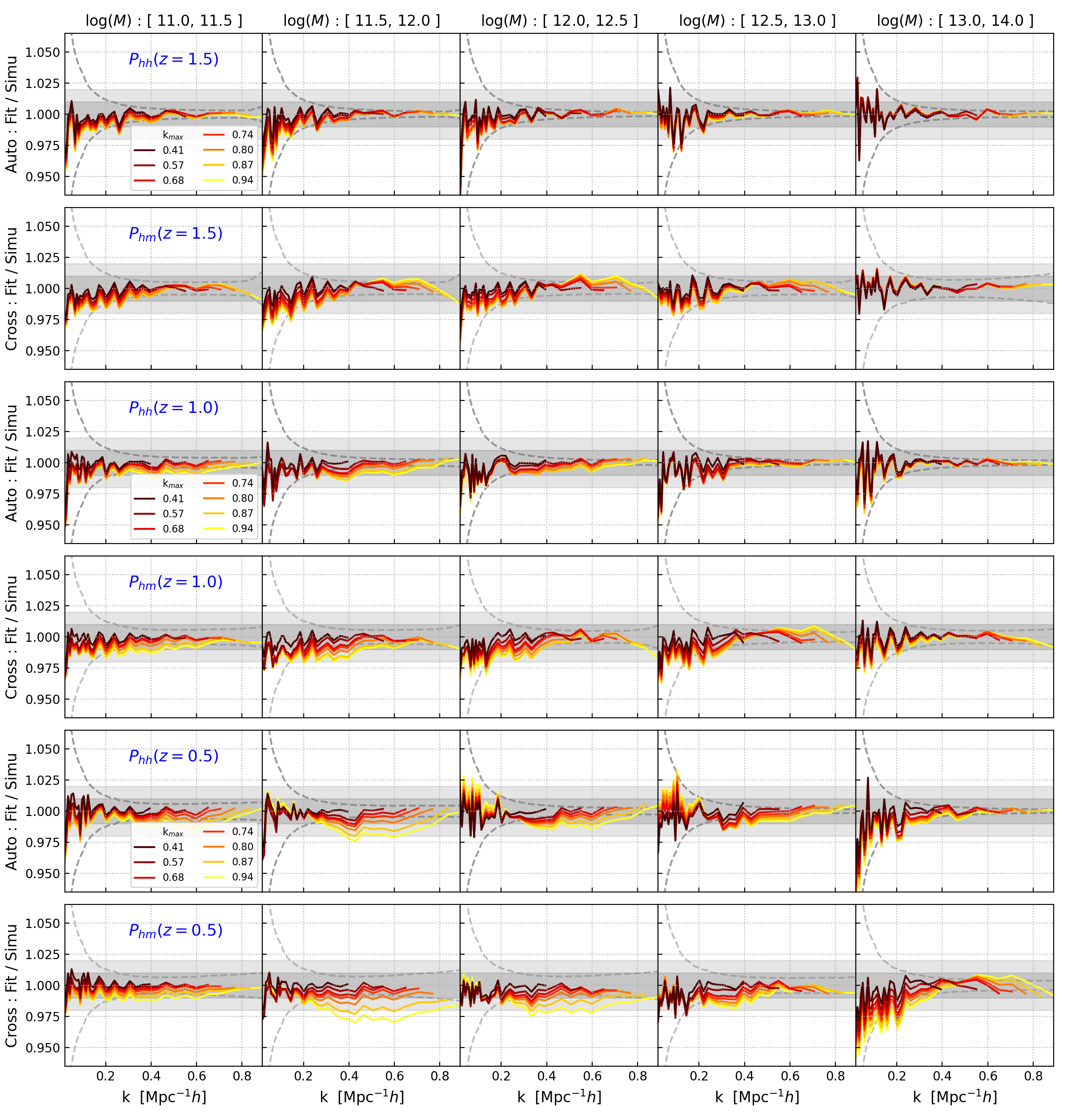}
\caption{ \label{fig:halo_fit}
The figure shows the ratio of the model prediction to the halo power spectrum measured from simulations. 
Here we perform a joint fit to the halo auto-power spectrum and the halo-matter cross-power spectrum, using halo samples randomly selected from 46 cosmologies that were not used in the emulator construction. 
The results of 3 redshift bins are shown, with $z=1.5$ at the upper 2 rows, $z=1.0$ at the middle 2 rows, and $z=0.5$ at the bottom 2 rows, respectively. Different columns represent different halo mass bins, with a total of 5 mass bins ranging $11<\log(M)<14$. 
For each redshift bin, the upper row presents the ratio of Lagrangian bias expansion model $\{b_1, b_2, b_s, b_{k^2}, \alpha_0\}$ to the measured halo auto power spectrum, and the lower row presents the ratio of model to the measured halo-matter cross power spectrum. The fitting is performed for 7 selected $k$-ranges, with maximum bin edges ranging $0.41 \leq k_{\rm max} \leq 0.94 \,{\rm Mpc}^{-1}h$. In each panel, the gray shades indicate the $\pm 1\%,\,\pm 2\%$ fractional error, and the gray dashed lines indicate $1\sigma$ uncertainty, the sum of Gaussian variance and emulator uncertainty. 
The approximately $2\sim 3\sigma$ deviation on small scales, e.g., the second mass bin at $z=0.5$, is likely caused by the inadequate approximation of data covariance in likelihood, and not due to the failure of the bias expansion model. 
}
\end{figure*}
%%%%%%%%%%%%%%%%%%%%%%%%%%%%%%%%%%%%%%%%%%%%%%%%%%%%%%%%%
%%%%%%%%%%%%%%%%%%%%%%%%%%%%%%%%%%%%%%%%%%%%%%%%%%%%%%%%%

However, it is still far from the $k_{\rm max}$ we obtain from the field level analysis, and including $\delta^3$ basis improves slightly, especially for the massive halo samples. The degradation is sourced by the poor fitting of the cross spectra, but not the auto spectra. We suppose the poor cross spectra fitting is due to the lack of enough degrees of freedom to absorb the impact arising from $P_{m\varepsilon}$, where such large impact is sourced by the strong matter clustering within the massive halo. It turns out to be beneficial to further include a $\beta k^2P_{11}$ term with free parameter $\beta$ to cross spectra, then all the issues on massive halo samples fitting disappear and $k_{\rm max}$ significantly increases. 
Another solution is replacing the counterterm $\nabla^2\delta$ in Lagrangian space with $\nabla^2\delta$ in Eulerian space, where the latter is simply the non-linear matter overdensity multiplied by $-k^2$ in Fourier space \cite{hadzhiyska2021hefty, kokron2021cosmology, derose2023aemulus}. These two counterterms agree with each other on the linear scale. While on a small scales, 
%such as $k\gtrsim 0.1\, {\rm Mpc}^{-1}h$ at $z\simeq0$, 
the spectra of Lagrangian space counterterm $P_{X\nabla^2}$ rise mildly, since we have selected only the modes $k\lesssim 1\, {\rm Mpc}^{-1}h$ in Lagrangian space. But the Eulerian counterterm $-k^2\,P_{X1}$ raises the small scales power radically because it directly amplifies the non-linear matter spectra. Such strong small scales power happens to absorb the non-linear clustering toward the halo profile dominating scale. Generally with 4 bias parameters $\{b_1, b_2, b_s, b_{k^2}, \alpha_0\}$, the unbiased fitting is able to reach $k_{\rm max} \simeq 0.8\, {\rm Mpc}^{-1}h$ for the halo samples at $z\gtrsim 1$, and $k_{\rm max} \simeq 0.7\, {\rm Mpc}^{-1}h$ for samples at $z\lesssim 1$. The smaller scale is allowed if we further include the $\delta^3$ basis.
In Fig.~\ref{fig:halo_fit}, we present the fitting results from one of the validation simulations, with 4 bias parameters utilizing the Eulerian space counterterm. Except for the massive halo in low redshift, which seems to bias the fitting at $k_{\rm max}\gtrsim 0.7\, {\rm Mpc}^{-1}h$, all the other fitting is still unbiased in the region $k_{\rm max} \gtrsim 0.8\, {\rm Mpc}^{-1}h$. 
%We are not concerned with the slight offset about a $1\sim 2$ percent level, since it has reached the emulator precision limit, as well as the affected by the violation of the Gaussian variance assumed in the likelihood construction. 
{We are not concerned with the $1\sim 2$ percent deviations in the fits at $k_{\rm max}\gtrsim\, 0.4 {\rm Mpc}^{-1}h$, as they are influenced by data covariance approximation. For example, there is an approximate $2\sigma$ deviation in the second mass bin at $z=0.5$ when $k_{\rm max}>\,0.8\,{\rm Mpc}^{-1}h$, but this discrepancy is alleviated when only the diagonal components of the covariance are used.}

Nevertheless, we are not aiming to determine a better counterterm in biased tracer modeling. In realistic galaxy samples, we do not expect a similar challenge for halos would happen for galaxies, i.e., the strong matter clustering around a massive halo extending a wide spatial region. We present the joint fitting here as a simple but general validation for the emulator, and it has exhibited the fitting performance for biased spectra though with approximated data covariance. While for galaxy clustering, the optimal bias model and appropriate scale cut depend on the details of galaxy sample properties, and it requires significant work in the future.

% ---------------------------------------------------------------------------------------
% ---------------------------------------------------------------------------------------

\section{Discussion and Conclusion}
\label{sec:discussion}

In this work, we present a hybrid Lagrangian bias expansion emulator utilizing the \Kun simulation suite, emulating across 8 cosmological parameters, including neutrino mass and dynamical dark energy parameters $w_0w_a$. We combine the Lagrangian bias expansion and accurate dynamical evolution in $N$-body simulation, to measure the Lagrangian basis spectra up to quadratic order, as well as a derivative bias term and a cubic order $\delta^3$ term. We employ the Zel'dovich variance control to suppress the sample variance in the basis spectra, achieving precise measurement and excellent agreement with theoretical predictions. We emulate the quadratic order spectra in one percent level accuracy, covering the scale $ k \leq 1\, {\rm Mpc}^{-1}h$ and redshift range $0\leq z\leq 3$. 
To validate the prediction ability of the emulator, we perform joint fitting of halo auto spectra and halo-matter cross spectra on the halo sample from 46 \Kun simulations not adopted to emulator training. We find that the quadratic order expansion based on the emulator output can achieve unbiased fitting up to $k \simeq 0.57\, {\rm Mpc}^{-1}h$, a typical scale reported in previous works. If we adopt an aggressive counterterm, we can alleviate the strong clustering feature in massive halo samples, and advance to smaller scale $k \simeq 0.7\, {\rm Mpc}^{-1}h$. We also verify that the third-order $\delta^3$ term can further suppress the stochasticity in small halo samples and optimize the model fitting. We emphasize that the halo spectra fitting serves as a validation of the emulator, testing the flexibility of the constructed emulator for general biased tracers, but not to explore the possibility of applying the emulator to real galaxy samples on a smaller scale. 
%The galaxy clustering modeling combined with weak gravitational lensing, including potential issues such as baryonic feedback, intrinsic alignment and observation systematics, requires further investigation and calibration in survey mock, which we will present in future works. 

Previous works have presented two HEFT power spectrum emulators constructed from \BACCO \cite{zennaro2023bacco} and \Aemulus \cite{kokron2022accurate} suites respectively, where \BACCO employs Gaussian smoothing for Lagrangian density field while \Aemulus does not adopt additional smoothing. 
We utilize only the Lagrangian density modes at $k\lesssim 1\, {\rm Mpc}^{-1}h$ to construct the Lagrangian basis fields, thus we are unable to make a point-by-point comparison. 
%\footnote{ The commonly matter power spectrum has been shown in the previous publication \cite{chen2025csst}. } 
\BACCO and \Kun simulation suites have similar parameter space design, both including the dynamical dark energy $w_0w_a$ parameters. But \BACCO employs the cosmology-rescaling algorithm to sample 400 cosmologies from only 4 $N$-body simulations, then calculates the basis spectra and replaces some noisy cross spectra with the analytic spectra at $k<0.1\, {\rm Mpc}^{-1}h$. While we calculate the basis spectra from high-resolution $N$-body simulations solely, and have precise measurement extending to a larger scale $k\sim 0.01\, {\rm Mpc}^{-1}h$ thanks to Zel'dovich variance control. Besides, we achieve excellent agreement between simulation outputs and the 1-loop theory prediction, exhibiting the advantageously theoretical control of the basis spectra. As for the emulator from \Aemulus, they simulate and emulate the impact of massive neutrino and $w$ dark energy. By contrast, \Kun simulations also cover neutrino mass but further $w_0w_a$ parameters. Both our construction and \Aemulus measure the precise spectra covering wide scale and emulate the quadratic order spectra at one percent level accuracy. 
We utilize the same technique for variance reduction and similar theoretical spectra to reduce the dynamic range of spectra. Additionally, we extend the Lagrangian basis fields with a derivative bias and a third order bias term, offering slightly more flexibility. 
Overall, on the one hand, these emulators are designed with different choices, and they possess complementary advantages, particularly covering different cosmological parameter spaces. On the other hand, multiple hybrid models enable the possibility of cross-validation, providing a theoretical toolkit support for many galaxy surveys going beyond CSST in the coming decades.

As the first step in emulating the biased tracer spectra for CSST, we confine the construction and validation to the tracer in real space. It is aimed at modeling the galaxy angular power spectrum and the galaxy-lensing cross angular power spectrum, which are the key science targets of the CSST mission.
Though the emulation has reached 1\% level accuracy, the projection statistics will further suppress the potential error compared to the designed precision shown in the main text, and further broaden the scale in the emulator application \cite{nicola2024galaxy}. 
Other impacts on the clustering of matter and galaxies, such as baryonic feedback, are not considered in the main text, since they can be incorporated at the power spectrum level, independently of emulator construction (e.g., \cite{2015JCAP...12..049S,2019JCAP...03..020S,2021MNRAS.506.4070A,2021JCAP...12..046G,2023MNRAS.523.2247S}). 
Furthermore, the redshift space distortion is required to model the full shape spectrum in redshift space, and we will incorporate it and extend the scope from projected statistics to the fully three-dimensional spectrum in subsequent works.

The emulator is publicly available at \url{https://github.com/ShurenZhou1999/csstlab}. It is implemented in pure Python and does not require additional compilation or heavily depend on external libraries. Generating all the basis spectra for a given cosmology takes approximately $\sim 20$ milliseconds, while interpolating the spectra into the typical $(k,z)$ bins takes about $\mO(10^{-2})$ seconds. The high efficiency meets the requirements of the future CSST cosmological analysis, as well as other ongoing and upcoming stage-IV galaxy surveys.

% ---------------------------------------------------------------------------------------
% ---------------------------------------------------------------------------------------
% ---------------------------------------------------------------------------------------
% ---------------------------------------------------------------------------------------

\Acknowledgements{
We thank Pengjie Zhang for the insightful discussions and his generous support for Shuren Zhou. 
This work is supported by the National Key R\&D Program of China (2023YFA1607800, 2023YFA1607801, 2020YFC2201602), the National Science Foundation of China (12273020), the China Manned Space Project with No. CMS-CSST-2021-A03 and CMS-CSST-2025-A04. 
This work was supported by the National Center for High-Level Talent Training in Mathematics, Physics, Chemistry, and Biology. This work was also supported by Yangyang Development Fund. 
The \textsc{Kun} simulation suite is run on Kunshan Computing Center, and 
the analysis uses the Gravity Supercomputer at the Department of Astronomy, Shanghai Jiao Tong University. 
}

%%%%%%%%%%%%%%%%%%%%%%%%%%%%%%%%%%%%%%%%%%%%%%%%%%%%%%%
%%% Conflict of interest.
%%%%%%%%%%%%%%%%%%%%%%%%%%%%%%%%%%%%%%%%%%%%%%%%%%%%%%%
\InterestConflict{The authors declare that they have no conflict of interest.}

%%%%%%%%%%%%%%%%%%%%%%%%%%%%%%%%%%%%%%%%%%%%%%%%%%%%%%%
%%% Supplements
%%%%%%%%%%%%%%%%%%%%%%%%%%%%%%%%%%%%%%%%%%%%%%%%%%%%%%%
%\Supplements{}

%%%%%%%%%%%%%%%%%%%%%%%%%%%%%%%%%%%%%%%%%%%%%%%%%%%%%%%
%%% Reference section.
%%% citation in the content using "some words~\cite{1,2}".
%%% ~ is needed to make the reference number is on the same line with the word before it.
%%%%%%%%%%%%%%%%%%%%%%%%%%%%%%%%%%%%%%%%%%%%%%%%%%%%%%%
%% Recommended: using scpma.bst file
\bibliographystyle{scpma}
\bibliography{citations}

\end{multicols}
% \begin{widetext}

% ---------------------------------------------------------------------------------------
% ---------------------------------------------------------------------------------------
% ---------------------------------------------------------------------------------------
% ---------------------------------------------------------------------------------------

%%%%%%%%%%%%%%%%%%%%%%%%%%%%%%%%%%%%%%%%%%%%%%%%%%%%%%%
%%% Appendix sections
%%%%%%%%%%%%%%%%%%%%%%%%%%%%%%%%%%%%%%%%%%%%%%%%%%%%%%%
\appendix

% ---------------------------------------------------------------------------------------
% ---------------------------------------------------------------------------------------

\section{Convolution Lagrangian Perturbation Theory}
\label{appendix:CLEFT-intro}

In this section, we briefly summarize the content of the perturbation theory in the emulator construction, and more detailed derivations can be found in the references. From the perspective of Lagrangian dynamics, the galaxy spectrum can be written as
\begin{equation}
(2\pi)^3\delta^D_{k,0} + P_g(k) = \int_\bfq e^{-i\bfk\cdot\bfq}\,\la [1+\delta_g(\bfq_1)][1+\delta_g(\bfq_2)]\,e^{i\bfk\cdot\bfDelta}\ra \ .
\end{equation}
In the Lagrangian bias expansion Eq.~\ref{equ:deltag_q} for galaxy overdensity, generally we have
\begin{equation}
F(\bfq) \equiv 1+\delta_g(\bfq) = \sum_i b_i \mO_i(\bfq) \ ,
\end{equation}
where the Lagrangian basis fields $\mO_i \in\{1,\delta,\delta^2,s^2,\delta^3, \cdots\}$, and we have suppressed the potentially stochastic contribution. The galaxy spectrum is accordingly expanded as
\begin{align}
& P_g(k) = \sum_{ij} b_ib_j P_{ij}(k)    \label{equ:cleft-key-1}  \ ,  \\
& P_{ij}(k) = \int_\bfq e^{-i\bfk\cdot\bfq} \, \mM_{ij}(\bfk,\bfq)     \label{equ:cleft-key-2}  \ , \\
& \mM_{ij}(\bfk,\bfq) = \la \mO_i(\bfq_1) \mO_j(\bfq_2) \,e^{i\bfk\cdot\bfDelta}\ra \,|_{\bfq=\bfq_1-\bfq_2}  \ ,   \label{equ:cleft-key-3}
\end{align}
where we have $\mO_i\,\in\{1,\delta,\delta^2,s^2\}$ up to quadratic order bias expansion. In a given perturbation order, the Lagrangian perturbation theory solves the kernel Eq.~\ref{equ:cleft-key-3} with the perturbation expansion of displacement $\bfDelta\equiv \mPsi(\bfq_1)-\mPsi(\bfq_2)$, and therefore obtains the basis spectra Eq.~\ref{equ:cleft-key-2}. 
The galaxy clustering is the tracer of the cold dark matter and baryon ($cb$), therefore we can simply plug the \CLASS output $P_{cb}(k)$ into the integration of Eq.~\ref{equ:cleft-key-3} and obtain the galaxy auto power spectrum. While the gravitational lensing is sensitive to the total matter ($tot$) including \textit{cb}+neutrino, thus in principle, one needs to account for the massive neutrino clustering for the matter side in the galaxy-lensing cross power spectrum.
Specifically, we have to separate the $P_{cb, cb}, P_{cb, tot}$ and $P_{tot, tot}$ contributions in the integral Eq.~\ref{equ:cleft-key-3} expression \cite{chen2022cosmological}, similar to the implementation for initial condition in Sec.~\ref{sec:cleft_theory}. However, our simulations do not treat neutrinos as extra particle species, and our direct measurement is \textit{cb} clustering but not \textit{tot} field. Therefore, to match the simulation results, we only consider \textit{cb} components and use $P_{cb}(k)$ as the theoretical calculation input. 
The neutrino clustering is incorporated in the basis spectrum emulation via Eq.~\ref{equ:tot-cb-convert-1}-\ref{equ:tot-cb-convert-3}.

\subsection{1-loop Basis Spectra}
\label{appendix:CLEFT-1loop}

For brevity, we suppress the exponential factor of infrared resummation, reexpressing the kernel $\mM_{ij}$ as $\mK_{ij}$ by 
\begin{equation}\label{equ:theory_kernel_redefine}
\mM_{ij} = e^{-{1\over2}\,k_ik_jA_{ij}^L} \mK_{ij} \ ,
\end{equation}
and all the disconnected terms are expressed in terms of the generalized correlation function, defined as \cite{schmittfull2016fast}
\begin{equation}
\xi_n^\ell(r) = \int_0^\infty {k^2\rmd k\over 2\pi^2}\, j_\ell(kr)\,k^n P_L(k) \ .
\end{equation}
Here $P_L(k)$ is the linear power spectrum of the initial density field, and $\xi\equiv\xi^0_0$ is the ordinary correlation function. In Eq.~\ref{equ:theory_kernel_redefine}, the exponential resummate the linear displacement correlator $k_ik_j A_{ij} \equiv \la\Delta^L\Delta^L\ra = k^2( X^L + \mu^2 Y^L )$, where $\mu=\bf \hatk\cdot\hatq$ and upper index $L$ denote linear order contribution. 
The kernels of infrared resummation basis spectra up to 1-loop order are given by \cite{white2014zel, carlson2013convolution, vlah2016gaussian, wang2014analytic, vlah2015lagrangian, vlah2019exploring, chen2020consistent}
\begin{align}
\mK_{11} &= 1 -{1\over2}\,k^2\left( X^\rmloop + \mu^2 Y^\rmloop \right) - {i\over6} k^3 \left( \mu V + \mu^3 T \right)   \ ,   \\
\mK_{1\delta} &=  -ik\mu\, \xi^1_{-1} + ik\mu\, U^{(3)} - {1\over2}k^2 (X^{10} +\mu^2 Y^{10} )    \ ,   \\
\mK_{1\delta^2} &=  i k\mu U^{20} -  k^2\mu^2 (\xi^1_{-1})^2   \ ,     \\
\mK_{1s^2} &= ik\mu U^{s2} -{1\over2}k^2 (X^{20} +\mu^2 Y^{20} )  \ ,   \\
\mK_{\delta\delta} &=   \xi +ik\mu\, U^{11} - k^2\mu^2 (\xi^1_{-1})^2  \ ,    \\
\mK_{\delta\delta^2} &=  - 2 i k\mu \xi\, \xi^1_{-1}  \ ,   \\
\mK_{\delta s^2} &= ik\mu\, V^{12}  \ ,  \\
\mK_{\delta^2\delta^2} &=  2 \left(\xi\right)^2  \ , \\
\mK_{\delta^2s^2} &=  {4\over 3}\left(\xi^2_0\right)^2  \ ,  \\
\mK_{s^2s^2} &=  {8\over45} \xi^2 + {16\over 63} \left(\xi^2_0\right)^2 + {16\over 35} \left(\xi^4_0\right)^2 \ .
\end{align}
The explicit expression of 1-loop integration $X^\rmloop$, $Y^\rmloop$, $V$, $T$, $U^{(3)}$, $X^{10}$, $Y^{10}$, $U^{20}$, $U^{s2}$, $X^{20}$, $Y^{20}$, $U^{11}$, $V^{12}$ have been detailed in previous works, e.g. Ref.~\cite{matsubara2008resumming, matsubara2008nonlinear}.

Apart from the basis spectra of leading and quadratic order fields, we also present the analysis of the Lagrangian space counterterm $\nabla^2\delta$ and the cubic order term $\delta^3$ in the main text. Their integration kernels up to 1-loop order are given by
\begin{align}
\mK_{1\nabla^2}  &=  ik\mu\, \xi^{1}_{1} - \frac{1}{2} \la[\nabla^2\delta_2] \Delta\Delta\ra_c    \ ,  \\
\mK_{\delta\nabla^2}   & =  - \xi_2^0  + k^2\mu^2 \xi^1_{-1}\, \xi^1_{1}  + i \la\delta_1 [\nabla^2\delta_2] \Delta\ra_c    \ ,   \\ 
\mK_{\delta^2\nabla^2}  & =  ik\mu\, \xi_{-1}^1\xi_2^0    \ ,   \\
\mK_{s^2\nabla^2}  & =  - ik\mu\; \xi^2_2\, \left( {8\over 15} \xi^1_{-1} - {4\over 5} \xi^3_{-1}\right)  \ , \\
\mK_{\nabla^2\nabla^2}  & =  \xi_4^0 -  k^2\mu^2 \left(\xi^1_{1}\right)^2 + i \la[\nabla^2\delta_1][\nabla^2\delta_2] \Delta\ra_c  \ ,   \\
\mK_{\delta^3 1} &= {1\over 6}ik^3\mu^3 \left(\xi^{1}_{-1}\right)^3  + {1\over 2}k^2\mu^2\, \xi^{1}_{-1}\, U^{20}    + {1\over 6} i \la\delta_1^3 \Delta\ra_c^{\rm(2-loop)}      \label{equ:oneloop-kernel-1delta3} \ ,  \\
\mK_{\delta^3\delta} &= - {1\over 2}k^2\mu^2\, \xi\,\left(\xi^1_{-1}\right)^2  + {1\over 2}ik\mu\, \xi\, U^{20}  \ ,  \\
\mK_{\delta^3\delta^2} &= - {1\over 2} ik \mu\; (\xi)^2\, \xi^1_{-1}  \ ,   \\
\mK_{\delta^3 s^2} &=  - {2\over 3} ik \mu\, \left(\xi^2_0 \right)^2  \xi^1_{-1}  \ ,    \\
\mK_{\delta^3\nabla^2} &=  {1\over 2}k^2\mu^2\, \xi^0_2\,\left(\xi^1_{-1}\right)^2  - {1\over 2}ik\mu\, \xi^0_2\, U^{20}    \ ,  \\
\mK_{\delta^3\delta^3} &=  {1\over 6}(\xi)^3  \ ,      
\end{align}
where the lower index in $\la\cdots\ra_c$ denotes the cumulant. The correlator can be formulated as double integrals, 
\begin{align}
\la\delta_1 [\nabla^2\delta_2] \Delta\ra_c
&= {3\over 7}\, \mu\, \int {kdk\over 2\pi^2}\, j_1(kq)\, P(k) \,\left[ k^2 G(k) + G^{(k2)}(k)\right]     \ ,   \\  
\la[\nabla^2\delta_1] [\nabla^2\delta_2] \Delta\ra_c 
&= - {6\over 7} \mu \int {kdk\over 2\pi^2}\, j_1(kq)\, P(k) \,\left[ k^2 G^{(k2)}(k)\right] \ ,      
\end{align}
where the inner integrations are
\begin{align}
G(k) &= k\, \int rdr \left[ {2\over 3}j_0(kr)\xi^0_0(r) - {2\over 3}j_2(kr)\xi^2_0(r) \right]
+ \int rdr \left[ -{2\over 5}j_1(kr)\xi^1_1(r) + {2\over 5}j_3(kr)\xi^3_1(r) \right]   \ ,   \\
G^{(k2)}(k) &= k\, \int rdr \left[ {2\over 3}j_0(kr)\xi^0_2(r) - {2\over 3}j_2(kr)\xi^2_2(r) \right]
+ \int rdr \left[ -{2\over 5}j_1(kr)\xi^1_3(r) + {2\over 5}j_3(kr)\xi^3_3(r) \right] \ .
\end{align}
Notice that $G(k)$ can be expressed in terms of the combination of familiar 1-loop integrations derived in Ref.~\cite{matsubara2008resumming, matsubara2008nonlinear,vlah2016gaussian,chen2020consistent}.
After similar and tedious derivation, one can obtain the expression of correlator $\la[\nabla^2\delta_2] \Delta\Delta\ra_c$. 

As manifested in Eq.~\ref{equ:oneloop-kernel-1delta3}, the leading order contribution of the basis $\mK_{1\delta^3}$ comes from three parts, the disconnected term, the 1-loop term and the 2-loop term $\la\delta_1^3 \Delta\ra_c$. Thus, keeping up to 1-loop contributions is not complete to account for the leading behavior for all the third-order basis spectra.

\subsection{$k$-expanded Basis Spectra}
\label{appendix:CLEFT-kexpand}

As discussed in the main text, the 1-loop spectra with fully linear displacement resummation achieve an accurate description of the long-wavelength behavior, but pay the price of damping spectra on small scales. 
%For example, the matter power spectrum $P_{11}(k)$ damps and cross zero even at $k\sim 0.3\, {\rm Mpc}^{-1}h$ in some cosmology we interest in, which is far from the non-linear scale we require. 
To maintain the stable behavior on small scales so that we can reduce the dynamical range of simulation spectra, we expand the infrared resummation exponential as the following integration kernels \cite{chen2020consistent, derose2023aemulus}

\begin{align}
\mM_{11} &= 1 - {1\over2}\,k^2\left( X^\rmloop + \mu^2 Y^\rmloop \right) - {i\over6} k^3 \left( \mu V + \mu^3 T \right)  
+ {1\over 8} \,k^4\left[ (X^L)^2 + 2 \mu^2 X^L Y^L+ \mu^4 (Y^L)^2 \right]  \ ,   \\
\mM_{1\delta} &=  -ik\mu\, \xi^1_{-1} + ik\mu\, U^{(3)} - {1\over2}k^2 (X^{10} +\mu^2 Y^{10} )
+ {1\over2}i k^3 \xi^1_{-1} (\mu X^L + \mu^3 Y^L )     \ ,  \\
\mM_{1\delta^2} &=  i k\mu U^{20} -  k^2\mu^2 (\xi^1_{-1})^2    \ ,   \\
\mM_{1s^2} &= ik\mu U^{s2} - {1\over2}k^2 (X^{20} +\mu^2 Y^{20} )    
+ {1\over4} k^4 (X^{20}X^L + \mu^2 Y^{20}X^L + \mu^2 X^{20}Y^L + \mu^4 Y^{20}Y^L )  \ ,  \\
\mM_{\delta\delta} &=   \xi +ik\mu\, U^{11} - k^2\mu^2 (\xi^1_{-1})^2  
- {1\over2}\,k^2 \xi\, \left( X^L + \mu^2 Y^L \right)  \ ,  \\
\mM_{\delta\delta^2} &=  - 2 i k\mu \xi\, \xi^1_{-1}   
+ i k^3\mu^3 \left(\xi^1_{-1}\right)^3   \ ,  \\
\mM_{\delta s^2} &= ik\mu\, V^{12}  
+  {1\over 2} i k^3 \xi^1_{-1}\, (\mu X^{s2} + \mu^3 Y^{s2}) \ , \\
% \mM_{\delta^2\delta^2} &=  2 \left(\xi\right)^2   \\
% \mM_{\delta^2s^2} &=  {4\over 3}\left(\xi^2_0\right)^2    \\
\mM_{s^2s^2} &=  \left[ {8\over45} \xi^2 + {16\over 63} \left(\xi^2_0\right)^2 + {16\over 35} \left(\xi^4_0\right)^2 \right] \left[  1 - {1\over2}\,k^2\left( X^L + \mu^2 Y^L \right)  \right]    \ ,  \\
\mM_{1\nabla^2}  &=  ik\mu\, \xi^{1}_{1} - \frac{1}{2} \la[\nabla^2\delta_2] \Delta\Delta\ra_c  - {1\over2} i\,k^3 \xi^{1}_{1}\, \left( \mu X^L + \mu^3 Y^L \right)    \ ,  \\
\mM_{\delta\nabla^2}   & =  - \xi_2^0  + k^2\mu^2 \xi^1_{-1}\, \xi^1_{1}  + i \la\delta_1 [\nabla^2\delta_2] \Delta\ra_c  + {1\over2}\,k^2 \xi_2^0 \left( X^L + \mu^2 Y^L \right)   \ ,  \\ 
% \mM_{\delta^2\nabla^2}  & =  ik\mu\, \xi_{-1}^1\xi_2^0       \\
% \mM_{s^2\nabla^2}  & =  - ik\mu\; \xi^2_2\, \left( {8\over 15} \xi^1_{-1} - {4\over 5} \xi^3_{-1}\right)   \\
% \mM_{\nabla^2\nabla^2}  & =  \xi_4^0 -  k^2\mu^2 \left(\xi^1_{1}\right)^2 + i \la[\nabla^2\delta_1][\nabla^2\delta_2] \Delta\ra_c     \\
\mM_{\delta^3 1} &= {1\over 6}ik^3\mu^3 \left(\xi^{1}_{-1}\right)^3  + {1\over 2}k^2\mu^2\, \xi^{1}_{-1}\, U^{20}     \ ,  \\
\mM_{\delta^3\delta} &= - {1\over 2}k^2\mu^2\, \xi\,\left(\xi^1_{-1}\right)^2  + {1\over 2}ik\mu\, \xi\, U^{20}   \ , \\
\mM_{\delta^3\delta^2} &= - {1\over 2} ik \mu\; (\xi)^2\, \xi^1_{-1}     \ .  
% \mM_{\delta^3 s^2} &=  - {2\over 3} ik \mu\, \left(\xi^2_0 \right)^2  \xi^1_{-1}      \\
% \mM_{\delta^3\nabla^2\delta} &=  {1\over 2}k^2\mu^2\, \xi^0_2\,\left(\xi^1_{-1}\right)^2  - {1\over 2}ik\mu\, \xi^0_2\, U^{20}      \\
% \mM_{\delta^3\delta^3} &=  {1\over 6}(\xi)^3        \\
\end{align}
Here, we apply the zeroth, first, or second order expansion for the infrared resummation factor, depending on the performance test for optimal emulation. We do not seek a consistent exponential expansion for different order basis spectra, instead, we aim to retain more power on small scales therefore these spectra serve as stable theoretical templates.
%given the smoothing scale $R=5\,h^{-1}{\rm Mpc}$ we adopt for the linear power spectrum. 
Thus, our artificial choice in the resummation expansion affects nothing.

\subsection{Zel'dovich Approximation}
\label{appendix:CLEFT-zel}

In the presence of only linear displacement $\mPsi\rightarrow\mPsi^{(1)}$, all the correction terms for non-linear displacement field are removed, and the cumulant expansion is truncated at a certain order since the linear displacement can not connect higher order correlations. Or mathematically, the only non-vanishing cumulants of a Gaussian field are the first two cumulants. Thus the convolutional Lagrangian perturbation theory with fully linear displacement resummation provides the analytically accurate results of the basis spectra \cite{vlah2016gaussian,vlah2015lagrangian, white2014zel, kokron2022accurate}. Apart from the leading and quadratic order expression presented in Ref.~\cite{kokron2022accurate}, we provide the result of power spectrum kernel for the $\nabla^2\delta$ and $\delta^3$ fields, 
\begin{align}
\mK_{1\nabla}  &=
ik\mu\,\xi^1_1  \ ,  \\
\mK_{\delta\nabla}  &=
-\xi^0_2 + k^2\mu^2\,\xi^1_1\, \xi^1_{-1}        \ ,  \\
\mK_{\delta^2\nabla}  &=
ik\mu\,\xi^0_2\xi^1_{-1}  \;-\; {1\over 2}ik^3\mu^3\,\xi^1_1\,(\xi^1_{-1})^2     \ ,   \\
\mK_{s^2\nabla}  &=
- \frac{1}{2} ik^3\,\xi^1_1\, ( \mu X^{s2} + \mu^3 Y^{s2} ) 
+ {4\over 3}ik\mu\, \xi^2_2\, \left({3\over 5}\xi^3_{-1} -{2\over 5}\xi^1_{-1}\right)       \ ,\\ 
\mK_{\nabla\nabla}  &=
\xi^0_4 - k^2\mu^2\,(\xi^1_1)^2       \ ,  \\
\mK_{\delta^3 1}  &=
{1\over 6} i k^3\mu^3 (\xi^1_{-1})^3     \ ,  \\
\mK_{\delta^3\delta}   &=
{1\over 6} k^4\mu^4 (\xi^1_{-1})^4   \;-\;  {1\over 2}k^2\mu^2 \xi^0_0\, (\xi^1_{-1})^2             \ ,  \\
\mK_{\delta^3\delta^2}  &=
- i\frac{1}{12} k^5\mu^5  \left(\xi^1_{-1}\right)^5  \;+\; {1\over 2}i\,k^3\mu^3\, \xi^0_0\, (\xi^1_{-1})^3   \;-\;  {1\over 2}i\,k\mu\, (\xi^0_0)^2 \xi^1_{-1}     \ ,\\
\mK_{\delta^3 s^2}   &=
{2\over 3}ik^3\mu^3\, \left(\xi^1_{-1}\right)^2\, \xi^2_0\, \left({3\over 5}\xi^3_{-1} -{2\over 5}\xi^1_{-1}\right)
- \frac{2}{3} ik\mu\, \xi^1_{-1}\, \left(\xi^2_0\right)^2 
\;-\; i {1\over 12} k^5 \left(\xi^1_{-1}\right)^3\, (\mu^3 X^{s2} + \mu^5 Y^{s2})  \ ,  \\    
\mK_{\delta^3\nabla}   &=
-{1\over 6} k^4\mu^4 (\xi^1_{-1})^3\, \xi^1_1   \;+\;  {1\over 2}k^2\mu^2 \xi^0_2\, (\xi^1_{-1})^2            \ ,   \\
\mK_{\delta^3\delta^3}   &=
{1\over 6}(\xi^0_0)^3 - {1\over 2}\, k^2\mu^2 (\xi^0_0\xi^0_{-1})^2  + \frac{1}{4} k^4\mu^4 \,\xi^0_0 \left(\xi^1_{-1}\right)^4   - \frac{1}{36} k^6\mu^6 \left(\xi^1_{-1}\right)^6  \ ,
\end{align}
where all the non-vanishing high order cumulants have been included in the kernel expressions. We have used the short notations for scalar tidal field \cite{white2014zel, vlah2016gaussian}
\begin{align}
\nonumber  X^{s2} &= 4 (J_3)^2 \ , \\
\nonumber  Y^{s2} &= 6(J_2)^2 +4(J_3)^2 + 2(J_4)^2 + 8 J_2J_3 + 4 J_2J_4 + 8 J_3J_4    \ , \\
\nonumber  J_2 &=  {2\over 15}\xi^1_{-1} - {2\over 5}\xi^3_{-1}    \;,\quad
J_3 = -{1\over 5}\xi^1_{-1} -{1\over 5}\xi^3_{-1}    \;,\quad
J_4 =  \xi^3_{-1}  \ .
\end{align}

% ---------------------------------------------------------------------------------------
% ---------------------------------------------------------------------------------------

%%%%%%%%%%%%%%%%%%%%%%%%%%%%%%%%%%%%%%%%%%%%%%%%%%%%%%%%%
%%%%%%%%%%%%%%%%%%%%%%%%%%%%%%%%%%%%%%%%%%%%%%%%%%%%%%%%%
%\begin{figure*}[htb!]
%\centering
%\begin{minipage}[b]{0.7\textwidth} 
%    \includegraphics[width=\textwidth]{./fig_LPTfield_maps_z1_zoomin.png} 
%\end{minipage}
%\hspace{0.1cm} 
%\begin{minipage}[b]{0.243\textwidth} 
%    \includegraphics[width=\textwidth]{./%fig_LPTfield_maps_z1_zoomin_DeltaTheta.png} 
%\end{minipage}
%\caption{ \label{fig:field_maps_zoomin}
%{\bf Left 3 columns }: Same as Fig.~\ref{fig:field_maps}, but zooming in the bottom-right region, with physical size $100\times 100\,h^{-2}{\rm Mpc}^2$.
%{\bf Right column }: The upper subfigure is the matter density contrast, same as the Lagrangian field $\mO_i(\bfq)=1$, but it is shown in higher grid resolution and the color-scaling is adjusted to emphasize the halo structure. The lower subfigure is the velocity divergence $\theta = - \nabla\cdot\bfv/ (afH)$, where the light yellow color indicates the region with infalling velocity toward the center of the dense structure. 
%}
%\end{figure*}
%%%%%%%%%%%%%%%%%%%%%%%%%%%%%%%%%%%%%%%%%%%%%%%%%%%%%%%%%
%%%%%%%%%%%%%%%%%%%%%%%%%%%%%%%%%%%%%%%%%%%%%%%%%%%%%%%%%

%%%%%%%%%%%%%%%%%%%%%%%%%%%%%%%%%%%%%%%%%%%%%%%%%%%%%%%%%
%%%%%%%%%%%%%%%%%%%%%%%%%%%%%%%%%%%%%%%%%%%%%%%%%%%%%%%%%
\begin{figure*}[htb!]
\includegraphics[width=0.99\textwidth]{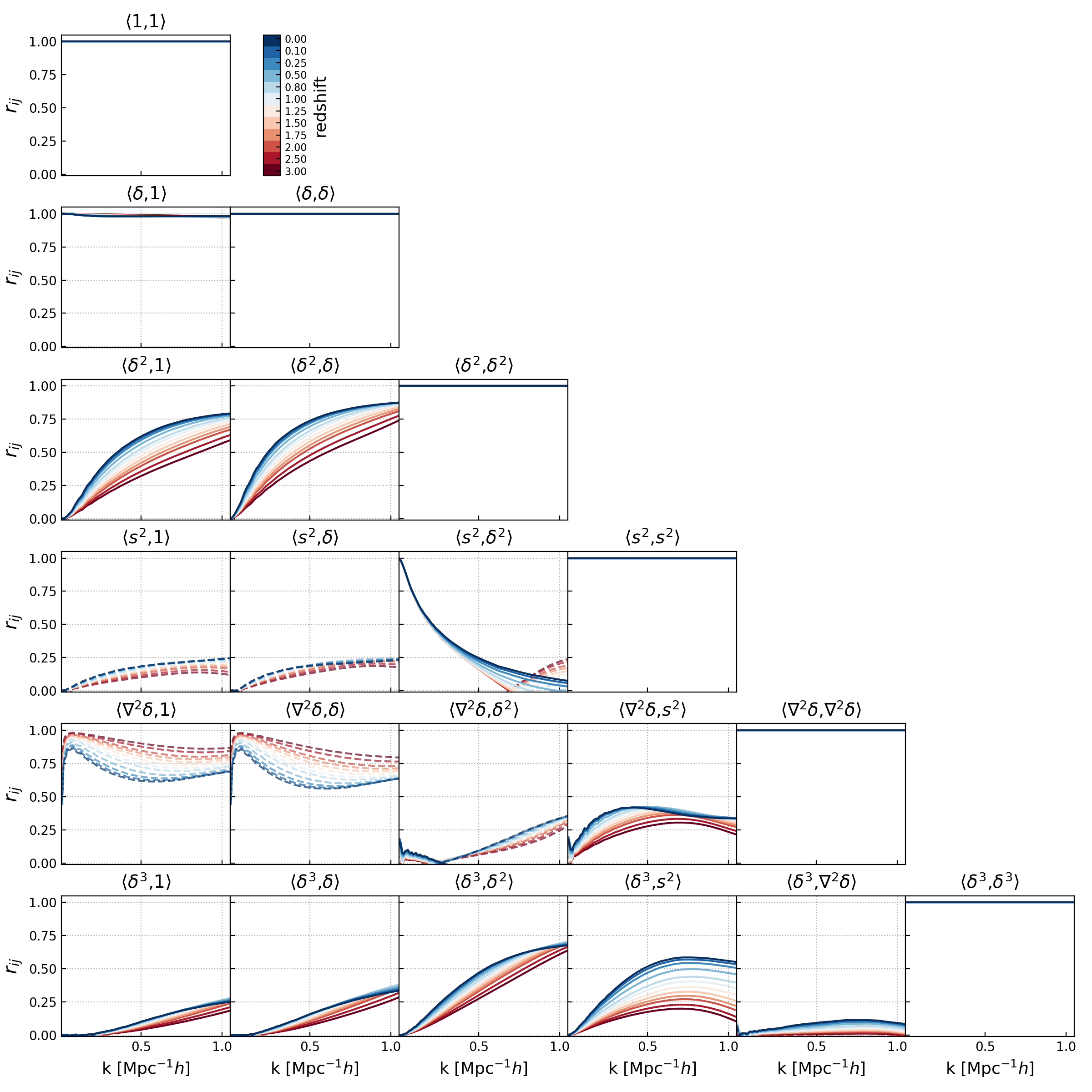}
\caption{ \label{fig:r2_ij}
The cross correlation coefficient {$r_{ij} = P_{ij} /\sqrt{ P_{ii}P_{jj}}$} between Lagrangian basis fields $\mO_{i}(\bfx, z)$ and $\mO_{j}(\bfx, z)$, measured in the first cosmological simulation of \Kun suite. 
Line colors range from red to blue, corresponding to redshifts from $z=3.0$ to $z=0.0$. 
For visualization, we take the absolute value of the $r_{ij}$, and the negative values are indicated as dashed lines. 
Though the measurement of basis spectra such as $P_{1\delta^3}$, $P_{\delta\delta^3}$, and $P_{\nabla\delta^3}$ is affected by the sample variance, their relative contribution compared to the auto spectra is sub-dominant. Therefore, the residual measurement noise or suboptimal emulation does not result in appreciable error in the bias expansion modeling. 
}
\end{figure*}
%%%%%%%%%%%%%%%%%%%%%%%%%%%%%%%%%%%%%%%%%%%%%%%%%%%%%%%%%
%%%%%%%%%%%%%%%%%%%%%%%%%%%%%%%%%%%%%%%%%%%%%%%%%%%%%%%%%

\section{Auxiliary Results of Measurement}
\label{appendix:more}

In this section, we list the details supporting the discussion on the measurement results in the main text.

\subsection{Measurement of Basis Spectra in Simulation}

%In Fig.~\ref{fig:field_maps_zoomin}, we present the visualization of Lagrangian basis fields same as Fig.~\ref{fig:field_maps}, but we zoom in on the bottom-right region for more careful identification of the structure. 
%We compare it with the high resolution matter density map, to confirm the correspondence between the massive halo and the bright points in $\delta^2 \,\&\, \delta^3$ maps, and show that $\delta^2 \,\&\, \delta^3$ map indeed enhances the feature of highly massive halos. From the visualization of the velocity field, we can identify that the central regions of the halo and filament are dominated by random motion with $\nabla\cdot\bfv\sim 0$, surrounded by the matter with infall velocity. It implies that the prominent positive/negative value in the center of the halo/filament of $\delta^2$ and $\delta^3$ fields form in early time, in other words, they are caused by the weighting value in Lagrangian space. 

In Fig.~\ref{fig:r2_ij}, we present the cross correlation coefficients $r_{ij}^2 = (P_{ij})^2 /\left(P_{ii}P_{jj}\right)$ between Lagrangian basis fields $\mO_i$ and $\mO_j$, to quantify the relative contribution of cross correlation terms compared to the auto correlation terms in the bias expansion modeling. Since in the bias expansion expression of power spectrum $P_{g} = \sum_{ij} b_ib_jP_{ij}$, the relative contribution of the cross correlation term to the auto correlation terms is just
$ {b_ib_j P_{ij} / \sqrt{ b_i^2P_{ii}\, b_j^2P_{jj} }} = r_{ij} $.
In the quadratic order, the only cross correlation term significantly contributing to the large scales is $P_{s^2\delta^2}$, where we have shown that it can reach percent level accuracy in emulation. The rest of the quadratic order terms become effective only in the non-linear scale. In particular, the emulation of spectra $P_{1\delta^2}$, $P_{\delta\delta^2}$, $P_{1s^2}$ and $P_{\delta s^2}$ seems to be large uncertain when $k$ toward zero, but their contribution compared to the auto spectra also approach zero as $k$ toward zero. 
It is a similar case for $\delta^3$ field cross correlation. The fractional contribution of $P_{1\delta^3}$ and $P_{\delta\delta^3}$ at $k < 0.2\, {\rm Mpc}^{-1}h$ is at the level of one percent, so we can safely ignore and drop their large scales contribution. 
While for $P_{\delta^3\nabla}$, it has $r^2_{\delta^3\nabla} \lesssim 10\%$ for all scales we consider, thus the error arising from the suboptimal emulation is also mild.

% ---------------------------------------------------------------------------------------
% ---------------------------------------------------------------------------------------

%%%%%%%%%%%%%%%%%%%%%%%%%%%%%%%%%%%%%%%%%%%%%%%%%%%%%%%%%
%%%%%%%%%%%%%%%%%%%%%%%%%%%%%%%%%%%%%%%%%%%%%%%%%%%%%%%%%
\begin{figure*}[htb!]
\includegraphics[width=0.99\textwidth]{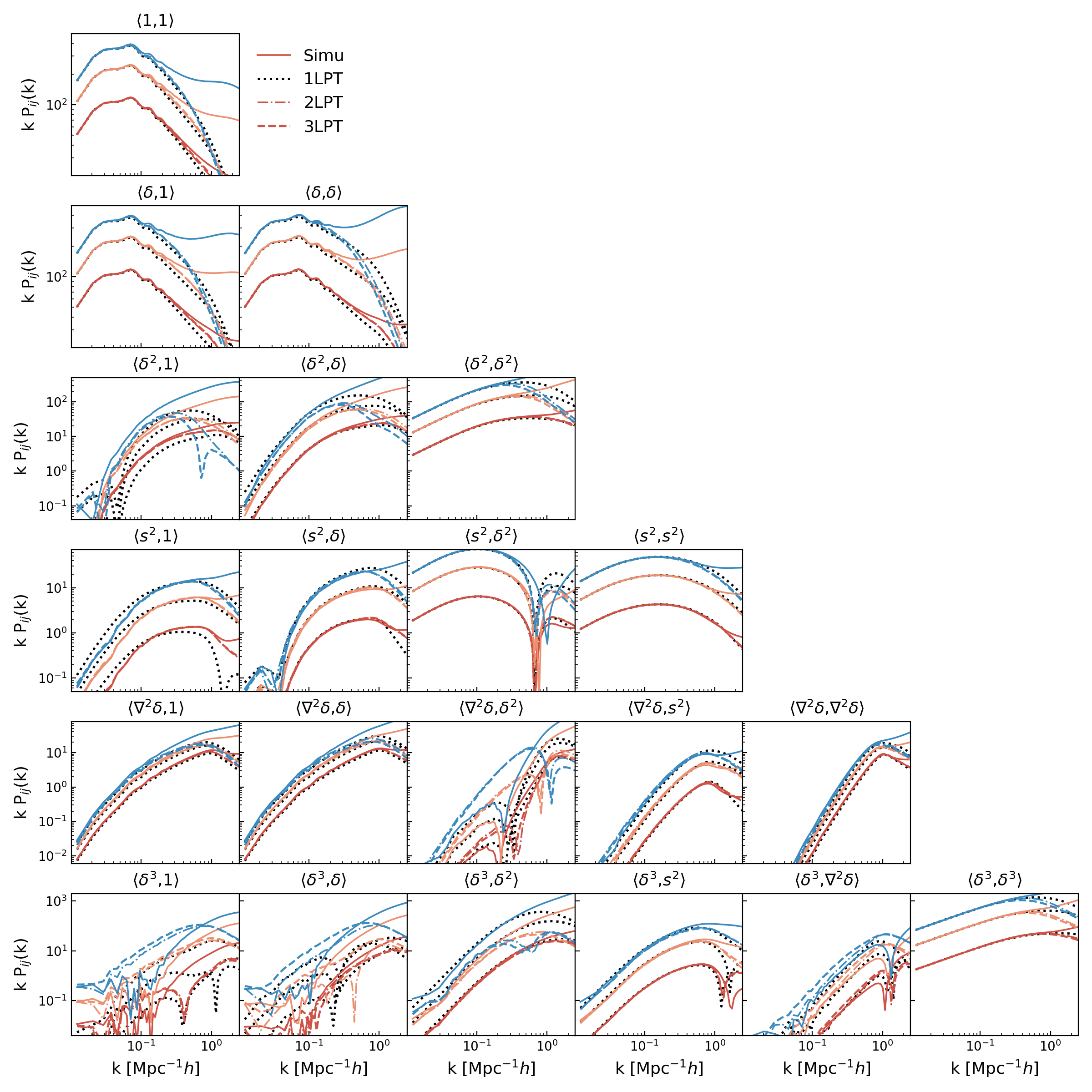}
\caption{ \label{fig:result_LPT}
Comparison of the basis spectra from \N-body simulations and grid-based nLPT calculations, using the first cosmology in the \Kun suite. 
The line colors indicate redshifts, with blue ($z=0.5$), orange ($z=1.0$), and red ($z=2.0$). 
Different line styles denote the N-body simulation (solid), 1LPT (dotted), 2LPT (dash-dotted), and 3LPT (dashed) results. 
We have taken the absolute value for all the power spectra, but differing from the main text, we do not present the $z=0$ results for better visualization. 
Similar to Fig.~\ref{fig:simu_1loop}, the simulated $P_{1\delta^3}(k)$ spectra are positive for $k \gtrsim 0.1\, h^{-1}{\rm Mpc}$, whereas the 2LPT and 3LPT predictions remain negative across the entire $k$-range. 
For $P_{\delta\delta^3}(k)$, the simulation spectra are also positive at $k \gtrsim 0.1\, h\,{\rm Mpc}^{-1}$, but the 2LPT calculation yields negative values throughout the $k$-range, while the 3LPT prediction exhibits oscillatory behavior.
}
\end{figure*}
%%%%%%%%%%%%%%%%%%%%%%%%%%%%%%%%%%%%%%%%%%%%%%%%%%%%%%%%%
%%%%%%%%%%%%%%%%%%%%%%%%%%%%%%%%%%%%%%%%%%%%%%%%%%%%%%%%%

\subsection{Grid-Based nLPT Calculation}
\label{appendix:nLPT}

The perturbation solution of the dynamical motion in Lagrangian space is well investigated up to fourth order \cite{catelan1995lagrangian, matsubara2008resumming, rampf2012lagrangian}. It supposes the spatial-time separation solution in the perturbation expansion of displacement, 
\begin{equation}
\mPsi(\bfq, \tau) = D^{(1)}(\tau) \mPsi^{(1)}(\bfq) + D^{(2)}(\tau) \mPsi^{(2)}(\bfq) + D^{(3)}(\tau) \mPsi^{(3)}(\bfq) + \cdots \ ,
\end{equation}
where $\tau$ denotes the conformal time and $D^{(n)}$ is the n-order time dependence factor usually solved in Einstein-de'Sitter model. The first order solution $\Psi^{(1)}_i$ is the Zel'dovich approximation, given by
\begin{equation}\label{equ:nLPT-1}
\Psi_{ii}^{(1)}(\bfq,\tau) = D(\tau) \delta(\bfq)\ ,
\end{equation}
where we denote $\Psi^{(n)}_{ij} \equiv \partial_{q_j}\Psi^{(n)}_i$ and $D^{(1)}(\tau) = D(\tau)$ is the linear growth factor. The $\delta(\bfq)$ is the overdensity in Lagrangian space, equivalently, the linear overdensity in initial redshift in our case. The second order solution is $\Psi^{(2)}_i$, with $D^{(2)}(\tau) = - 3/7\, D^2(\tau)$, and the spatial part given by
\begin{equation}\label{equ:nLPT-2}
\Psi^{(2)}_{ii}(\bfq) = {1\over 2}\left( \Psi^{(1)}_{ii} \Psi^{(1)}_{jj} - \Psi^{(1)}_{ij} \Psi^{(1)}_{ij} \right)    \ .
\end{equation}
The third order solution consists of two parts, 
\begin{equation}\label{equ:nLPT-3}
\Psi^{(3)}_i(\bfq,\tau) = D^{(3)}_a(\tau)\Psi^{(3)}_{a,i}(\bfq) + D^{(3)}_b(\tau)\Psi^{(3)}_{b,i}(\bfq)\ ,
\end{equation}
where the spatial solutions are given by
\begin{align}
& \Psi_{a,ii}^{(3)} = {\rm Det}(\Psi_{ij}^{(1)}) = {1\over 3!}\; \varepsilon_{ijk}\varepsilon_{mnl}  \Psi_{im}^{(1)}\Psi_{jn}^{(1)}\Psi_{kl}^{(1)}     \ ,  \\
& \Psi_{b,ii}^{(3)}  = {1\over 2}\left( \Psi^{(1)}_{ii} \Psi^{(2)}_{jj} - \Psi^{(1)}_{ij} \Psi^{(2)}_{ij} \right) \ ,
\label{equ:nLPT-3x}
\end{align}
and the time dependence factor are $D^{(3)}_a(\tau) = -1/3\, D^{3}(\tau)$ and $D^{(3)}_b(\tau) = 10/21\, D^{3}(\tau)$. We have neglected the rotation part in the third order displacement. In Fourier space, they are expressed in a unified formulation, and the convolutional kernel $L^{(n)}$ in integration couples the modes for higher order correction, 
\begin{align}
\mPsi^{(n)}(\bfk, \tau)  
&= {i D^n(\tau)\over n!} \int_{\bfk_1\bfk_2\cdots\bfk_n} (2\pi)^3\delta^D_{\bfk_{1..n},\bfk}  \;L^{(n)}(\bfk_1,\bfk_2\cdots,\bfk_n) \;   \delta(\bfk_1) \delta(\bfk_2)\cdots\delta(\bfk_n)  \ ,  \\
L^{(1)}(\bfk)  &=  {\bfk\over k^2}   \ ,   \\
L^{(2)}(\bfk_1,\bfk_2)  &=  {3\over 7} {\bfk\over k^2} \left[ 1 - \left( {\bfk_1\cdot\bfk_2\over k_1k_2 }\right)^2  \right]   \ ,  \\
L^{(3)}(\bfk_1,\bfk_2,\bfk_3)\; &= 
-{1\over 3} {\bfk\over k^2} \left[{ 1 + 2 { (\bfk_1\cdot\bfk_2)(\bfk_2\cdot\bfk_3)(\bfk_3\cdot\bfk_1) \over (k_1k_2k_3)^2} - 3\left({\bfk_1\cdot\bfk_2\over k_1k_2}\right)^2   }\right]  
+  {5\over 7}\,{\bfk\over k^2} \left[ 1 - \left({\bfk_3\cdot(\bfk_1+\bfk_2)\over k_3 |\bfk_1+\bfk_2|}\right)^2 \right] \left[1 - \left( {\bfk_1\cdot\bfk_2\over k_1k_2 }\right)^2  \right]  \ ,
\end{align}
where $\bfk=\sum_i^{n}\bfk_i$ in the kernel expressions. 
We obtain the nLPT displacement by solving equations Eq.~\ref{equ:nLPT-1} - Eq.~\ref{equ:nLPT-3x}. In particular, we utilize 1LPT to generate the Zel'dovich realizations for the sample variance suppression in the main text. 

We present the grid-based nLPT calculation of basis spectra in Fig.~\ref{fig:result_LPT}, compared to the measurement in the \N-body simulation. The 1LPT calculation captures the primary features of basis spectra, and they are accurate enough on the large scales for the leading order spectra. The 2LPT $\&$ 3LPT enhance the accuracy for predicting the high redshift spectra, but worsen low redshift results since LPT breaks down in the extremely non-linear scenario. In details, for spectrum $P_{1\delta^2}$ in region $k\lesssim 0.1\, {\rm Mpc}^{-1}h$, the 2LPT calculation is precise at $z\simeq 2$, but significantly deviates when non-linearity become prominent in $z\simeq 1$. For spectrum $P_{1s^2}$, we have seen that the 1-loop theory mismatches simulation by an overall amplitude, shown in Fig.~\ref{fig:simu_1loop}. While in the 2LPT $\&$ 3LPT calculation, the mismatching is removed, and they agree with each other $k\lesssim 0.6\, {\rm Mpc}^{-1}h$, because the 1-loop theory does not take account of all the 2LPT displacement correlation, which is a prominent contribution. 
For the $\delta^3$ terms, the LPT calculation does not describe the simulation well for the spectra dominated by the mode-coupling, such as $P_{1\delta^3}$ and $P_{\delta\delta^3}$. But for the spectra dominated by the disconnected part, such as $P_{\delta^3\delta^3}$, LPT is still an excellent description. 

%The blue, orange and red lines represent the result at redshfit $z=0.5, 1$ and $2$. We do not present the $z=0$ results here for better visualization. 
%Similar to Fig.~\ref{fig:simu_1loop}, the simulation $P_{1\delta^3}(k)$ spectra is positive at range $k \gtrsim 0.1 h^{-1}{\rm Mpc}$ while 2LPT$\&$3LPT prediction are negative within the entire $k$-range.  For $P_{\delta\delta^3}(k)$, the simulation spectra is also positive at range $k \gtrsim 0.1 h^{-1}{\rm Mpc}$, and 2LPT prediction is negative within the entire $k$-range while the 3LPT prediction oscillates. 

% ---------------------------------------------------------------------------------------
% ---------------------------------------------------------------------------------------

%%%%%%%%%%%%%%%%%%%%%%%%%%%%%%%%%%%%%%%%%%%%%%%%%%%%%%%%%
%%%%%%%%%%%%%%%%%%%%%%%%%%%%%%%%%%%%%%%%%%%%%%%%%%%%%%%%%
\begin{figure*}[htb!]
\includegraphics[width=0.99\textwidth]{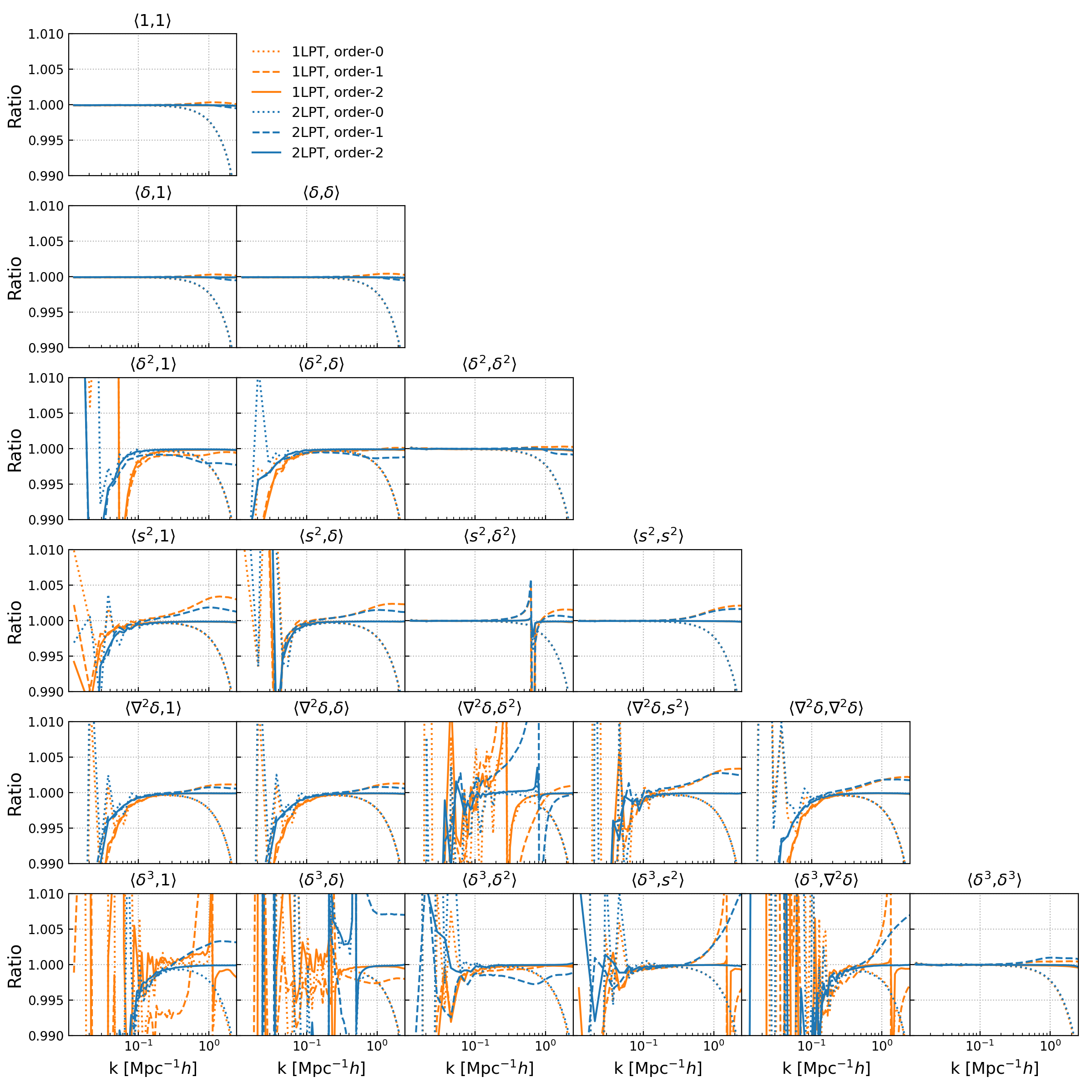}
\caption{ \label{fig:validate_compLPT}
The ratio of nLPT spectra pre-initialized with the glass-like distribution to those pre-initialized with the regular grid distribution, evaluated at redshift $z=1.0$, using the same initial condition as \Kun simulations. 
The regular-grid initialization yields unbiased LPT power spectra, free from any bias induced by particle initial positions, thus providing a ground truth for validating the compensation method. 
The 1LPT and 2LPT simulations are shown in orange and blue, respectively. The derivative compensation from Eq.~\ref{equ:compensate_expand} is represented by dotted (order-0), dashed (order-1), and solid (order-2) lines, where order-0 corresponds to no compensation. 
It shows that the order-2 compensation method is sufficient to eliminate the particle-mesh mismatching bias in the basis spectra measurement. 
}
\end{figure*}
%%%%%%%%%%%%%%%%%%%%%%%%%%%%%%%%%%%%%%%%%%%%%%%%%%%%%%%%%
%%%%%%%%%%%%%%%%%%%%%%%%%%%%%%%%%%%%%%%%%%%%%%%%%%%%%%%%%

\subsection{Validation of Derivative Compensation Method}
\label{append:validation_taylor_compensation}

Compared to previous works, where the particles are pre-initialized in a uniform grid, the \Kun simulation initially places the particles in a glass-like distribution. So we can not associate each particle with a grid cell exactly, and the mismatch between particle and grid cell potentially results in a bias when assigning the value from the field distributed in a uniform grid to the particle positions. 
In our Lagrangian basis fields construction, we need to assign the Lagrangian field to the particle resting in the initial position at $z_{\rm ini}$. 
The direct assignment of the grid value to interior particles suppresses the power of cross-correlation, especially for high order basis fields. We compensate the mismatching with Eq.~\ref{equ:compensate_expand}, where we plus the derivative terms since the fields are smooth enough with long-wavelength modes only.

We use the nLPT calculation to validate the compensation method and determine the truncation of Taylor expansion orders. The reason for choosing nLPT as validation is that we can fast generate the realization with pre-initialization in the uniform grid as the true answer, without rerunning the expensive \N-body simulation. In Fig.~\ref{fig:validate_compLPT}, we present the validation results at $z=1$, where we utilize the Lagrangian field mesh size $N_{\rm ini}^3=6144^3$ as same as the simulation calculation. Compared to direct assignment, the order-2 derivative compensation gives an almost accurate result, removing the mismatching bias. 
Though only $z=1$ is shown, we have also calculated the validation results at other redshift snapshots. The suppression is more severe at high redshift, while milder at low redshift, since the non-linear clustering becomes dominant at lower redshift and overwhelms the Lagrangian weighting. 
Although the fine grid number $N_{\rm ini}^3=6144^3$ appears to cause slight suppression, we apply the compensation method to the basis spectra calculation presented in the main text, to prevent any amplification of potential error during fully non-linear evolution.

% ---------------------------------------------------------------------------------------
% ---------------------------------------------------------------------------------------

%%%%%%%%%%%%%%%%%%%%%%%%%%%%%%%%%%%%%%%%%%%%%%%%%%%%%%%%%
%%%%%%%%%%%%%%%%%%%%%%%%%%%%%%%%%%%%%%%%%%%%%%%%%%%%%%%%%
\begin{figure*}[htb!]
\includegraphics[width=0.99\textwidth]{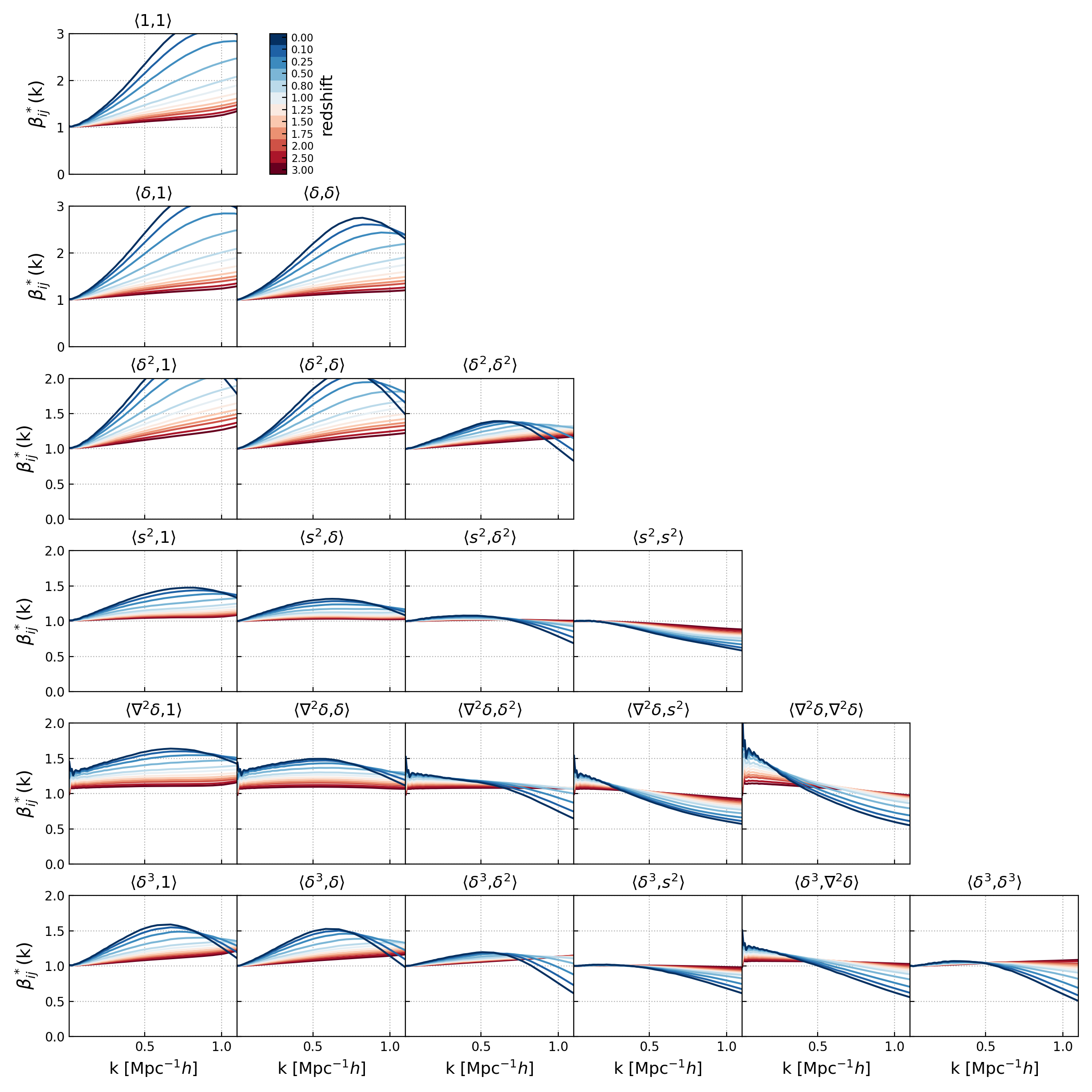}
\caption{ \label{fig:validate_Zel_beta}
The optimal weight $\beta^*_{ij}$ in Eq.~\ref{equ:zel_weight} for Zel'dovich variance control, derived from the first cosmological simulation in the \Kun simulation suite.
}
\end{figure*}
%%%%%%%%%%%%%%%%%%%%%%%%%%%%%%%%%%%%%%%%%%%%%%%%%%%%%%%%%
%%%%%%%%%%%%%%%%%%%%%%%%%%%%%%%%%%%%%%%%%%%%%%%%%%%%%%%%%

%%%%%%%%%%%%%%%%%%%%%%%%%%%%%%%%%%%%%%%%%%%%%%%%%%%%%%%%%
%%%%%%%%%%%%%%%%%%%%%%%%%%%%%%%%%%%%%%%%%%%%%%%%%%%%%%%%%
\begin{figure*}[htb!]
\includegraphics[width=0.99\textwidth]{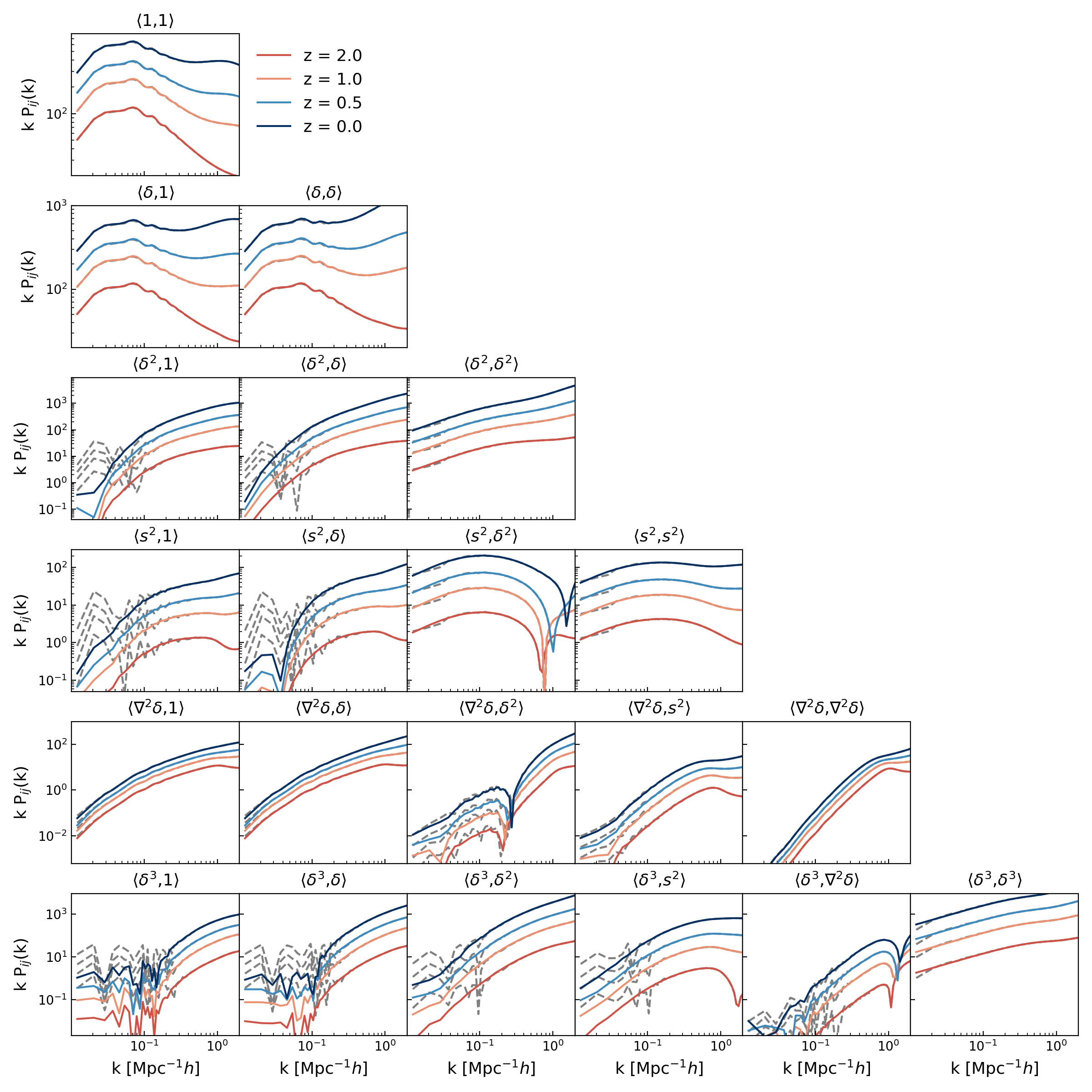}
\caption{ \label{fig:validate_Zel_result}
Comparison between the raw simulation measurement (gray dashed lines) and the results after applying Zel'dovich variance control (solid lines), obtained from the first cosmological simulation of the \Kun suite.
}
\end{figure*}
%%%%%%%%%%%%%%%%%%%%%%%%%%%%%%%%%%%%%%%%%%%%%%%%%%%%%%%%%
%%%%%%%%%%%%%%%%%%%%%%%%%%%%%%%%%%%%%%%%%%%%%%%%%%%%%%%%%

\subsection{Validation of Zel'dovich variance control}
\label{append:validation_ZVC}

We apply the technique of Zel'dovich variance control to reduce the sample variance of basis spectra (Sec.~\ref{sec:zvc}). 
For the phase matching realization, we generate the Lagrangian basis fields and the linear displacement seeded by the initial density field with smoothing window $W(k)$, same as the simulation calculation described in Sec.~\ref{sec:measure_Pkij}. The smoothing window ensures all the isotropic Fourier modes are captured by the Zel'dovich realization, so that we can match the 1LPT simulation and linear Lagrangian theory exactly. 

In Fig.~\ref{fig:validate_Zel_beta}, we present the optimal weights $\beta^*_{ij}$ utilized for variance suppression, which are obtained from one cosmology and fixed for all simulation spectra.
In Fig.~\ref{fig:validate_Zel_result}, we compare the basis spectra before and after variance control. Especially, the spectra $P_{1\delta^2}$, $P_{1s^2}$, $P_{1s^2}$, $P_{\delta s^2}$, $P_{\delta^3\delta^2}$ and $P_{\delta^3 s^2}$ are dominated by the sample variance on large scales $k\lesssim 0.1\, {\rm Mpc}^{-1}h$, while after variance reduction, they are signal dominant. 

We apply the variance reduction to the power spectrum presented in this work unless otherwise specified, including the simulation and nLPT calculation. 
In particular, we present the halo spectra fitting in Fig.~\ref{fig:halo_fit}, and we also apply the variance reduction to the halo spectra to partially reduce the fluctuation on large scales \cite{kokron2022accurate, hadzhiyska2023mitigating}. It is more numerically stable to search for the optimal bias with the variance-suppressed halo spectra, because the Gaussian covariance estimation is not true. The effectiveness of the reduction strongly depends on the cross-correlation coefficients between halo and matter overdensity, slightly effective for the massive sample due to the decorrelation caused by shot noise. Nevertheless, the halo power spectrum $P_{hh}\,\& \,P_{hm}$ fitting is a simple illustration of emulator performance, but not for a rigorous theoretical interpretation.

{
\subsection{Covariance Estimation for Validation with Halo Power Spectrum}
\label{append:gaussian_sample_variance}

As a validation of the biased power spectrum fitting performance of the emulator, we perform a joint fit to the halo auto power spectrum $P_{hh}$ and halo-matter cross power spectrum $P_{hm}$. It requires an estimation of covariance that accounts for all the uncertainties in both the data and the theoretical power spectra. Generally, it is given by
\begin{align*}
{\rm\bf Cov}_{\rm tot}[P_{\alpha\beta}(k), P_{\gamma\eta}(k')] = 
{\rm\bf Cov}_{\rm data}[P_{\alpha\beta}(k), P_{\gamma\eta}(k')] +
{\rm\bf Cov}_{\rm emu}[P_{\alpha\beta}(k), P_{\gamma\eta}(k')] 
\end{align*}
where $\alpha,\beta,\gamma,\eta \in \{\delta_h, \delta_m\}$ denote the halo and matter overdensity field. The ${\rm\bf Cov}_{\rm data}$ term is the data covariance of the halo power spectrum, approximated as 
\begin{align*}
{\rm\bf Cov}_{\rm data}[P_{\alpha\beta}(k), P_{\gamma\eta}(k')] = 
{\delta^D_{k, k'}\over N_{\bfk}} \,\left( P_{\alpha\gamma} P_{\beta\eta} + P_{\alpha\eta} P_{\beta\gamma} \right)   
+ {\delta^D_{k\neq k'}\over N_{\bfk}} \, P_{\alpha\beta} P_{\gamma\eta} \;
{ {\rm\bf Cov}_{mm}(k, k') \over  \sqrt{ {\rm\bf Cov}_{mm}(k, k)\, {\rm\bf Cov}_{mm}(k', k')} }  \quad.
\end{align*}
The diagonal $k=k'$ components are estimated by the Gaussian variance approximation. The off-diagonal $k\neq k'$ components are estimated as the cross correlation coefficients rescaled by the data spectrum, where ${\rm\bf Cov}_{mm}(k, k') \equiv {{\rm\bf Cov}[P_{mm}(k), P_{mm}(k')]}$ is the matter power spectrum covariance from 25 pairs of FastPM simulation \cite{chen2025csst}. 
To reduce the sample variance, we apply a Savitzky-Golay filter to smooth the ${\rm\bf Cov}_{mm}$ ratio. The emulator uncertainty is estimated by
\begin{align*}
{\rm\bf Cov}_{\rm emu} [P_{\alpha\beta}(k), P_{\gamma\eta}(k')] 
= \sum_{ijmn}\, b_ib_jb_mb_n\, {\bf\rm Cov}_{\rm emu} \left[ P_{ij}(k), P_{mn}(k')\right] 
\end{align*}
where ${\bf\rm Cov}_{\rm emu} \left[ P_{ij}(k), P_{mn}(k')\right]$ is the basis spectra covariance between $P_{ij}$ and $P_{mn}$, given by Eq.~\ref{equ:emu-err-tot}. The bias parameters are fixed as the best-fit result when setting $k_{\rm max}=0.62\, {\rm Mpc}^{-1}h$ without emulator uncertainty. 

We do not consider any realistic survey noise, as we aim to test the fitting performance. 
The off-diagonal data covariance from ${\rm\bf Cov}_{mm}(k, k')$ already captures the approximate mode-coupling behavior arising from non-linear clustering. It provides a sufficient penalty for small scales during the likelihood maximization. Without this term, the off-diagonal of the whole covariance ${\rm\bf Cov}_{\rm tot}$ would be dominated by the emulator covariance, which significantly contributes at $k\gtrsim\, 0.3\, {\rm Mpc}^{-1}h$ and could overwhelm the diagonal contribution from ${\rm\bf Cov}_{\rm data}$. 
Moreover, a proper weighting scheme to balance the contributions from different scales in the likelihood is sufficient for the purpose of validating the emulator fitting. 
%The accurate estimation of power spectrum covariance is challenging and requires vast mock samples, 
}

% ---------------------------------------------------------------------------------------
% ---------------------------------------------------------------------------------------

\section{Auxiliary Parameters in Surrogate Model}
\label{appendix:fine_tuning_parameters}

\subsection{Smoothing and Interpolation of Basis Spectra}

In Sec.~\ref{sec:emulation}, we introduce our emulation scheme and the procedure to fine-tune the surrogate model to maximize the emulation accuracy. We describe it in more detail in this section. Notably, all the artificial operations here aim to optimize the interpolation performance, but not to modify any physical characteristic in the basis spectra. 

To further reduce the noise in the simulation measurement after Zel'dovich variance control, we apply the Savitzky-Golay filter with an order-2 polynomial to smooth the ratio before emulation. The smoothing spectrum is 
\begin{equation}
\hat{P}_{ij}(k) \,\rightarrow\, P_{ij}^T(k)\; {\rm\bf Filter}\left[ \hat{P}_{ij}(k)\over P_{ij}^T(k) \right]  \ , 
\end{equation}
where $\hat{P}_{ij}(k)$ is the spectrum after the variance control, and $P_{ij}^T(k)$ is the theoretical template to reduce the dynamic range. The Savitzky-Golay filter uses a window length of 7 $k$-points, and we have tested that this choice does not introduce any bias or signal distortion. The specific $k$ region we smooth depends on the spectra type, as described following. 
\begin{itemize}
\item We do not smooth $P_{11}$, $P_{1\delta}$ and $P_{\delta\delta}$ where the BAO features are significant. Also, we do not smooth $P_{\delta^2 s^2}$ and $P_{\delta^2\nabla^2}$ because the transition of sign results in a sharp change of the ratio, and the critical points strongly depend on cosmology and redshift. 
\item For the basis spectra $P_{1\delta^2}$, $P_{1s^2}$, $P_{\delta\delta^2}$, $P_{1\nabla^2}$, $P_{\delta\nabla^2}$ and $P_{\nabla^2\nabla^2}$, we apply smoothing operation within range $0.05 < k < 0.8\, {\rm Mpc}^{-1}h$, where the larger scale is noisy due to the decorrelation of Zel'dovich and simulation calculation, while the smaller scale exhibit sharp change due the inaccuracy prediction of $P_{ij}^T$. 
\item For the basis spectra $P_{\delta s^2}$, we adopt more conservative smoothing range $0.1 < k < 0.8\, {\rm Mpc}^{-1}h$, because the spectrum $P_{\delta s^2}$ is negative at $k > 0.1\, {\rm Mpc}^{-1}h$, but may become positive values at larger scale. The sign transition is limited to $k < 0.1\, {\rm Mpc}^{-1}h$ for the cosmology and redshift we consider. 
\item For the rest of the basis spectra, we choose a smoothing range $0.008 < k < 0.8\, {\rm Mpc}^{-1}h$, and smaller scales are not included for the same reason as above. 
\end{itemize}

As shown in the main text, the 1-loop and 2-loop order terms dominate the $P_{1\delta^3}$ and $P_{\delta\delta^3}$, thus the Zel'dovich variance control only partially removes the noise correlated with linear displacement. The noise coupled with high order displacement is still significant and overwhelms the tiny signal. We mask these noisy regions before performing SVD to avoid contaminating other $k$-bins, 
\begin{equation}
P_{ij}(k) \;\rightarrow\; {1\over 2} \left[ 1+ \tanh\left(50(k-k_{\rm drop})\right)\right] P_{ij}(k)\ ,
\end{equation}
where $k_{\rm drop}=0.2\,{\rm Mpc}^{-1}h$ for $P_{1\delta^3}$, and $k_{\rm drop}=0.1\,{\rm Mpc}^{-1}h$ for $P_{\delta\delta^3}$. 

In Eq.~\ref{equ:surrogate_rescaled}, we introduce the estimated cross correlation coefficient $\tilde{r}_{ij}$ to suppress the contribution of auxiliary terms compared to $P_{ij}$. The estimated values are given by
\begin{align*}
\tilde{r}_{\delta^2\nabla^2}(k) = 8\, e^{-(k-2)^2}    \;\;;\quad
\tilde{r}_{s^2\delta^3}(k) = 2\, e^{-(k-1.5)^2}     \;\;;\quad
\tilde{r}_{\nabla^2\delta^3}(k) = 2\, e^{-(k-1.5)^2}  \ .
\end{align*}

\subsection{Prediction of Lagrangian Basis Spectra Beyond the Simulation Box}

%%%%%%%%%%%%%%%%%%%%%%%%%%%%%%%%%%%%%%%%%%%%%%%%%%%%%%%%%
%%%%%%%%%%%%%%%%%%%%%%%%%%%%%%%%%%%%%%%%%%%%%%%%%%%%%%%%%
\begin{figure*}[htb!]
\includegraphics[width=0.99\textwidth]{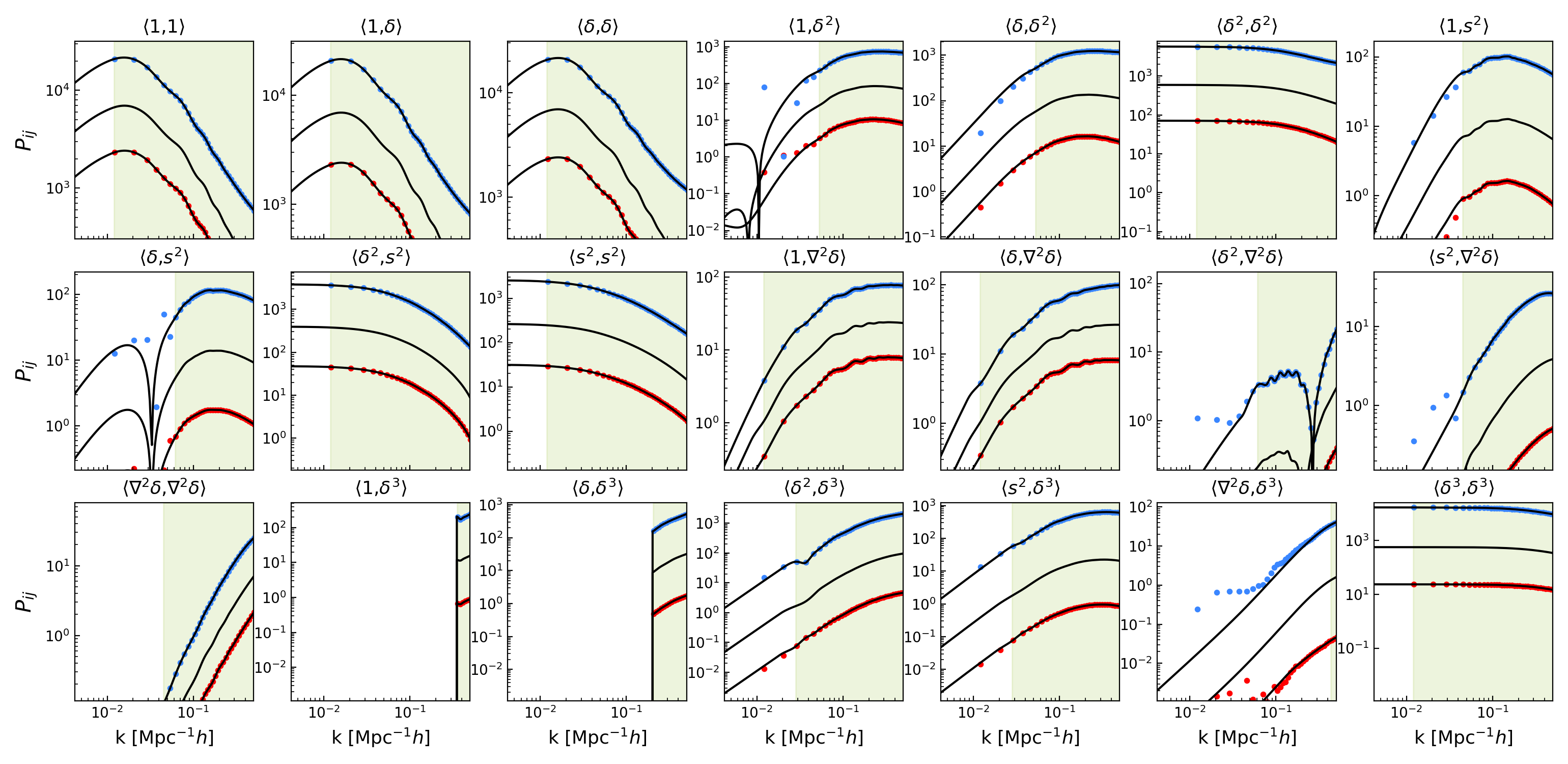}
\caption{ \label{fig:hybrid_transition}
Comparison of the emulator predictions with and without including the 1-loop theory prediction on very large scales. The circular markers show the $k$-bins measured directly from simulations and subsequently interpolated. Black curves represent the combined emulator prediction, which incorporates simulation results within the green shaded region and 1-loop theory outside this range.
Three redshift bins are shown in each panel, with $z=0.0,\, 1.0,\, 3.0$ from top to bottom. For visual clarity, the $z=1.0$ results are shown without circular markers to emphasize the smooth transition between 1-loop theory predictions and simulation measurements. 
As an example, the input cosmology for the emulator was randomly selected, with $\{\,\Omega_b, \Omega_m, h, n_s, A_s, w_0, w_a, M_\nu \,\}$ $=$ $\{\, 0.048, 0.31, 0.67, 0.9665, 2.105\times 10^{-9}, -0.7, -0.4, 0.2 \,\}$. 
}
\end{figure*}
%%%%%%%%%%%%%%%%%%%%%%%%%%%%%%%%%%%%%%%%%%%%%%%%%%%%%%%%%
%%%%%%%%%%%%%%%%%%%%%%%%%%%%%%%%%%%%%%%%%%%%%%%%%%%%%%%%%

The emulator provides the option to predict the power spectrum from the direct interpolation of simulation measurement, however, it is limited by the simulation box size $k \gtrsim 0.01\,{\rm Mpc}^{-1}h$ and suffers from the sample variance in the high-order spectra. To benefit from the precise theoretical calculation, we combine the 1-loop theory prediction on large scales, $0.001\,{\rm Mpc}^{-1}h  \,\leq\,k  \,<\, k_{\rm trans}$, where the transition scale $k_{\rm trans}$ depends on the SNR of the spectrum measurement, and the combination enables a wider scale of emulation wavenumber, $0.001 \,\leq\,k  \,<\, 1\, {\rm Mpc}^{-1}h$. For the precise simulation measurement, such as the leading order spectra $P_{11}, P_{1\delta}, P_{\delta\delta}$, we choose $k_{\rm trans}$ as the smallest $k$-bin in the measurement, $k_{\rm trans} = 0.012 \,{\rm Mpc}^{-1}h$. While for the noisy measurement on large scales, such as $P_{1\delta^2}$, we use a conservative $k_{\rm trans} = 0.046 \,{\rm Mpc}^{-1}h$. The specific values of $k_{\rm trans}$ are fine-tuned to avoid noisy measurement and guarantee the smooth transition between two parts. 

Operationally, for a given cosmology, we compute the 1-loop theory at $k<k_{\rm trans}$ and simulation results at $k\geq k_{\rm trans}$ for 12 redshift bins ranging $0<z<3$, then combine them into a 2D array where the two dimensions correspond to wavenumber and redshift respectively. We then interpolate the 2D array utilizing the third-order bivariate spline to sample the desired $(k, z)$ bins, implemented via \texttt{RectBivariateSpline} in \texttt{scipy}. The entire procedure takes about $\mathcal{O}(10^{-2})$ seconds. 
Notably, the 1-loop theory and simulation results agree well on large scales, which enables the smooth transition between two components, even when simply stacking the arrays. 

In Fig.~\ref{fig:hybrid_transition}, we compare the combination prediction and the simulation-only prediction, and the green shadow regions indicate the scales where we adopt the simulation measurements. Thanks to the precise 1-loop theory calculation, the transition between the 1-loop theory and simulation measurement is smooth, particularly for the cross spectra of the leading and quadratic order spectra. Moreover, though not shown here, the second order derivative with respect to the cosmological parameter does not present an anomaly around the transition scale. Therefore, the smoothness of the 3-order spline is sufficient, and the extra process such as forcing the smooth transition \cite{hadzhiyska2021hefty} is not necessary.

%%%%%%%%%%%%%%%%%%%%%%%%%%%%%%%%%%%%%%%%%%%%%%%%%%%%%%%%%
%%%%%%%%%%%%%%%%%%%%%%%%%%%%%%%%%%%%%%%%%%%%%%%%%%%%%%%%%
\begin{figure*}[phtb!]
\includegraphics[width=0.95\textwidth]{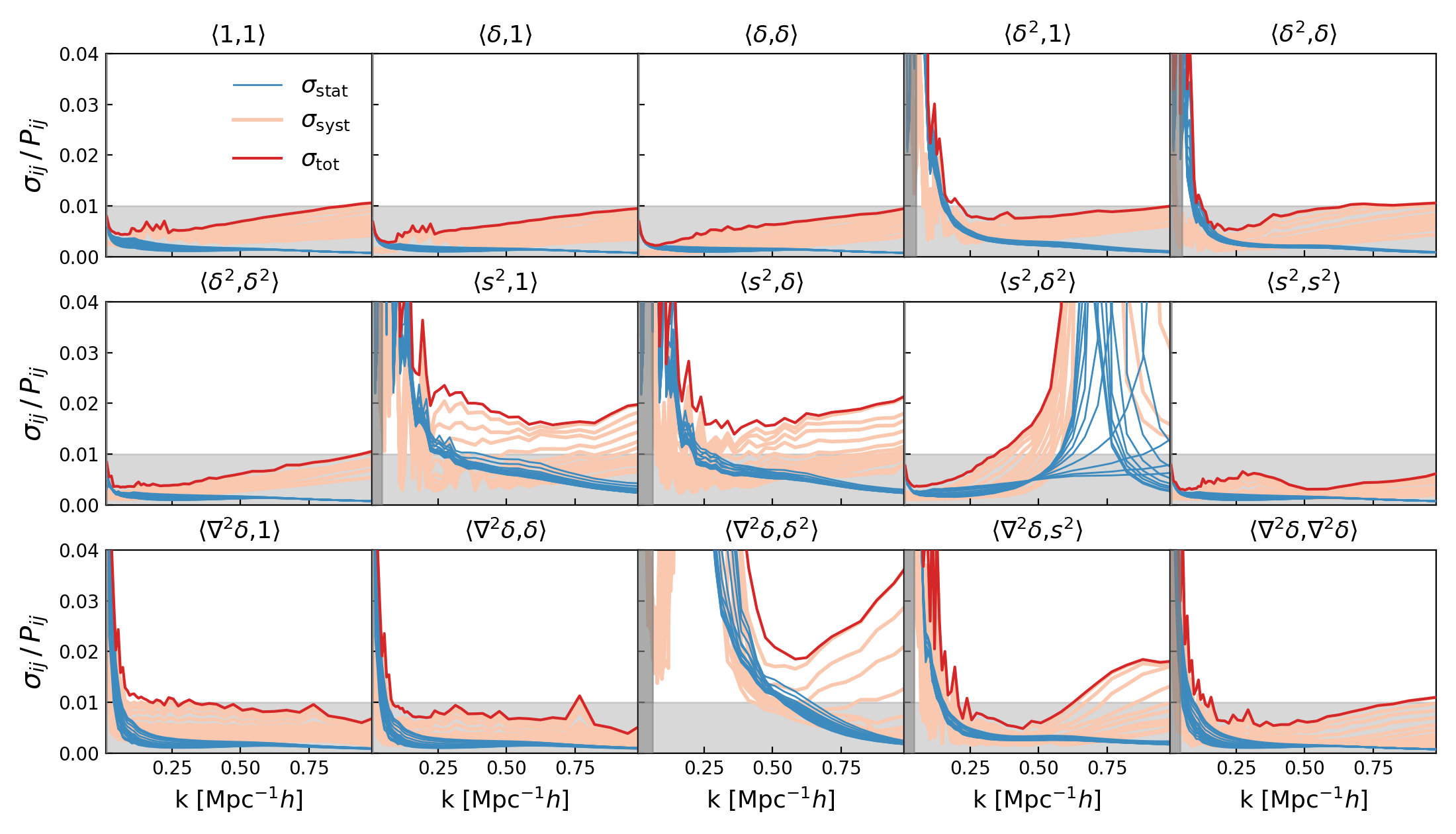}
\caption{ \label{fig:error_estimate}
Uncertainty estimation of the basis spectra emulation for the simulation measurement. The vertical gray regions indicate the scale using 1-loop theory predictions, which can be highly precise. 
The statistical uncertainties $\sigma_{\rm stat}$ (Eq.~\ref{equ:emu-err-stat}) sourced by the sample variance are indicated as blue lines. The systematic uncertainties $\sigma_{\rm syst}$ (Eq.~\ref{equ:emu-err-syst-1}) sourced by the interpolation error are indicated as orange lines. 
The upper limit of $\sigma_{\rm tot} = \sqrt{ \sigma_{\rm stat}^2 + \sigma_{\rm syst}^2 }$ for 12 redshift bins is indicated as red line. 
Different lines of $\sigma_{\rm stat}$ or $\sigma_{\rm syst}$ denote 12 redshift bins ranging $0<z<3$. 
}
\end{figure*}

\begin{figure*}[phtb!]
\includegraphics[width=0.95\textwidth]{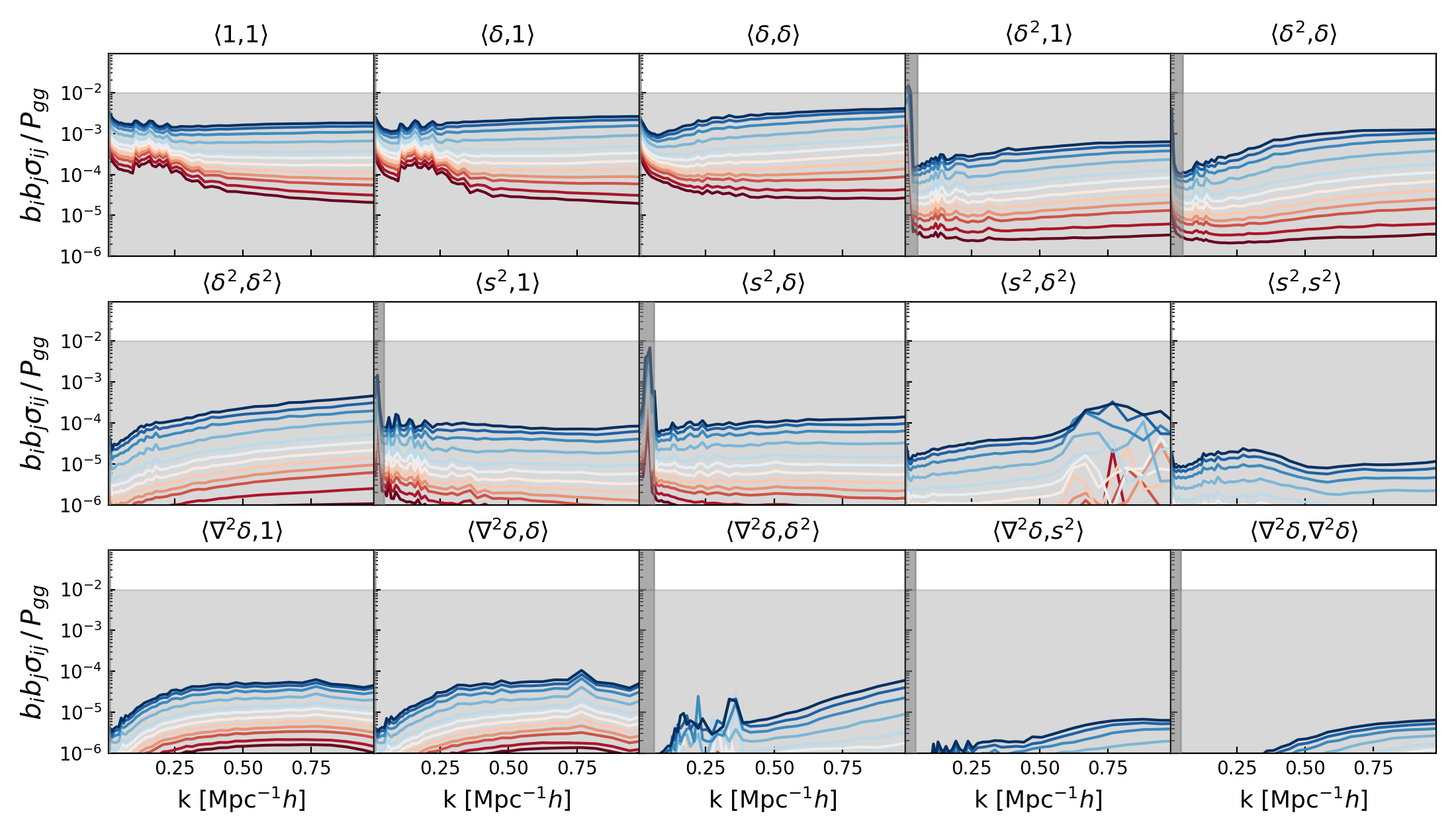}
\caption{ \label{fig:error_estimate-2}
Same as the Fig.~\ref{fig:error_estimate}, but this figure shows the relative contribution of the total basis spectra uncertainty $\sigma_{\rm tot}$ to the galaxy power spectrum. We assume the galaxy power spectra is measured at $z=0.5$, with density $\bar{n}_g = 10^{-3} \,{\rm Mpc}^{-3}h^3$ and bias parameters $b_1=1$, $b_2=0.2$, $b_s=0.2$, $b_\nabla=0.2$. The dominant uncertainty arises from the leading order terms $P_{11}, P_{1\delta}, P_{\delta\delta}$, contributing approximately a fraction of $10^{-3}$. In contrast, the higher-order terms are sub-dominant, with $\sigma_{ij}/P_{gg}\lesssim 10^{-3}$ relative to the total galaxy power spectrum. 
}
\end{figure*}
%%%%%%%%%%%%%%%%%%%%%%%%%%%%%%%%%%%%%%%%%%%%%%%%%%%%%%%%%
%%%%%%%%%%%%%%%%%%%%%%%%%%%%%%%%%%%%%%%%%%%%%%%%%%%%%%%%%

{
\subsection{Estimation of Emulation Uncertainty}
\label{append:uncertainty}

The sources of basis spectra emulation uncertainty are the sample variance in the simulation measurements and the interpolation error across the parameter space. 

Though we have applied the variance control technique to suppress the sample variance, the residual uncertainty may arise from the decorrelation of the Zel'dovich realization. This statistical uncertainty is estimated by the Gaussian covariance approximation,
\begin{align}\label{equ:emu-err-stat}
{\bf\rm Cov}_{\rm stat} \left[ P_{ij}(k), P_{mn}(k')\right] = 
{\delta^D_{k, k'}\over N_{\bfk}}\, \left[1 - F_{\beta*}^2\,r^2_{ZN}\right] \,\left( P_{im} P_{jn} + P_{in} P_{jm} \right)   \;,
\end{align}
where the factor $1 - F_{\beta*}^2 \,r^2_{ZN}$ accounts for the variance reduction using Zel'dovich variance control. 
%$r^2_{ZN} \equiv { {\rm\bf Cov}_{\rm ZN}^2  \over {\rm\bf Cov}_{\rm ZZ} {\rm\bf Cov}_{\rm NN} }$ 
The $r_{ZN}$ is the cross correlation coefficient between the N-body simulation and Zel'dovich basis spectra, estimated by the disconnected covariance approximation. The $F_{\beta*}$ is the estimation of the breakdown of the approximation applied in Eq.~\ref{equ:zel_weight}, well characterized by \cite{kokron2022accurate}
\begin{align*}
F_{\beta*}(k) = {1\over 2} \left[ 1 - \tanh\left({k-k_0 \over \Delta_k}\right)\right]   \; ;\quad
k_0 = 0.618 \,{\rm Mpc}^{-1}h \,,\;  \Delta_k = 0.167 \,{\rm Mpc}^{-1}h  \quad. 
\end{align*}
In Fig.~\ref{fig:error_estimate}, we show the fraction uncertainty of the diagonal component $\sigma_{ij}(k) \equiv {\bf\rm Cov} \left[ P_{ij}(k), P_{ij}(k)\right] $ up to the quadratic order spectra. Particularly for the dominant spectra $P_{11}, P_{1\delta}, P_{\delta\delta}$, the sample variance is non-negligible only on large scales, which is caused by the limited simulation box size. While the potential shot noise on small scales is significantly suppressed because of the high simulation mass resolution.

The major emulation uncertainty is the interpolation error. This uncertainty is systematic, since it depends on the position in the parameter space, with a minor interpolation error in the central region while a large error in the marginal region. In principle, the Gaussian process regression naturally predicts an interpolation uncertainty, but it does not capture the cross-correlation between different scales $k\neq k'$ or different basis spectra, ${\rm\bf Cov}[P, P']$ with $P\neq P'$. This cross block in covariance could be significant, especially for the leading order spectra. 
To obtain the full covariance, we utilize the Leave-one-out (LOO) approach to estimate the uncertainty \cite{derose2023aemulus}, given by
\begin{align}
\label{equ:emu-err-syst-1}
& {\bf\rm Cov}_{\rm syst} \left[ P_{ij}(k), P_{mn}(k')\right] =  P_{ij}(k) P_{mn}(k') \, \rho^{\rm LOO}_{ijmn}\left( k, k'\right)      \quad, 
\end{align}
where the fraction uncertainty $\rho^{\rm LOO}_{ijmn}\left( k, k'\right)$ is estimated by $N_{\rm LOO}=83$ samples from LOO estimation. 
\begin{align}
\label{equ:emu-err-syst-2}
& \rho^{\rm LOO}_{ijmn}\left( k, k'\right) = {1+\alpha \over N_{\rm LOO}} 
\sum_{n\in {\rm LOO}} { \left[P^n_{ij}(k) - \tilde{P}^n_{ij}(k) \right] \left[ P^n_{mn}(k') - \tilde{P}^n_{mn}(k') \right] \over |P^n_{ij}(k) P^n_{mn}(k')| + \alpha\, \sigma_{ijmn}(k,k') }      \quad,     \\
\label{equ:emu-err-syst-3}
& \sigma_{ijmn}(k,k') = {1 \over N_{\rm LOO}} 
\sum_{n\in {\rm LOO}} { \left[P^n_{ij}(k) - \tilde{P}^n_{ij}(k) \right] \left[ P^n_{mn}(k') - \tilde{P}^n_{mn}(k') \right]  }      \quad.
\end{align}
Here $P^n_{ij}$ is the true simulation measurement, and $\tilde{P}^n_{ij}$ is prediction in the $n$-th LOO estimation. We introduce additional $\sigma_{ijmn}(k,k')$ to avoid large uncertainties when $P_{ij}$ in the denominator is closed to zero, which would be unreasonable since their absolute deviation $\delta P_{ij}$ is always small. We choose $\alpha=0.01$, but the result is insensitive to the $\alpha$ value and the $\alpha\,\beta_{ijmn}$ term only works as $P_{ij}\rightarrow 0$. By setting $\alpha=0$ and removing the absolute value operation in the denominator, Eq.~\ref{equ:emu-err-syst-2} revert to the LOO error definition in Fig.~\ref{fig:LOO}. We further assume that the systematic and statistical uncertainties are independent, so that the total covariance is 
\begin{align}
\label{equ:emu-err-tot}
{\bf\rm Cov}_{\rm emu} \left[ P_{ij}(k), P_{mn}(k')\right] 
= {\bf\rm Cov}_{\rm stat} \left[ P_{ij}(k), P_{mn}(k')\right] 
+ {\bf\rm Cov}_{\rm syst} \left[ P_{ij}(k), P_{mn}(k')\right]   \quad .
\end{align}
In Fig.~\ref{fig:error_estimate}, we show the fraction of systematic uncertainty up to the quadratic order. The systematic uncertainty is similar to the LOO error in Fig.~\ref{fig:LOO}, as we have only removed the extremely large uncertainty arising purely from the definition. It dominates over the sample variance, particularly on small scales, while the total uncertainty remains at $1\sim 2\%$ level. 
Though the total uncertainty seems large in some high order spectra, such as $P_{s^2 1}$, they are subdominant in the final galaxy power spectrum emulation, shown in Fig.~\ref{fig:error_estimate-2}. The leading order spectra dominate the uncertainty in the galaxy power spectrum, but the relative contribution remains below one percent. The uncertainty from quadratic order spectra only contributes a fraction of $10^{-3}\sim 10^{-5}$ even for the lowest redshift. 

}

% ---------------------------------------------------------------------------------------
% ---------------------------------------------------------------------------------------

\section{Gaussian Process Emulation}
\label{sec:introduction_GP}

The high-dimensionality interpolation of the principal components is realized by the Gaussian process regression \cite{GPR-MIT}. We summarize the contents in this section. 

The general formulation states that the mapping function $f$ obeys the multi-dimensional Gaussian distribution
\begin{equation}
f(\hat\theta) \sim \mathcal{GP}(\, m(\hat\theta), K(\hat\theta,\hat\theta')\,) \ ,
\end{equation}
where $m(\hat\theta)$ is the mean function and $K(\hat\theta,\hat\theta')$ is the covariance function of data at points $\hat\theta$ and $\hat\theta'$. Without loss of generality, we can always normalize the dataset so that $m(\theta)=0$. Given the training data $w=f(\hat\theta)+\hatn$ at point $\hat\theta$ with assumed Gaussian noise $\hatn\sim \mathcal{N}(0, \sigma_n^2)$, we can obtain the joint distribution of training data and the predicted data $y_*=f(\hat\theta_*)$ at given test point $\hat\theta_*$, 
\begin{equation}
\label{equ:gpr-pri}
\begin{bmatrix} w \\  y_* \end{bmatrix}
\sim \mathcal{N} 
\left( 
0, 
\begin{bmatrix}
K(\hat\theta, \hat\theta) + \sigma_n^2 I & K(\hat\theta, \hat\theta_*) \\ 
K(\hat\theta_*, \hat\theta) & K(\hat\theta_*, \hat\theta_*)
\end{bmatrix}
\right)\ .
\end{equation}
The conditional distribution and its realization mean and variance are
\begin{align}
& \label{equ:gpr-1} 
y_* \mid \hat{\theta}, w, \hat{\theta}_* \sim \mathcal{N} \left( \mu_*, \sigma_*^2 \right)    \quad,  \\
& \label{equ:gpr-2}
\mu_* = K(\hat{\theta}, \hat{\theta}_*)^\top \left[ K(\hat{\theta}, \hat{\theta}) + \sigma_n^2 I \right]^{-1} w(\hat{\theta})    \quad,  \\
& \label{equ:gpr-3}
\sigma_*^2 = K(\hat{\theta}_*, \hat{\theta}_*) - K(\hat{\theta}_*, \hat{\theta}) \left[ K(\hat{\theta}, \hat{\theta}) + \sigma_n^2 I \right]^{-1} K(\hat{\theta}, \hat{\theta}_*)   \quad. 
\end{align}
After selecting the appropriately parameterized kernel function, we can determine the optimal hyperparameters by maximizing the logarithm marginal likelihood
\begin{equation}
\ln \mathcal{L} = 
- \frac{1}{2} w(\hat{\theta})^T \left[ K(w(\hat{\theta}), w(\hat{\theta})) + \sigma_n^2 I \right]^{-1} w(\hat{\theta})
- \frac{1}{2} \log \left| K(w(\hat{\theta}), w(\hat{\theta})) + \sigma_n^2 I \right|
- \frac{n}{2} \log 2\pi \ .
\end{equation}
{However, applying Eq.~\ref{equ:gpr-2} for emulation with a rigorous interpretation of Eq.~\ref{equ:gpr-pri} is challenging in this work, as it requires a proper estimation of the noise level in the SVD coefficients Eq.~\ref{equ:PCA}, which demands an additional large suite of simulations with varying random phases. However, since Eq.~\ref{equ:gpr-2} represents a linear weighting of training data, it can be interpreted as the linear decomposition into basis functions $K(\hat\theta, \hat\theta')$ \cite{bakx2025optimal}, where $\sigma_n^2$ is introduced to maintain the numerical stability of matrix inversion. It abandons the interpretation of Eq.~\ref{equ:gpr-3} as the emulation uncertainty, as we have estimated the uncertainty in Appendix~\ref{append:uncertainty}. Moreover, all the off-diagonal components of the basis spectra covariance are neglected in Eq.~\ref{equ:gpr-3}, but estimated in Eq.~\ref{equ:emu-err-tot}. 
}

The commonly used kernel functions are the constant kernel, radial basis function (RBF), dot-product kernel (DP), and Matérn kernel, specified as follows, 
\begin{align}
& K_{\text{const}}(\hat{\theta}_i, \hat{\theta}_j) = {\rm c}   \ , \\
& K_{\text{RBF}}(\hat{\theta}_i, \hat{\theta}_j) = \exp\left( -\frac{r^2}{2l^2} \right)    \ , \\
& K_{\text{DP}}(\hat{\theta}_i, \hat{\theta}_j) = \sigma^2_0 + \hat\theta_i\cdot\hat\theta_j     \ ,\\
& K_{\text{M3/2}}(\hat{\theta}_i, \hat{\theta}_j)  =  \left( 1+ {\sqrt{3}r\over l}\right) \exp\left(-{\sqrt{3}r\over l}\right)\ ,
\end{align}
where $r=d(\hat{\theta}_i, \hat{\theta}_j)$ is the Euclidean distance between two data points and $l$ is the length scale. Here $k_{\text{M3/2}}$ is Matérn kernel with $\nu=3/2$. 
\begin{equation}
k_{\text{Matern}}(\hat{\theta}_i, \hat{\theta}_j)  =  {2^{1-\nu}\over\Gamma(\nu)} \left({\sqrt{2\nu}\, r \over l}\right)^\nu K_\nu \left({\sqrt{2\nu}\, r\over l}\right)\ ,
\end{equation}
where $K_\nu$ is the modified Bessel function with parameter $\nu$. 

The training and prediction process is operated in Python library \scikitlearn . For emulating the basis spectra from simulation measurement, we adopt the product of constant kernel and the radial basis function kernel, $K(\hat{\theta}_i, \hat{\theta}_j) = K_{\text{const}}K_{\text{RBF}}(\hat{\theta}_i, \hat{\theta}_j)$.
%constant kernel parameter as ${\rm c} = 2$, and the RBF kernel as length scale as $l=5$. 
For emulating the theoretical template basis spectra, we adopt the kernel $K(\hat{\theta}_i, \hat{\theta}_j) = K_{\text{DP}}(\hat{\theta}_i, \hat{\theta}_j) K_{\text{M3/2}}(\hat{\theta}_i, \hat{\theta}_j)$.
% where the length scale kernel are fixed as $l=0.1$ and $l=3$ respectively. We find the emulation accuracy is insensitive to the hyperparameters, so that not further fine-tuning is operated. 
{The interpolation parameter is fixed as the default value $\sigma^2_n=10^{-10}$ for all the basis spectra emulation.}

% ---------------------------------------------------------------------------------------
% ---------------------------------------------------------------------------------------
% ---------------------------------------------------------------------------------------
% ---------------------------------------------------------------------------------------
% \end{widetext}

% % Create the reference section using BibTeX:
% \bibliographystyle{apsrev4-2}
% \bibliography{citations}

\end{document}